\documentclass[aps,amsmath,twocolumn,floatfix,superscriptaddress,prb,footinbib,10pt]{revtex4-2}

\usepackage{amsmath,amssymb,bm,mathtools,mathrsfs,comment}
\usepackage{setspace}[10]
\usepackage[italicdiff]{physics}
\usepackage{siunitx}
\AtBeginDocument{\RenewCommandCopy{\qty}{\SI}}
\usepackage{physics2}
\usepackage{graphicx}
\usepackage[usenames, dvipsnames]{color}
\usepackage{xcolor}
\usepackage{tabularray}
\PassOptionsToPackage{hyphens}{url}%url wrapping
\usepackage{hyperref}
\usepackage[nameinlink]{cleveref}% autorefよりベター
\usepackage[version=3]{mhchem}

\usephysicsmodule{ab}
\usephysicsmodule{ab.braket}
\hypersetup{% hyperref
	setpagesize=false,
	bookmarksnumbered=true,%
	bookmarksopen=true,%
	colorlinks=true,%
	linkcolor=blue,
	citecolor=red,
}

\graphicspath{{fig/}{./fig/}}

\allowdisplaybreaks

\crefname{equation}{Eq.}{Eqs.}% {環境名}{単数形}{複数形} \crefで引くときの表示
\crefname{figure}{Fig.}{Figs.}% {環境名}{単数形}{複数形} \crefで引くときの表示

\Crefname{equation}{Equation}{Equations}% {環境名}{単数形}{複数形} \crefで引くときの表示
\Crefname{figure}{Figure}{Figures}% {環境名}{単数形}{複数形} \crefで引くときの表示

\newcommand{\bk}{{\bm{k}}}
\newcommand{\bK}{{\bm{K}}}
\newcommand{\br}{{\bm{r}}}

\newcommand{\bd}{{\bm{d}}}
\newcommand{\ah}{\hat{a}}
\newcommand{\bh}{\hat{b}}

\newcommand{\bj}{{\bm{J}}}
\newcommand{\hbj}{{\hat{\bm{J}}}}

\newcommand{\brone}{{\br+\bd_{1}}}
\newcommand{\brj}{{\br+\bd_{j}}}

\newcommand{\dc}{{\mathrm{dc}}}
\newcommand{\ac}{{\mathrm{ac}}}

\newcommand{\edc}{E_\dc}
\newcommand{\eac}{E_\ac}

\newcommand{\ddc}{\delta_\dc}
\newcommand{\ddcv}{\bm{\delta}_\dc}

\newcommand{\tini}{{t_{\mathrm{ini}}}}

\newcommand{\bkd}{{\bm{k}+\ddcv}}

\begin{document}
\title{High-harmonic generation in graphene under the application of a DC electric current: \texorpdfstring{\\
        From perturbative to nonperturbative regimes}{From perturbative to nonperturbative regimes}}
\author{Minoru Kanega}
\email{m.kanega.phys@chiba-u.ac.jp}
\affiliation{Department of Physics, Chiba University, Chiba 263-8522, Japan}
\author{Masahiro Sato}
\email{sato.phys@chiba-u.ac.jp}
\affiliation{Department of Physics, Chiba University, Chiba 263-8522, Japan}

\date{\today}
\begin{abstract}
    We theoretically investigate high-harmonic generation (HHG) in honeycomb-lattice graphene models when subjected to a DC electric field.
    By integrating the quantum master equation with the Boltzmann equation,
    we develop a numerical method to compute laser-driven dynamics in many-electron lattice systems under DC electric current.
    The method enables us to treat both the weak-laser (perturbative) and intense-laser (nonperturbative) regimes in a unified way,
    accounting for the experimentally inevitable dissipation effects.
    From it, we obtain the HHG spectra and analyze their dependence on laser frequency, laser intensity, laser-field direction, and DC current strength.
    We show that the dynamical and static symmetries are partially broken by a DC current or staggered potential term,
    and such symmetry breakings drastically change the shape of the HHG spectra,
    especially in terms of the presence or absence of $(2n+1)$th-, $2n$th-, or $3n$th-order harmonics ($n\in \mathbb Z$).
    The laser intensity, frequency, and polarization are also shown to affect the shape of the HHG spectra.
    Our findings indicate that HHG spectra in conducting electron systems can be quantitatively or qualitatively controlled by tuning various external parameters,
    and DC electric current is used as such an efficient parameter.
\end{abstract}
\maketitle

%%%%%%%%%%%%%%%%%%%%%%%%%%%%%%%%%%%%%%%%%%%%%%%%%%%%%%%%%%%%%%%%%
%%%%%%%%%%%%%%%%%%%%%%%%%%%%%%%%%%%%%%%%%%%%%%%%%%%%%%%%%%%%%%%%%
%%%%%%%%%%%%%%%%%%%%%%%%%%%%%%%%%%%%%%%%%%%%%%%%%%%%%%%%%%%%%%%%%
\section{Introduction}\label{Sec:Intro}

\begin{figure}[tb]
    \centering
    \includegraphics[width=\linewidth]{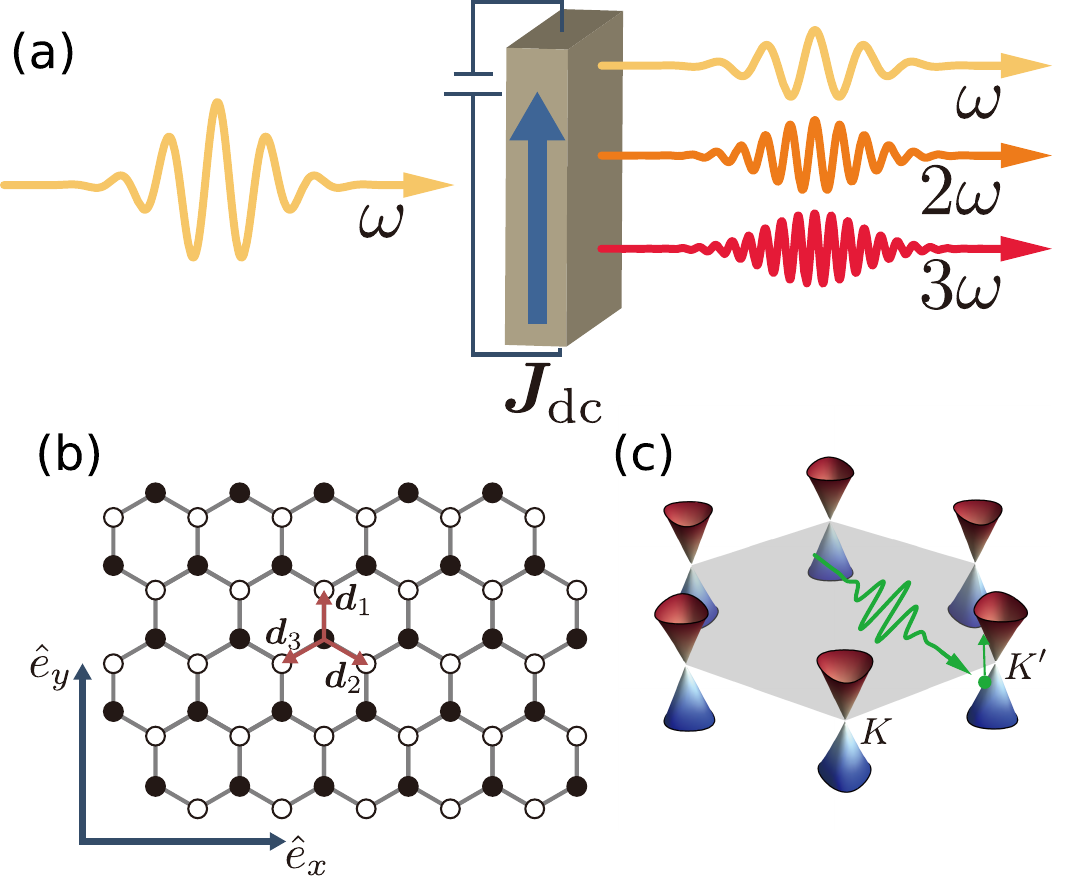}
    \caption{
        (a) Schematic diagram of high-harmonic generation in solids under the application of a DC current.
        (b) Top view of the two-dimensional honeycomb lattice of graphene, whose unit cell includes A- and B-sublattice sites (black and white circles). $x$ and $y$ axes are respectively defined by the unit vectors ${\hat e}_x$ and ${\hat e}_y$.
        Vectors $\bd_1$, $\bd_2$, and $\bd_3$ connect a site on A sublattice and one of the three nearest neighboring sites.
        (c) A typical electronic band structure of graphene in the wave-vector space.
        Blue and red Dirac cones respectively represent the valence and conduction bands, and they are located at $\bm{K} = \frac{2\pi}{3\sqrt 3 a}(1,\sqrt 3)$ and $\bm{K}' = \frac{2\pi}{3\sqrt 3 a}(-1,\sqrt 3)$ points.
    }
    \label{fig:figure1}
\end{figure}

In the past few decades,
nonlinear optical responses in solid-state electronic systems have seen remarkable growth thanks to the development of laser techniques.
Various laser-driven nonequilibrium phenomena have been explored
including high-harmonic generation (HHG)~\cite{Ghimire2019Highharmonic, Yue2022Introduction, Li2023High, Bhattacharya2023Strong, Hirori2024HighOrder},
photo-rectification effects~\cite{Young2012First, Tan2016Shift, Morimoto2016Topological, Tokura2018Nonreciprocal, Ishizuka2019Rectification, Sturman1992Photovoltaic, Watanabe2021Chiral, Ishizuka2024Peltier},
Floquet engineering~\cite{Eckardt2015Highfrequency, Mikami2016BrillouinWigner, Mori2016Rigorous, Kuwahara2016Floquet, Eckardt2017Colloquium, Oka2019Floquet, Sato2021Floquet},
and others.
Among them, HHG is a simple phenomenon in which a system subjected to intense light of frequency $\Omega$ emits light with different frequencies $n\Omega$, as shown in Fig.~\ref{fig:figure1}(a).
It is relatively easily detectable in experiments compared to other nonlinear optical effects.
Though the HHG research had
focused on atomic gas systems in the 1990s~\cite{McPherson1987Studies, Ferray1988Multipleharmonic, Krause1992Highorder, Corkum1993Plasma, Schafer1993threshold, Macklin1993Highorder},
its targets have been expanded to solid-state systems since the 2010s, such as semiconductors~\cite{Ghimire2011Observation, Schubert2014Subcycle, Vampa2015Semiclassical, Luu2015Extreme, Liu2017Generation, You2017Anisotropic, Vampa2018Strongfield, Xia2018Nonlinear, Yoshikawa2019Interband, Lakhotia2020Laser},
superconductors~\cite{Matsunaga2014Lightinduced},
semimetals~\cite{Yoshikawa2017Highharmonic, Hafez2018Extremely, Kovalev2020Nonperturbative, Cheng2020Efficient},
strongly correlated electrons~\cite{Murakami2018HighHarmonic,Bionta2021Tracking,Granas2022Ultrafast,Uchida2022HighOrder},
magnetic insulators~\cite{Baierl2016TerahertzDriven, Lu2017Coherent, Ikeda2019Highharmonic, Kanega2021Linear, Zhang2023Generation, Takayoshi2019Highharmonic},
etc.

It is well known that
even-order harmonics of $n=2, 4, 6, \ldots$
are all generally suppressed in solid-state electronic systems with spatial inversion symmetry~\cite{Neufeld2019Floquet}.
This fact leads to an intriguing question:
How can one observe/control these suppressed responses?
HHG also provides a means to extract light of particular beneficial frequencies. Therefore, addressing the above question becomes vital from both scientific and application perspectives.
One viable strategy is to use inversion-asymmetric systems,
like $p$--$n$ junctions~\cite{Sze2006Physics},
perovskite ferroelectrics~\cite{Li2021Photoferroelectric},
and Weyl semimetals~\cite{Wu2017Giant, Patankar2018Resonanceenhanced, Osterhoudt2019Colossal, Sirica2019Tracking}.
The inversion asymmetry in these systems generates even-order harmonics in general, and
this research direction has long been thriving.
For instance, $p$--$n$ junctions show potential for solar energy conversion through the $n=0$ light-induced electric potential~\cite{Sze2006Physics}.
Similarly, the nonlinear optical responses in Weyl semimetals are subjects of intensive
study~\cite{Wu2017Giant, Patankar2018Resonanceenhanced, Osterhoudt2019Colossal, Sirica2019Tracking}.

On the other hand, even for inversion-symmetric materials,
some extrinsic means can be employed to break the inversion symmetry.
Applying a DC current is an effective way to achieve the breakdown.
Building on this idea,
current-induced second-harmonic generation has been investigated theoretically~\cite{Khurgin1995Current, Wu2012QuantumEnhanced, Cheng2014DC, Takasan2021Currentinduced, Gao2021Currentinduced}
and reported experimentally in materials like \ce{Si}~\cite{Aktsipetrov2009DCinduced},
\ce{GaAs}~\cite{Ruzicka2012SecondHarmonic},
graphene~\cite{Bykov2012Second, An2013Enhanced},
superconducting \ce{NbN}~\cite{Nakamura2020Nonreciprocal},
and others.

It is noteworthy that both DC current and laser light push the system out of equilibrium.
Namely, the application of DC current generally increases complexity in laser-driven systems, making computational predictions daunting.
Consequently, so far, only second-harmonic generation spectra have been %perturbatively
computed within perturbative ways in most of the previous works for systems under the application of both laser and DC current~\cite{Khurgin1995Current, Wu2012QuantumEnhanced, Cheng2014DC, Takasan2021Currentinduced, Gao2021Currentinduced}.
However, the perturbation theories generally become less feasible when laser intensity grows.
Therefore, for HHG in DC-current-driven systems, it is significant to develop a theoretical method that analyzes both perturbative (weak laser) and nonperturbative (strong laser) ranges.

In this paper,
motivated by the above backgrounds,
we theoretically investigate HHG in honeycomb-lattice graphene models when subjected to a DC electric field.
Combining the quantum master equation~\cite{Gorini1976Completely, Lindblad1976generators, Breuer2007Theory, Ikeda2019Highharmonic, Sato2020Twophoton, Kanega2021Linear}
with the Boltzmann equation~\cite{Abrikosov2017Fundamentals},
we develop numerical methods to quantitatively compute the laser-driven time evolution of observables in many-electron lattice systems under DC electric current,
and we demonstrate the HHG spectra and their dependence on laser frequency, laser intensity, and DC current strength,
while accounting for the experimentally inevitable dissipation effects.
Our findings indicate that the HHG spectra undergo significant modifications due to dynamical symmetry breaking induced by the applied DC current.
Additionally, we observe that nonperturbative effects become pronounced when the laser is strong,
leading to a marked alteration in the laser frequency dependence of the HHG spectra.

The remaining part of the paper is organized as follows.
In Sec.~\ref{Sec:Model},
we introduce the formalism of the combination with the master equation and the Boltzmann equation,
which describes the time evolution of laser and DC-current induced nonequilibrium states in many-electron systems.
We comment on some advantages of the master equation in Sec.~\ref{subsec:advantage}.
The main numerical results based on the master equation are in Sec.~\ref{sec:result}.
Section~\ref{subsec:DC-strength} reveals that the HHG spectrum is modulated by a DC current,
leading to even-order harmonic responses.
The section also delves into the DC current dependence of the HHG spectra.
In Sec.~\ref{subsec:frequency},
we explore the influence of laser frequency on the HHG spectra,
emphasizing the pronounced differences between the perturbative and nonperturbative regimes.
Furthermore, we underscore the significant role of intra-band dynamics in the nonperturbative regime.
Section~\ref{subsec:intensity} examines the relationship between laser intensity and the HHG spectra,
showing that the crossover between the perturbative and nonperturbative regimes occurs depending on the chemical potential.
Section~\ref{subsec:bandgap} discusses the characteristics of the HHG spectrum in the presence of a staggered potential (i.e., the effects of a finite band gap).
When the system is subjected to an intense laser,
the interplay between inter- and intra-band dynamics in different polarization directions results in intricate shifts in the HHG spectra.
Section~\ref{subsec:strong-field} outlines the variations in the HHG spectra under extremely high laser intensities.
Finally, in Sec.~\ref{sec:conclusions}, we summarize our results and make concluding remarks. We discuss some theoretical details associated with dynamical symmetry and the master equation in the Appendixes.

%%%%%%%%%%%%%%%%%%%%%%%%%%%%%%%%%%%%%%%%%%%%%%%%%%%%%%%%%%%%%%%%%
%%%%%%%%%%%%%%%%%%%%%%%%%%%%%%%%%%%%%%%%%%%%%%%%%%%%%%%%%%%%%%%%%
%%%%%%%%%%%%%%%%%%%%%%%%%%%%%%%%%%%%%%%%%%%%%%%%%%%%%%%%%%%%%%%%%
\section{Model and Method}\label{Sec:Model}

\subsection{Model and observable}
\label{subsec:modelObs}

We focus on a model of single-layered graphene~\cite{CastroNeto2009electronic, Aoki2014Physics} with A and B sublattices [black and white circles in Fig.~\ref{fig:figure1}(b)].
The tight-binding Hamiltonian is given by
\begin{align}
    \hat{H}_0= & -t_0\sum_{\bm{r}}\sum_{j=1,2,3}\ab(\bh_\brj^\dag\ah_\br + \ah_\br^\dag\bh_\brj)\notag \\
               & + \Delta\sum_\br\ab(\ah^\dag_\br\ah_\br - \bh^\dag_\brone\bh_\brone),
\end{align}
where $\bd_j$, the position vectors pointing to the three nearest-neighbor sites from each A site of the hexagonal plaquette, are given by
\begin{align}
    \begin{aligned}
        \bd_1 & = a\ab(0, 1),                                             \\
        \bd_2 & = a\ab(\cos\ab(\frac{\pi}{6}), -\sin\ab(\frac{\pi}{6})),  \\
        \bd_3 & = a\ab(-\cos\ab(\frac{\pi}{6}), -\sin\ab(\frac{\pi}{6})),
    \end{aligned}
\end{align}
with $a$ being a lattice constant.
The vector $\br$ represents each position of sublattice A.
The fermionic operator $\hat a_\br (\hat a_\br^\dag)$ annihilates (creates) an electron at position $\br$ for the sublattice A.
The fermionic one $\hat b_\br (\hat b_\br^\dag)$ is defined similarly for the sublattice B.
They satisfy the anti-commutation relations
$\acomm{\hat{a}_\br}{\hat{a}_{\br'}}=\acomm{\hat{b}_\br}{\hat{b}_{\br'}}=\acomm{\hat{a}_\br}{\hat{b}_{\br'}}=\acomm{\hat{a}_\br}{\hat{b}^\dag_{\br'}}=0$
and
$\acomm{\hat{a}_\br}{\hat{a}^\dag_{\br'}}=\acomm{\hat{b}_\br}{\hat{b}^\dag_{\br'}}=\delta_{\br,\br'}$.
The first term in $\hat{H}_0$ describes the nearest-neighboring electron hopping between two sublattices A and B with transfer integral $t_0$,
and the second term represents an onsite staggered potential with a band-gap energy $\Delta$.
For graphene, transfer integral $t_0$ is estimated as $t_0=\SI{2.7}{eV}$~\cite{Reich2002Tightbinding, CastroNeto2009electronic} and $\Delta$ is usually negligible.

By the Fourier transformation,
$\hat a_\br = \frac{1}{\sqrt{N}}\sum_{\bk}\tilde a_\bk e^{-i\bk\cdot\br},\;\hat b_\br = \frac{1}{\sqrt{N}}\sum_{\bk}\tilde b_\bk e^{-i\bk\cdot\br}$,
where $N$ is the total number of unit cells,
the Hamiltonian $\hat{H}_0$ is expressed in the following bilinear form:
$\hat{H}_0=\sum_\bk\bm{C}^\dag_\bk M(\bk)\bm{C}_\bk$,
where $\bm{C}_\bk=\mqty(\tilde{a}_\bk & \tilde{b}_\bk)^\top$
and $M(\bk)=h^x_\bk\sigma_x+h^y_\bk\sigma_y+h^z_\bk\sigma_z$ with $(h^x_\bk,h^y_\bk,h^z_\bk)=(-t_0\sum_{j=1}^3\cos(\bk\cdot\bm{d}_j),t_0\sum_{j=1}^3\sin(\bk\cdot\bm{d}_j),\Delta)$ and $\sigma_{x,y,z}$ being Pauli matrices.
Through unitary transformation $\mqty(\xi_\bk & \zeta_\bk)=U^\dag_\bk\bm{C}_\bk$ with new fermion operators $\xi_\bk$ and $\zeta_\bk$,
the Hamiltonian is diagonalized as $\hat{H}_0=\sum_\bk \varepsilon (\bk)(\xi^\dag_\bk\xi_\bk-\zeta^\dag_\bk\zeta_\bk)$,
with the energy dispersion $\varepsilon (\bk)=\sqrt{(h^x_\bk)^2+(h^y_\bk)^2+(h^z_\bk)^2}$.
This dispersion has the Dirac cones at the $\bm{K} = \frac{2\pi}{3\sqrt 3 a}(1,\sqrt 3)$ and $\bm{K}' = \frac{2\pi}{3\sqrt 3 a}(-1,\sqrt 3)$ points when $\Delta=0$ as shown in Fig.~\ref{fig:figure1}(c).
We note that around the $\bm{K}$ ($\bm{K}'$) point,
the energy dispersion is approximated to $E_{\bm{K}(\bm{K}')+\bk}\approx\pm\hbar v_F\abs{\bm{k}}$ with the Fermi velocity $v_F=\sqrt{3}at_0/(2\hbar)$ and Dirac's constant $\hbar$.
We set $\hbar=1$ throughout the paper.

The operators $\ab\{\xi_\bk\}$ and $\ab\{\zeta_\bk\}$ respectively correspond to fermionic annihilation operators on the conduction and the valence bands.
When a chemical potential $\mu=0$ [see the left panel of Fig.~\ref{fig:figure2}(a)],
the ground state is given by $\ket | \mathrm{gs} > =\otimes_\bk\ket | g_\bk > $,
where $\ket | g_\bk > =\zeta^\dag_\bk\ket | 0_\bk >=(-v_\bk\tilde{a}^\dag_\bk+u_\bk\tilde{b}^\dag_\bk)\ket | 0_\bk >$,
with $(u_\bk,v_\bk)=(\sqrt{(1+h^z_\bk/\varepsilon (\bk))/2},\Delta_\bk\sqrt{(1-h^z_\bk/\varepsilon (\bk))/2}/\abs{\Delta_\bk})$, $\Delta_\bk=h^x_\bk-ih^y_\bk$,
and $\ket | 0_\bk >$ is the Fock vacuum for electrons $(\zeta_\bk, \xi_\bk)$.
The one-electron state occupying the conduction band at wave vector $\bk$ is given by $\ket | e_\bk > =\xi^\dag_\bk\ket | 0_\bk >=(u_\bk\tilde{a}^\dag_\bk+v^*_\bk\tilde{b}^\dag_\bk)\ket | 0_\bk >$.
As one will see later,
we use $(\ket |e_\bk>,\ket |g_\bk>)$ as the basis of time evolution.
In the present work, as we will explain in more detail in Sec.~\ref{subsec:boltzmann},
we focus on the optical response in a low- or intermediate-energy regime with $0.2\lesssim \mu/t_0 \lesssim 1.0$, in which ``Dirac'' electrons around $\bm{K}$ and $\bm{K}'$ are mainly photo-excited.

We adopt the Peierls-phase formalism to consider the laser-driven dynamics.
The time-dependent Hamiltonian with light-matter coupling is given by
\begin{align}
    \hat{H}(t)= & -\sum_{\bm{r},j}\ab(t_{\br,\brj}(t)\bh_\brj^\dag\ah_\br + t_{\brj,\br}(t)\ah_\br^\dag\bh_\brj)\notag \\
                & + \Delta\sum_\br\ab(\ah^\dag_\br\ah_\br - \bh^\dag_\brone\bh_\brone),
    \label{eq:time-dependent-hamiltonian}
\end{align}
where $t_{\br,\brj}(t)$ is the hopping amplitude with Peierls substitution
\begin{equation}
    t_{\br,\brj}(t) = t_0\exp\ab(-ie\int_\br^\brj \bm{A}(t
        )\cdot d\br).
\end{equation}
Here, $e$ is the elementary charge, $\bm{A}(t)$ is a vector potential,
and the AC electric field of laser $\bm{E}(t)$ in the Coulomb gauge % velocity gauge
is related to $\bm{A}(t)$ via the relation $\bm{E}(t)=-\pdv{\bm{A}(t)}{t}$.
The Fourier-space representation of the Hamiltonian $\hat{H}(t)$ is given by
$\hat{H}(t)=\sum_\bk\hat{H}_\bk(t)=\sum_\bk\bm{C}^\dag_\bk M(\bk+e\bm{A}(t))\bm{C}_\bk$.
We note that, though under the irradiation of a laser, the Hamiltonian $\hat{H}(t)$ is still $\bk$-diagonal like $\hat{H}_0$. This is because we now simply apply a spatially uniform laser to the graphene model.
Therefore, as we will discuss in Sec.~\ref{subsec:time-evolution},
we can independently compute the time evolution of the density matrix at each $\bk$ space,
and sum them up in the whole Brillouin zone to represent the time evolution of the entire system.

In this paper,
we focus on the HHG spectra driven by laser pulse (not continuous wave) with angular frequency $\Omega$ because intense laser pulses are usually used in experiments.
Hereafter, we simply refer to the angular frequency as frequency unless otherwise noted. In the Coulomb gauge, the vector potential of a laser pulse with frequency $\Omega$ is defined as
\begin{equation}
    \bm{A}(t)=\frac{E_\ac}{\Omega\sqrt{1+\epsilon^2}}f_{\mathrm{env}}(t)\mqty(\cos(\Omega t)\\\epsilon\sin(\Omega t)),
    \label{eq:VectorPotential}
\end{equation}
where $E_\ac$ is the strength of the AC electric field of the pulse,
and $f_{\mathrm{env}}(t)$ is a Gaussian envelope function $f_{\mathrm{env}}(t)=\exp[-2(\ln 2)(t^2/t^2_{\mathrm{FWHM}})]$ with full width at half-maximum $t_{\mathrm{FWHM}}$.
To fix the pulse width,
we adopt the five-cycle period of laser at $\Omega=0.2t_0$ as the standard of $t_{\mathrm{FWHM}}$.
The dimensionless parameter for the field strength $E_\ac$ is given by $e\eac a/t_0$:
For graphene,
$e\eac a/t_0=0.01$ corresponds to $\eac\sim\SI{1.1}{MV/cm}$.
The laser ellipticity $\epsilon$ denotes the degree of laser polarization:
$\epsilon=0$ means a linear polarization along the $x$-axis, while $\epsilon=\pm1$ denotes a circular one.

To analyze the nonlinear optical response,
we consider the electric current in the whole system,
\begin{align}
    \hbj(t) & = \pdv{\hat{H}(t)}{\bm{A}(t)} \notag                                                                                                         \\
            & = \sum_{\bk}\sum_{\alpha,\beta}C^\dag_{\alpha,\bk} \bm{\mathcal J}_{\alpha\beta}(\bk+e\bm{A}(t))C_{\beta,\bk}\eqqcolon\sum_{\bk}\hbj_\bk(t),
\end{align}
and the expectation value per unit cell,
$\bj(t) = \frac{1}{N}\sum_\bk \bm{J}_\bk(t)=\frac{1}{N}\sum_\bk\braket < \hbj_\bk(t) >_t$,
as an observable of interest.
Here, $N$ is the system size,
$\bm{\mathcal J}_{\alpha\beta}(\bk+e\bm{A}(t)) = \pdv{\bm A}M_{\alpha\beta}(\bk+e\bm{A}(t)) = \bm j^x_{\bk+e\bm{A}(t)}(\sigma_x)_{\alpha\beta}+\bm j^y_{\bk+e\bm{A}(t)}(\sigma_y)_{\alpha\beta}+\bm j^z_{\bk+e\bm{A}(t)}(\sigma_z)_{\alpha\beta}$ with $(\bm j^x_\bk,\bm j^y_\bk,\bm j^z_\bk)=(et_0\sum_{\ell=1}^3\bm d_\ell\sin(\bk\cdot\bm{d}_\ell),et_0\sum_{\ell=1}^3\bm d_\ell\cos(\bk\cdot\bm{d}_\ell),\bm{0})$,
and $\braket < \cdots >_t={\rm Tr}[\hat\rho(t)\cdots]$
denotes the expectation value for a density matrix $\hat\rho(t)$.
Note that this paper concentrates on laser application to nonequilibrium steady states with a steady current in the graphene model.
We are therefore interested in the difference between the electric currents of a laser-irradiated state and the initial steady one.
When $\bj(t)$ evolves in time,
it becomes a source of electromagnetic radiation.
The radiation is known to be proportional to $d\bj(t)/dt$ within the dipole radiation approximation, and
the normalized radiation power spectrum of high-order harmonics (i.e., HHG spectrum) at frequency $\omega$ is given by~\cite{Jackson1998}
\begin{align}
    I_{x,y}(\omega) = \abs{\omega J_{x,y}(\omega)}^2,
\end{align}
where $J_{x}(\omega)$ and $J_y(\omega)$ are the Fourier components of $\bj(t)=(J_x(t), J_y(t))$ in the temporal direction [see Fig.~\ref{fig:figure1}(b)].
To find characteristic features of the HHG spectra, we will also estimate the power spectrum of the $n$th-order harmonics, which is defined as
\begin{align}
    \tilde I_{x,y}(n\Omega) = \int_{(n-\frac{1}{2})\Omega}^{(n+\frac{1}{2})\Omega}\dd\omega I_{x,y}(\omega).
\end{align}

%%%%%%%%%%%%%%%%%%%%%%%%%%%
%%%%%%%%%%%%%%%%%%%%%%%%%%%
%%%%%%%%%%%%%%%%%%%%%%%%%%%
\subsection{Current-induced steady state}\label{subsec:boltzmann}

\begin{figure}[tb]
    \centering
    \includegraphics[width=\linewidth]{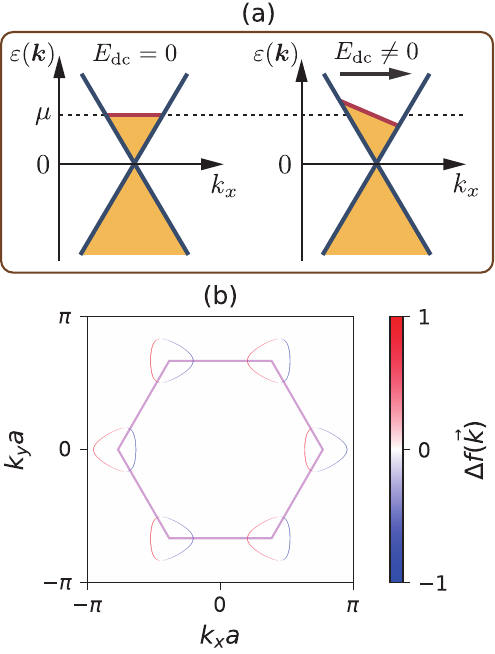}
    \caption{
    (a) $\bk$-space schematic image of DC-current-induced steady state and equilibrium state near a Dirac point on $k_y=0$ in the case of zero temperature, $\mu>0$ and $\Delta=0$.
    DC electric field ${\bm E}_{\rm dc}$ is parallel to the $x$ axis.
    Yellow region means that electrons are occupied.
    (b) The difference $\Delta f(\bk)=f_{\mathrm{ss}}(\bk)-f_0(\bk)$ in the momentum space at zero temperature.
    Parameters are set to $\mu/t_0=0.7$,  $\delta_\dc a=0.01$, and $\Delta=0$.
    }
    \label{fig:figure2}
\end{figure}

To consider the current-induced steady state,
we employ the Boltzmann equation approach~\cite{Abrikosov2017Fundamentals}.
The Boltzmann equation under the application of a static electric field $\bm{E}_\dc$ is given by
\begin{equation}
    \pdv{f(\bk,t)}{t}-e\bm{E}_\dc\cdot\pdv{f(\bk,t)}{\bk}=\ab(\pdv{f}{t})_{\mathrm{col}},
\end{equation}
where $f(\bk,t)$ is a nonequilibrium distribution function for electrons,
and $\ab(\pdv{f}{t})_{\mathrm{col}}$ is a collision term.
Using the relaxation-time approximation and assumption of a system being steady state (i.e., $\pdv{f}{t}=0$),
the Boltzmann equation leads to the steady-state distribution function
\begin{equation}\label{eq:boltzmann}
    f_{\mathrm{ss}}(\bk)=f_0(\bk)+e\tau\bm{E}_\dc\cdot\pdv{f_{\mathrm{ss}}(\bk)}{\bk}\approx f_0(\bk+\ddcv),
\end{equation}
where $f_0(\bk)$ is the Fermi distribution function $f_0(\bk)=\ab(e^{\beta(\epsilon_\bk-\mu)} + 1)^{-1}$ with $\beta$ being the inverse temperature,
$\tau$ denotes the relaxation time of an electron,
and $\ddcv=e\tau\bm{E}_\dc$.
The schematic image of the steady-state distribution is given in Fig.~\ref{fig:figure2}(a).
Hereafter, we take the zero-temperature limit (i.e., $\beta\rightarrow\infty$)
and assume $\ddcv\parallel\hat e_x$ for simplicity.

When we utilize the result of the relaxation-time approximation of Eq.~\eqref{eq:boltzmann}, we should be careful about the following two points.
The first one is that we have assumed the condition of $\ddc a\ll 1$ with $\ddc=\abs{\ddcv}$ in Eq.~\eqref{eq:boltzmann}.
For graphene,
$\tau$ is estimated as $\tau\sim\SI{1}{ps}$~\cite{Hwang2008Singleparticle}, and therefore the condition $\ddc=\abs{\ddcv}$ holds if the DC electric field satisfies the inequality $\edc\ll10^4$~\si{V/cm}.
If the DC conductivity of graphene near $\mu=0$ can be estimated as the universal value $e^2/(4\hbar)$~\cite{Nair2008Fine, Aoki2014Physics},
the relation $\edc\ll10^4$~\si{V/cm} is equivalent to the condition that DC electric current is much smaller than \SI{1}{A/cm}.
In experiments, the maximum value of the observed DC current in monolayer graphene approaches $\sim\SI{30}{A/cm}$~\cite{Moser2007Currentinduced}, which is clearly outside the condition of $\edc\ll10^4$~\si{V/cm}.
However, the experimental result indicates that the condition of $\edc\ll10^4$~\si{V/cm} is easily satisfied by applying a weak DC electric field (i.e., weak voltage) to graphene.
We have performed all calculations in this weak-electric-field regime with the relation of Eq.~\eqref{eq:boltzmann}, in which the laser response is linearly proportional to the DC field, as we will show in Sec.~\ref{subsec:DC-strength}.

The second point is that
the Boltzmann equation approach, including Eq.~\eqref{eq:boltzmann}, is valid only when the chemical potential is sufficiently far from the Dirac point $\mu=0$~\cite{Takasan2021Currentinduced} and a sufficiently large Fermi surface exists.
The Boltzmann equation is a one-band effective theory that contains the intraband dynamics,
while neglecting the interband one.
When the chemical potential is close to the Dirac point or the DC current is strong,
the interband transition is not negligible,
and the Boltzmann equation approximation breaks down.
Therefore, the minimum value $\abs{\mu_0}$ exists when we apply the Boltzmann equation to graphene. The value is given by $\abs{\mu_0}\sim\abs{v_F\delta_{\rm dc}}\sim ev_F\tau \edc$.
For graphene, it is estimated as $\abs{\mu_0}\sim\SI{0.2}{eV}$ under the application of a DC electric field $\edc\sim\SI{100}{V/cm}$.

Figure~\ref{fig:figure2}(b) shows the difference between the distribution functions of nonequilibrium steady and equilibrium states,
\begin{equation}
    \Delta f(\bk)=f_{\mathrm{ss}}(\bk)-f_0(\bk),
\end{equation}
at $\ddc a=0.01$, $\mu/t_0=0.7$ and $\Delta=0$ in $\bk$ space.
This is induced by the application of a DC electric field.
The red region shows the DC-field driven change of electron occupation from a valence-band one-particle state to a two-particle one at the subspace with a wave vector $\bk$, while the blue region shows the reverse change at $\bk$, i.e., the change from a two-particle occupied state to
a one-particle one.
As we will discuss later in Sec.~\ref{subsec:frequency},
the even-order harmonics are generated only from photo excitations in the blue area of Fig.~\ref{fig:figure2}(b).

\subsection{Time evolution and photo excitations}\label{subsec:time-evolution}
When we consider the laser-driven dynamics in graphene, we set the initial state to the current-induced steady state of $f_{\mathrm{ss}}(\bk)$. In this paper, we compute the time evolution of the density matrix (not the quantum state) to describe such laser-driven dynamics, taking dissipation effects into account.

As we mentioned in Sec.~\ref{subsec:modelObs},
since the time-dependent Hamiltonian has a $\bk$-diagonal form,
we can independently solve the time evolution for each wave vector $\bk$ under the assumption that dissipation effects at $\bk$ and $\bk'$ are simply independent of each other.
We thereby introduce the $\bk$-decomposed master equation of the GKSL form~\cite{Gorini1976Completely, Lindblad1976generators, Breuer2007Theory, Ikeda2019Highharmonic, Sato2020Twophoton, Kanega2021Linear} as
the equation of motion:
\begin{align}
    \dv{\hat{\rho}_\bk(t)}{t}= & -i\comm{\hat{H}_\bk(t)}{\hat{\rho}_\bk(t)}\notag
    \\
                               & +\gamma\ab(\hat{L}_\bk\hat{\rho}_\bk(t)\hat{L}_\bk^\dag-\frac{1}{2}\acomm{\hat{L}_\bk^\dag\hat{L}_\bk}{\hat{\rho}_\bk(t)}),
    \label{eq:quantum-master-eqre}
\end{align}
where $\hat{\rho}_\bk(t)$ and $\hat{L}_\bk$ are the density matrix and the jump operator for the subspace with wave vector $\bk$, respectively. The first and second terms on the right-hand side describe the unitary and dissipative time evolutions, respectively.
The phenomenological relaxation rate $\gamma$ represents the typical relaxation time of system $\tau\sim 1/\gamma$, where we simply neglect the $\bk$ dependence of $\gamma$ in this paper.
We set $\gamma=0.1t_0$, corresponding to $\tau\sim\SI{2.4}{fs}$.
The initial current-induced steady state is described by $\hat{\rho}_\bk(\tini)=\hat{\rho}^{\mathrm{dc}}_\bk=\dyad{g_\bkd}{g_\bkd}$ at the initial time $t=\tini$.
To calculate realistic HHG spectra under the application of DC current, we set the jump operators at each $\bk$ to relax to
the steady state $f_{\rm ss}(\bk)$.
To this end, we define the jump operator as $\hat{L}_\bk=\hat{L}^{\mathrm{dc}}_\bk=\dyad{g_\bkd}{e_\bkd}$,
which induces an interband electron transition from the conduction
to the valence band.
This jump operator satisfies the detailed balance condition at zero temperature when we consider the equilibrium limit of $\ddcv\to{\bm 0}$, i.e., the absence of a DC electric field.

We note that the master equation with the above jump operator can be mapped to a so-called optical Bloch equation (see Appendix~\ref{app:relation}).
The jump operator induces both longitudinal and transverse relaxation processes in the Bloch-equation picture.

%%%%%%%%%%%%%%%%%%%%%%%%%%%%%
%%%%%%%%%%%%%%%%%%%%%%%%%%%%%
%%%%%%%%%%%%%%%%%%%%%%%%%%%%%
\subsection{Reduction of the density matrix size}

Here, we discuss how to combine the steady state of the Boltzmann equation and the master equation in the computation of the density matrix.
The main target of the present study is the case with a finite $\mu$, but first, we briefly touch on the case of $\mu=0$ and $\ddcv={\bm 0}$,
in which the valence band is completely occupied and the conducting band is empty in the initial state, namely, each subspace with $\bk$ has one electron.
Since both the graphene's tight-binding Hamiltonian and the light-matter coupling do not change the electron number, it is enough to consider
two basis states $\ket |g_\bk>$ and $\ket |e_\bk>$ at each $\bk$.
Therefore, the density matrix $\hat{\rho}_\bk(t)$ is given by a $2\times2$ form.

On the other hand, when we study the case with a Fermi surface and $\mu\neq0$ in the presence or absence of a DC electric field,
the density matrix $\hat{\rho}_\bk(t)$ seems to be a $4\times4$ form
because two-electron or completely empty states exist in a certain regime of the $\bk$ space in addition to the one-electron state.
For $\mu>0$, we have two-electron states, while empty states appear for $\mu<0$ [see Fig.~\ref{fig:figure2}(a)].
A natural set of the bases at each $\bk$ is given by
$(\ket |0_\bk>,\ket |g_\bk>,\ket |e_\bk>,\ket |2_\bk>)$, in which $\ket |0_\bk>$ is the empty state and
$\ket |2_\bk> = \xi^\dag_\bk\zeta^\dag_\bk\ket | 0_\bk >$ is the two-electron state.
However,
the Hamiltonian $\hat{H}_\bk(t)$,
the electric current $\hat{\bm{J}}_\bk(t)$,
and the jump operator $\hat{L}_\bk$ are represented as a $2\times2$ matrix,
namely,
$\hat{H}_\bk(t)\ket |\alpha_\bk>=\hat{\bm{J}}_\bk(t)\ket |\alpha_\bk> = \hat{L}_\bk\ket |\alpha_\bk>=0$ with $\alpha=\{0,2\}$.
In other words,
there is no unitary dynamics in the subspace of $\ket |0_\bk>$ and $\ket |2_\bk>$.
In the absence of a DC field, we thus obtain $\bm{J}_\bk(t)=\bm{0}$ for the range of $\varepsilon (\bk)<\abs{\mu}$,
and we can still use the $2\times2$-form master equation for $\varepsilon (\bk)>\abs{\mu}$.
Even if we consider our main target of the case with a finite DC field and a finite wave-vector shift $\bk+{\bm\delta}_{\rm dc}$,
a similar structure still holds.
Namely, we can use the $2\times 2$ master equation for one-electron states, while it is not necessary to the time evolution for
a subspace of $\ket |0_\bk>$ and $\ket |2_\bk>$.

The above discussion can easily be extended to the case of finite temperatures.
At finite temperatures, the distribution function of electrons becomes smooth, and the states $\ket |0_\bk>$ and $\ket |2_\bk>$ exist with a finite probability.
However, there is no unitary dynamics for $\ket |0_\bk>$ and $\ket |2_\bk>$,
and it is still enough to consider only the two bases $(\ket |g_\bk>,\ket |e_\bk>)$.
For the relaxation process,
we should use the two jump operators
\begin{align}
     & \hat L^{(1)}_\bk = \dyad{g_\bkd}{e_\bkd}, \\
     & \hat L^{(2)}_\bk = \dyad{e_\bkd}{g_\bkd},
\end{align}
to meet the detailed balance condition.
In the absence of lasers, the system should relax to the equilibrium state after a sufficiently long time.
One may hence determine the relaxation rates $\gamma^{(1,2)}_{\bkd}$ for $\hat L^{(1,2)}_\bk$ such that they satisfy the detailed balance condition
\begin{equation}
    \gamma^{(1)}_{\bkd} e^{-\beta E_{\bkd}} = \gamma^{(2)}_{\bkd} e^{\beta E_{\bkd}}.
    \label{eq:detailed}
\end{equation}
We note that for $\delta_{\rm dc}=0$,
Eq.~(\ref{eq:detailed}) makes the system approach a finite-temperature thermal state in the canonical ensemble (not the grand-canonical ensemble). Even in many-electron systems, if we focus on a subspace with a fixed electron number, it is enough to consider the canonical distribution. In fact, as we mentioned,
we may concentrate on only one-particle states in the full Hilbert space at each $\bm k$ to analyze the time evolution of our model.

%%%%%%%%%%%%%%%%%%%%%%%%%%%%%%%%%%%
%%%%%%%%%%%%%%%%%%%%%%%%%%%%%%%%%%%
%%%%%%%%%%%%%%%%%%%%%%%%%%%%%%%%%%%
\subsection{Time-evolution of observable}\label{subsec:observable}

Within the formalism of the Markovian master equation,
the expectation value of the electric current at time $t$ is given by
\begin{equation}
    \bm{J}(t)=\frac{1}{N}\sum_\bk\bm{J}_\bk(t)=\frac{1}{N}\sum_\bk\Tr[\hat{\rho}_\bk(t)\hat{\bm{J}}_\bk(t)].
    \label{eq:current_exp}
\end{equation}
In this paper, the computation is always done in $\bk$ space, and we take $1280\times1280$ points in an equally spaced fashion in the full Brillouin zone, which corresponds to the system size $N=1280\times1280$.
The current $\bj(t)$ can be divided into a contribution of an interband transition and an intraband one as $\bj(t)=\bj_{\mathrm{inter}}(t) + \bj_{\mathrm{intra}}(t)$.
The interband (intraband) transition contribution arises from the time evolution of off-diagonal (diagonal) density matrix elements.

%%%%%%%%%%%%%%%%%%%%%%%%%%%%%%%%%%%
%%%%%%%%%%%%%%%%%%%%%%%%%%%%%%%%%%%
%%%%%%%%%%%%%%%%%%%%%%%%%%%%%%%%%%%
\subsection{Advantages of the master equation}\label{subsec:advantage}

Here, we comment on two important aspects of the numerical method we use in the present study.
A significant advantage of the use of Eqs.~\eqref{eq:quantum-master-eqre} and~\eqref{eq:current_exp} is that one can directly obtain realistic HHG spectra from their solution without any additional process because the master equation can take an experimentally inevitable dissipation effect.
On the other hand, if we solve the standard Schr\"odinger equation for laser-driven electron systems, an artificial procedure like the use of a ``window function'' is often necessary to obtain proper HHG spectra.
This is because the energy injected by a laser pulse always remains in the system within the Schr\"odinger equation formalism, and such a setup of an isolated system differs considerably from real experiments.

Another point is that the numerical analysis of the master equation enables us to compute laser ``pulse'' driven HHG spectra directly.
This is in contrast with analytical computation methods of HHG spectra (e.g., linear and nonlinear response theories), in which one usually considers the optical response to an ideal ``continuous wave'' ($t_{\mathrm{FWHM}}\to\infty$).
Since a finite-width pulse is used in experiments, the ability to directly compute pulse-driven dynamics could be another advantage of the master equation approach.

%%%%%%%%%%%%%%%%%%%%%%%%%%%%%%%%%%%
%%%%%%%%%%%%%%%%%%%%%%%%%%%%%%%%%%%
%%%%%%%%%%%%%%%%%%%%%%%%%%%%%%%%%%%
\section{Computation of HHG spectra}\label{sec:result}

\begin{figure*}[tb]
    \centering
    \includegraphics[width=\linewidth]{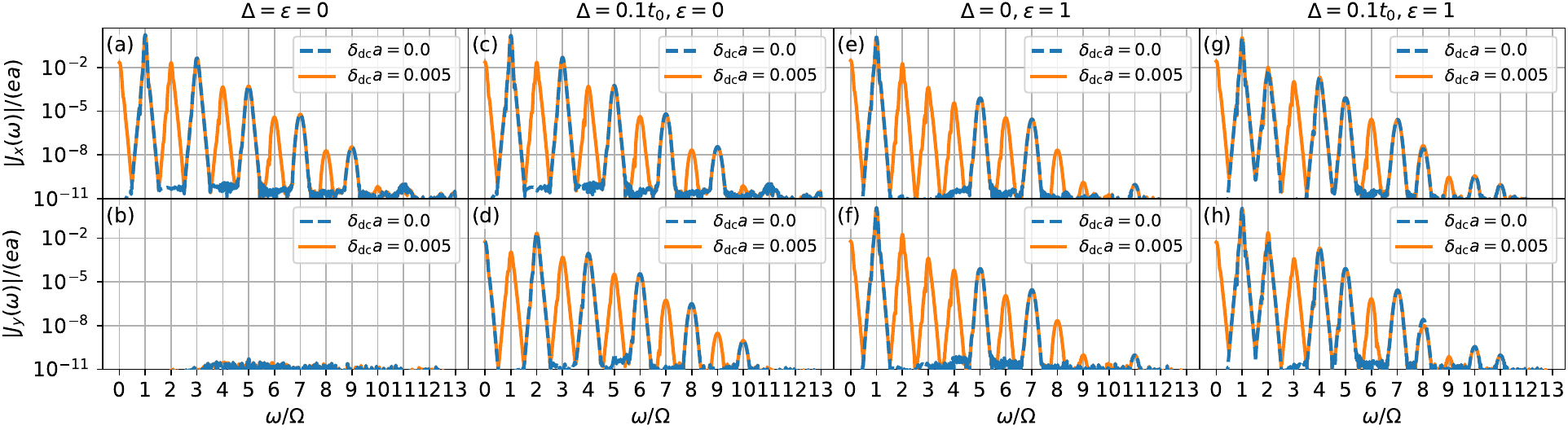}
    \caption{
        Laser-pulse driven HHG spectra of the current along the $(x,y)$-directions in graphene models with/without DC electric field $\edc$ at $\mu=0.4t_0=\Omega$.
            [(a)--(h)]  Comparative plots of $\abs{J_x(\omega)}$ (top row) and $\abs{J_y(\omega)}$ (bottom row) at $\ddc a=0$ and $0.005$ for different conditions of the staggered potential $\Delta$ and the laser ellipticity $\epsilon$.
        Panels (a) and (b) show data for $(\Delta, \epsilon) = (0, 0)$,
        (c) and (d) are for $(\Delta, \epsilon) = (0.1t_0, 0)$,
        (e) and (f) are for $(\Delta, \epsilon) = (0, 1)$,
        and (g) and (h) are for $(\Delta, \epsilon) = (0.1t_0, 1)$.
        Other parameters are $e\eac a/t_0=0.05$ and $\gamma=0.1t_0$.
    }
    \label{fig:figure3}
\end{figure*}
%
%
%

%%%%%%%%%%%%%%%%%%%%%%%%%%%%%%%%%%
\begin{table*}[tb]
    \centering
    \caption{
        Various selection rules for HHG in the graphene model of Eq.~\eqref{eq:time-dependent-hamiltonian} without DC current.
        These rules are derived from the (dynamical) symmetries.
        Each pair $(\hat U, t+\Delta t)$ represents the static symmetry operation and the time translation $t\to t+\Delta t$ of a dynamical symmetry.
        We list four cases [(i)--(iv)] of $\Delta=0$ (e.g., graphene) and $\Delta\neq0$ (e.g., transition-metal dichalcogenides; TMDC) with linear polarized light (LPL) along the $x$ direction or circularly polarized light (CPL).
        These dynamical symmetries are all broken by the application of DC current, except for $\hat U_{\sigma_{zx}}$.
    }
    \SetTblrInner{colsep=12pt, rowsep=5pt}
    \begin{tblr}{colspec={Q[c]|Q[c]Q[c]Q[c]Q[c]},rowspec={||Q[m]|Q[m]|[dotted]Q[m]||}}
         & (i) LPL $\parallel \hat e_x$, $\Delta=0$ & (ii) LPL $\parallel \hat e_x$, $\Delta\neq0$ & (iii) CPL, $\Delta=0$ & (iv) CPL, $\Delta\neq0$ \\
        {(Dynamical)\vspace{5pt}                                                                                                                     \\symmetry}             & {$(\hat U_{\sigma_{yz}},\;t+T/2)$ \vspace{3pt}                                                                                                            \\ $(\hat U_{2},\;t+T/2)$ \vspace{3pt}\\ $(\hat U_{\sigma_{zx}},\;t)$} & $(\hat U_{\sigma_{yz}},\;t+T/2)$ & {$(\hat U_{2},\;t+T/2)$ \vspace{3pt}\\ $(\hat U_{3},\;t+T/3)$} & $(\hat U_{3},\;t+T/3)$ \\
        {Selection \vspace{5pt}                                                                                                                      \\rule ($n\in\mathbb Z$)} & {$J_x(2n\Omega)=0$ \vspace{3pt}                                                                                               \\ $J_y(t)=0$} & {$J_x(2n\Omega)=0$ \vspace{3pt}\\ $J_y((2n+1)\Omega)=0$} & {$\bm J(2n\Omega)=0$ \vspace{3pt}\\ $\bm J(3n\Omega)=0$} & $\bm J(3n\Omega)=0$ \\
    \end{tblr}
    \label{tab:sym}
\end{table*}
%%%%%%%%%%%%%%%%%%%%%%%%%%%%%%%%%%

This section is the main content of the present study.
We show several numerical results and the essential properties of HHG spectra, especially even-order harmonics.

%%%%%%%%%%%%%%%%%%%%%%%%%%%%%%%%%%%
%%%%%%%%%%%%%%%%%%%%%%%%%%%%%%%%%%%
%%%%%%%%%%%%%%%%%%%%%%%%%%%%%%%%%%%
\subsection{HHG spectra and DC-electric-field dependence}\label{subsec:DC-strength}

First, we focus on the typical HHG spectra and the DC-electric-field dependence of HHG in graphene.
Figure~\ref{fig:figure3} shows typical HHG spectra of the current along the $x$ and $y$ directions $\abs{J_{x,y}(\omega)}$ with or without DC current ($\ddc a=0$ and $0.005$).
The laser intensity is chosen to be moderate ($e\eac a/t_0=0.05$), and the frequency is equal to the chemical potential ($\Omega=\mu$).
Different panels correspond to different values of the staggered potential $\Delta$ and the laser ellipticity $\epsilon$.
Figure~\ref{fig:figure3} shows that
$2n$th- or $3n$th-order harmonics
($n\in \mathbb Z$) are forbidden in the absence of DC current, depending on the existence or absence of $\Delta$ and $\epsilon$, except for the panel (b).
It also shows that a weak DC current with $\ddc a=0.005$, which breaks the inversion symmetry, is enough to obtain $2n$th- or $3n$th-order harmonics, whose intensity is comparable with that of neighboring $(2n+1)$th- or $(3n+1)$th-order harmonics.

These features, i.e.,
the appearance or absence of
$n$th-order harmonics,
can be understood by finding a dynamical symmetry~\cite{Alon1998Selection,Morimoto2017Floquet,Neufeld2019Floquet,Ikeda2019Highharmonic,Chinzei2020Time,Ikeda2020Highorder,Kanega2021Linear} of the system.
The dynamical symmetry is a sort of symmetry including a time translation as well as a usual symmetry operation in time-periodic systems like the present system-laser complex (see Appendix~\ref{app:dynamical-symmetry}).
It is defined by the following relation:
\begin{equation}
    \hat{U}^\dag\hat{H}(t+\Delta t)\hat{U}=\hat{H}(t),
    \label{eq:dynamical_symm}
\end{equation}
where $\hat{H}(t)=\hat{H}(t+T)$ is the time-periodic Hamiltonian of the target system with a time period $T$, $\Delta t
$ $(0< \Delta t <T)$ is a time shift, and
$\hat U$ is a unitary (or anti-unitary) operator.
In laser-driven systems, $\Omega$ is the laser frequency and $T=2\pi/\Omega$ is the period of the laser.
For such a dynamical symmetric system, if a vector operator $\hat{\bm O}(t)=(\hat O_1(t),\hat O_2(t),\ldots)$ satisfies a similar relation
\begin{equation}
    \hat{U}^\dag\hat{\bm O}(t+\Delta t)\hat{U}=\mathcal R\hat{\bm O}(t),
\end{equation}
then we can lead to a selection rule of $\bm O(m \Omega)=0$ with
a certain integer $m$.
Here, $\mathcal R$ is a matrix acting only on the vector $\hat{\bm O}$,
and the vector $\bm O(\omega)$ is the Fourier transform of the expectation value $\langle\hat{\bm O}(t)\rangle_t$ along the time direction.
In the above symmetry argument, we have assumed that $\hat U^\dag\hat{\rho}(t+\Delta t)\hat U=\hat{\rho}(t)$ holds, namely, not only the Hamiltonian but also the density matrix (quantum states) is dynamical symmetric~\cite{Ikeda2019Highharmonic, Chinzei2020Time, Ikeda2020Highorder, Kanega2021Linear}.
When we consider HHG spectra, the operator $\hat{\bm O}(t)$ should be chosen to the current $\hat {\bm J}(t)$.
For instance, if a system has a dynamical symmetry with $\Delta t=T/2$ and the current satisfies $\hat{U}^\dag\hat{J}_\alpha(t+\Delta t)\hat{U}=-\hat{J}_\alpha(t)$ ($\alpha=x$ or $y$), one can prove that $J_\alpha(2n\Omega)=0$ with arbitrary integer $n$, i.e., even-order harmonics all vanish.

Table~\ref{tab:sym} summarizes dynamical symmetries and the resulting selection rules that hold for some panels of Fig.~\ref{fig:figure3} in the absence of an extrinsic DC current.
The pair $(\hat U, t+\Delta t)$ represents the symmetry operation and the time translation of dynamical symmetry we consider.
An operation without time translation, i.e., $(\hat U, t)$, corresponds to a standard static symmetry.
In the case of $\Delta=0$ and linear polarization ($\epsilon=0$) as in Figs.~\ref{fig:figure3}(a) and~\ref{fig:figure3}(b),
the dynamical symmetry $(\hat U_{\sigma_{yz}},\;t+T/2)$ prohibits even-order harmonics of $J_x(2n\Omega)$ (see Appendix~\ref{app:dynamical-symmetry}).
Here, $\hat U_{\sigma_{yz}}$ and $\hat U_{\sigma_{zx}}$ are, respectively, the mirror operations across the $y$-$z$ plane ($x\to -x$) and the $z$-$x$ plane ($y\to -y$).
In addition to this dynamical symmetry, the system possesses the (static) mirror symmetry of $\hat U_{\sigma_{zx}}$. From these dynamical and static symmetries, optical responses in the $y$ direction are all shown to be prohibited.
A finite $\Delta\neq0$ breaks
a dynamical symmetry with in-plane $C_2$ rotation, $(\hat U_{2},\;t+T/2)$, allowing even-order harmonics in the $y$ direction as shown in Fig.~\ref{fig:figure3}(d).
On the other hand,
$(\hat U_{\sigma_{yz}},\;t+T/2)$ is preserved regardless of the value of $\Delta$, and as a result, even-order harmonics in the $x$ direction are all suppressed even if $\Delta$ exists [see Fig.~\ref{fig:figure3}(a) and (c)].

In the case of circular polarization ($\epsilon=1$), we should note the following two dynamical symmetries: (i) $(\hat U_{2},\;t+T/2)$ and (ii) $(\hat U_{3},\;t+T/3)$, where $\hat U_{3}$ is the in-plane $C_3$ rotation.
For $\Delta=0$, both (i) and (ii) hold, and they prohibit even-order and $3n$th-order harmonics in both $x$ and $y$ directions ($n\in \mathbb Z$).
On the other hand, for $\Delta\neq0$,
(i) is violated,
and only $3n$th-order harmonics are prohibited.

The application of a DC current breaks all these dynamical symmetries,
allowing the appearance of all the harmonics.
However, the static mirror symmetry $\hat U_{\sigma_{zx}}$ is not broken by the DC current in the $x$ direction,
and therefore, the response in the $y$ direction is suppressed.

These clear properties of HHG spectra are consistent with our numerical results of Fig.~\ref{fig:figure3}.
We emphasize that several HHG signals can be activated and controlled by an external DC current, as shown in Fig.~\ref{fig:figure3}.
Hereafter, we mainly focus on the linear polarized light of $\epsilon=0$, which has often been used in experiments.

\begin{figure}[tb]
    \centering
    \includegraphics[width=\linewidth]{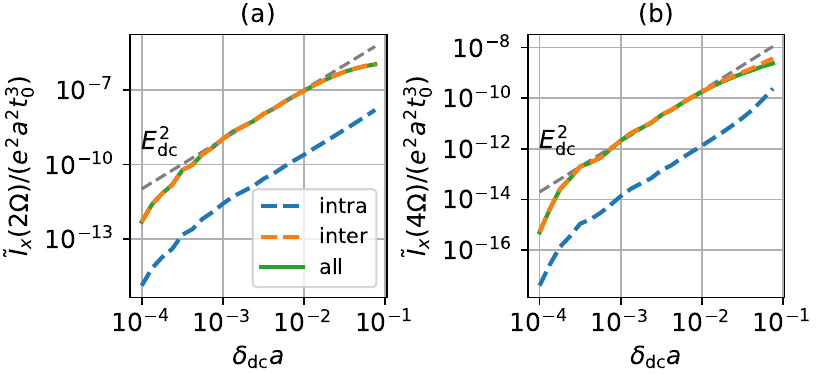}
    \caption{
        $\edc$ dependence of (a) SHG [$\tilde I_x(2\Omega)$] and (b) FHG [$\tilde I_x(4\Omega)$] spectra in pulse-driven graphene at $\mu=\Omega=0.4t_0$ and $\mu=\Omega/2=0.4t_0$.
        Each graph simultaneously displays the intraband dynamics component (blue line), the interband dynamics component (orange line), and the total spectra (green line).
        The gray dashed line is the fitting line of the spectra $\propto \edc^2$.
        The other parameters are set to $e\eac a/t_0=0.01$, $\gamma=0.1t_0$, and $\Delta=\epsilon=0$.
    }
    \label{fig:figure4}
\end{figure}

Before ending this subsection, we remark on two things.
First, as we discussed in Sec.~\ref{subsec:boltzmann}, the current-induced steady state obtained by the Boltzmann equation is valid only when the current (or DC electric field) is sufficiently small, i.e.,
$\delta_\dc a \ll 1$.
In this condition, the intensities of DC-field-induced harmonics are expected to be proportional to the DC-field power $\edc^2$ because the DC-field induced current $J_\alpha(\omega)$ would linearly respond to $\edc$.
Figures~\ref{fig:figure4}(a) and (b) show that
intensities of the DC-field induced second- and fourth-order harmonic generations (SHG and FHG) are both almost proportional to $\edc^2$ in a moderate-$\edc$ region.
In a sufficiently weak field regime of $\delta_\dc a\lesssim 10^{-3}$, they deviate from the $\edc^2$ curves. This is because a small $\edc$ causes a very small $\Delta f(\bk)$ and the accurate numerical detection of the effects of such a small $\Delta f(\bk)$ is beyond our resolution in $\bk$ space ($N=1280\times1280$).
On the other hand, in the strong $\edc$ regime,
SHG and FHG intensities again deviate from the
$\edc^2$ line since their nonlinear $\edc$ dependence is activated.
We therefore focus on the mid $\edc$ region and set $\delta_\dc a=0.005$ in the following sections unless otherwise noted.

The second point concerns the time periodicity in the above argument of dynamical symmetry.
When we argue the dynamical symmetry, we usually assume that the Hamiltonian satisfies $\hat H(t)=\hat H(t+T)$. Namely, we implicitly consider a system irradiated by a ``continuous wave.''
On the other hand, a short laser pulse is generally used in experiments.
Therefore, the argument based on dynamical symmetry does not seem to be applicable to discussing pulse-induced HHG spectra.
However, empirically, selection rules based on dynamical symmetry work at least at a qualitative level, even if the applied laser pulse contains only a few cycles.
As we discussed above, our results on laser pulses in Fig.~\ref{fig:figure3} are indeed explained by dynamical symmetries in Table~\ref{tab:sym}.

%%%%%%%%%%%%%%%%%%%%%%%%%%%
%%%%%%%%%%%%%%%%%%%%%%%%%%%
%%%%%%%%%%%%%%%%%%%%%%%%%%%
%%%%%%%%%%%%%%%%%%%%%%%%%%%
\subsection{Laser frequency dependence}\label{subsec:frequency}

\begin{figure}[tb]
    \centering
    \includegraphics[width=\linewidth]{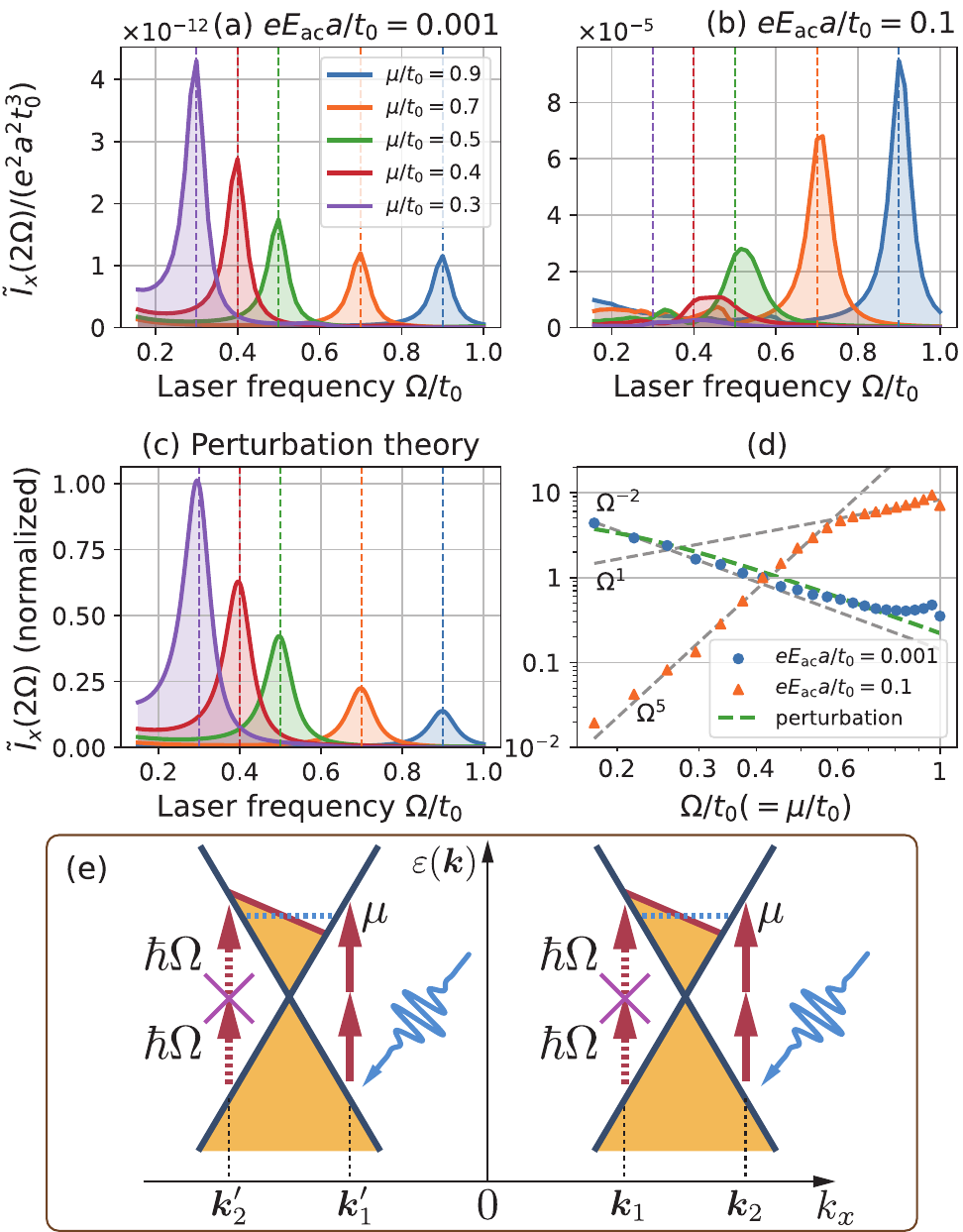}
    \caption{
        [(a),(b)] $\Omega$ dependence of SHG [$\tilde I_x(2\Omega)$] spectra of the current along the $x$ direction in graphene with DC current at various conditions of chemical potentials $\mu/t_0=\{0.3,0.4,0.5,0.7,0.9\}$ under the irradiation of a weak laser (a) ($e\eac a/t_0=0.001$) and a strong laser (b) ($e\eac a/t_0=0.1$).
        The dashed lines show the positions of $\Omega=\mu$.
        (c) DC-current driven SHG spectra computed by second-order perturbation theory in Ref.~\cite{Cheng2014DC}. Parameters are the same as those of panel (b).
        (d) Scaling behavior of the SHG peak intensities as a function of $\Omega$.
        The points are numerically calculated for the cases of weak (blue points) and strong (orange points) lasers.
        The dashed line shows the result of the perturbation theory.
        (e) $\bk$-space image of SHG photo excitations in DC-current driven graphene with ${\bm E}_{\rm dc}\parallel \hat e_x$ on $k_y=0$.
        Red arrows show the two-photon absorption processes, where the laser frequency is chosen to $\Omega=\mu$, and we temporally revive $\hbar$. The symbol $\times$ means the impossibility of the absorption process.
        The contribution from the wavevector pairs $\bk$ and $\bk'$,
        which should cancel each other in inversion-symmetric graphene,
        becomes finite due to the DC-current driven shift of the Fermi distribution, and it activates even-order harmonics.
        The other parameters are set to $\delta_\dc a=0.005$, $\gamma=0.1t_0$, and $\Delta=\epsilon=0$.
    }
    \label{fig:figure5}
\end{figure}

Next, we show a laser-frequency dependence of HHG spectra,
especially SHG, FHG, and sixth-order harmonic generation (sixth HG).
We show the characteristic resonance structure and how the frequency dependence changes with the growth of laser strength.

Figures~\ref{fig:figure5}(a) and~\ref{fig:figure5}(b),
respectively,
show the laser-frequency $\Omega$ dependence of the SHG intensities $\tilde I_x(2\Omega)$ for different chemical potentials $\mu$ in the application of weak and strong laser pulses [$e\eac a/t_0=(0.001,0.1)$].
In the weak pulse case of Fig.~\ref{fig:figure5}(a),
we find a peak of SHG at $\Omega=\mu$,
and its intensity is proportional to $\Omega^{-2}$ as in Fig.~\ref{fig:figure5}(d).
These results are consistent with the previous predictions based on the analytical perturbation theory for a Dirac electron model~\cite{Cheng2014DC}, which is valid in a weak AC-field limit.
The divergent $\Omega^{-2}$ behavior in the low-frequency regime is due to the singular Berry connection of the Dirac cone~\cite{Cheng2014DC}.
Figure~\ref{fig:figure5}(c) and the green dashed line in Fig.~\ref{fig:figure5}(d) are the analytical results, which correspond, respectively, to Fig.~\ref{fig:figure5}(a) and the blue circles in
Fig.~\ref{fig:figure5}(d).
We note that in a large-$\Omega$ regime of Fig.~\ref{fig:figure5}(d),
the SHG peak intensity deviates from the $\Omega^{-2}$ line.
This is because higher-energy photo-excited electrons, which cannot be described by the $\bk$-linear Dirac electron model, become dominant in the large-$\Omega$ regime.
The perturbation theory also predicts  $J(2\Omega)\propto \edc\eac^2$, and we confirm that this power-law relation holds in our numerical result in the weak laser case of $e\eac a/t_0=0.001$ (see Sec.~\ref{subsec:intensity}).
From these results, we can conclude that our numerical method well reproduces the previous analytical predictions.

In the strong laser case of Fig.~\ref{fig:figure5}(b), on the other hand, different features of SHG spectra are observed. We again find a peak structure around $\Omega=\mu$, but its intensity is no longer proportional to $\Omega^{-2}$ and it clearly increases with growing $\Omega$.
Figure~\ref{fig:figure5}(d) shows that there are two scaling regimes of the SHG peak intensity: It is proportional to $\Omega^5$ (i.e., nondivergent) in the low-$\Omega$ regime, whereas it is proportional to $\Omega$ in the high-$\Omega$ regime.
These scaling laws ${\tilde I}_x(2\Omega)\propto \Omega^5$ and $\propto \Omega^1$ are both nontrivial, and we cannot find any simple explanation for them.
We stress that these properties in the nonperturbative, strong-laser regime were first captured by the present numerical method based on the master equation.

Let us consider these resonance-like structures in Fig.~\ref{fig:figure5} from a microscopic viewpoint.
Even-order optical responses are generally prohibited in spatial-inversion symmetric electron systems, whereas (as mentioned in Sec.~\ref{subsec:boltzmann})
a weak DC electric field (i.e., a weak DC current) causes a shift of the electron distribution and breaks the inversion symmetry.
We can prove that in $C_2$ rotation-symmetric systems, the even-order harmonics generated by electrons in the $k_\alpha$ and $-k_\alpha$ points cancel each other out, and as a result, even-order ones all vanish (see Appendix~\ref{app:selection-ruleB}).
When a DC electric field $\bm{E}_\dc$ is applied along the $x$ axis in graphene, the electron distribution changes from a usual distribution with a Fermi surface to an asymmetric one, as shown in Fig.~\ref{fig:figure5}(e). For such a shifted distribution of Fig.~\ref{fig:figure5}(e),
we have one-electron states around $\bk=(k_x,k_y)$, while we have two-electron states around $\bk'=(-k_x,k_y)$. Therefore, when we tune the laser frequency to $\Omega\sim\mu$, two-photon absorption can take place around $\bk$, whereas it cannot around $\bk'$, as shown in Fig.~\ref{fig:figure5}(e).
As a result, the cancellation between $k_x$ and $-k_x$ is broken and
the SHG peak appears at $\Omega\sim\mu$.
This scenario is easily extended to the cases of generic $2n$th-order HHG in DC-current-driven steady states.
In this way, the $2n$th-order HHG spectra of DC-current-driven graphene are predicted to have a peak when the laser frequency satisfies the following inequality:
\begin{equation}
    \mu/n-v_F\ddc/(2n)\leq\Omega\leq\mu/n+v_F\ddc/(2n).
\end{equation}

\begin{figure}[tb]
    \centering
    \includegraphics[width=\linewidth]{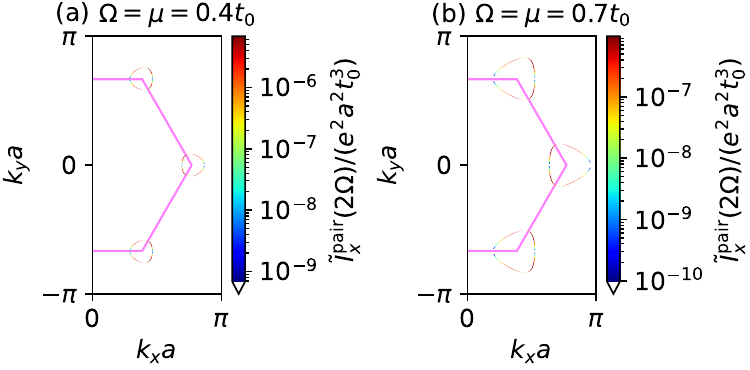}
    \caption{
    Numerically computed symmetrical pair of $\bk$-resolved SHG intensity $\tilde I_x^{\mathrm{pair}}(\bk,2\Omega)$ of Eq.~\eqref{eq:kresolvedSHG} at (a) $\mu=\Omega=0.4t_0$ and (b) $\mu=\Omega=0.7t_0$.
    The region of a finite $\tilde I_x^{\mathrm{pair}}(\bk,2\Omega)$ coincides with that of a finite deviation $\Delta f_{\rm ss}(\bk)$ in Fig.~\ref{fig:figure2}(b).
    Parameters are set to $e\eac a/t_0=0.001$, $\ddc a=0.01$, $\gamma=0.1t_0$, and $\Delta=\epsilon=0$.
    }
    \label{fig:figure6}
\end{figure}

\begin{figure*}[tb]
    \centering
    \includegraphics[width=\linewidth]{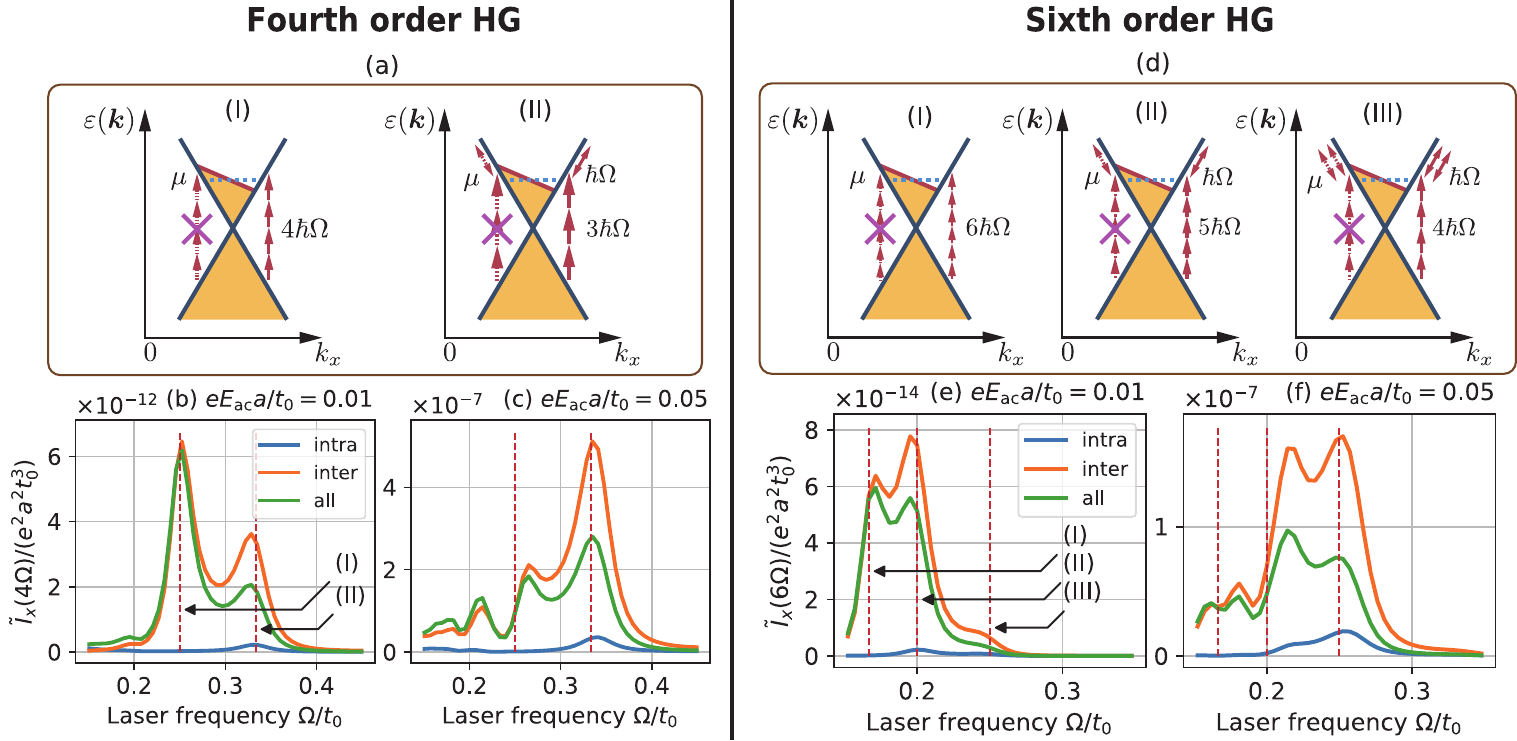}
    \caption{
        Laser-frequency $\Omega$ dependence of even-order harmonics intensities (bottom row) and schematic images of the photon absorption processes and intraband dynamics corresponding to the bottom row (top row).
        Each graph in the bottom row simultaneously displays the intraband dynamics intensity (blue line), the interband dynamics one (orange line), and the total one (green line) at $e\eac a/t_0=0.01$ and $e\eac a/t_0=0.05$.
        Left panels (a)-(c) represent FHG, and right panels (d)-(f) do sixth-order HG.
        In cartoons (a) and (d), we have revived the symbol $\hbar$.
        Other parameters are set as follows: $\mu=0.5t_0$, $\delta_\dc a=0.005$, $\gamma=0.1t_0$, and $\Delta=\epsilon=0$.
    }
    \label{fig:figure7}
\end{figure*}

To see the validity of the above argument about
even-order HHG, we define the following $\bk$-resolved SHG intensity
\begin{equation}
    \tilde I_{x}^{\mathrm{pair}}(\bk, 2\Omega) = \int_{(2-\frac{1}{2})\Omega}^{(2+\frac{1}{2})\Omega}\dd\omega\abs{\omega (J_{x,\bk}(\omega) + J_{x,\bk'}(\omega))}^2,
    \label{eq:kresolvedSHG}
\end{equation}
where $\bk=(k_x,k_y)$, $\bk'=(-k_x,k_y)$, and this value is defined in half of the Brillouin zone ($k_x>0$).
If the $\bk$- and $\bk'$-point SHG intensities perfectly cancel each other out, the value of $\tilde I_{x}^{\mathrm{pair}}(\bk, 2\Omega)$ becomes zero. However, it has a finite value when the cancellation is broken.
That is, Eq.~\eqref{eq:kresolvedSHG} can detect the degree of breakdown of the cancellation between $\bk$ and $\bk'$ points.
Figure~\ref{fig:figure6} gives the numerically computed results of $\tilde I_{x}^{\mathrm{pair}}(\bk, 2\Omega)$ in half of the Brillouin zone ($k_x>0$) at $\Omega=\mu$.
It shows that $\tilde I_{x}^{\mathrm{pair}}(\bk, 2\Omega)$ takes a finite value around $\bK$ and $\bK'$ Dirac points.
Especially, comparing Fig.~\ref{fig:figure6}(b) with Fig.~\ref{fig:figure2}(b),
one sees that $\tilde I_{x}^{\mathrm{pair}}(\bk, 2\Omega)$ is enhanced in the region where
the electron distribution is shifted [i.e., $\Delta f_{\rm ss}(\bk)\neq 0$].
We thus conclude that the argument in the above paragraph is indeed correct.

We further discuss the higher even-order harmonics (FHG and sixth-order HG) in graphene under the application of DC current.
As we will soon see, in these harmonics, multiple resonance peaks appear,
and the response driven by not only interband but also intraband dynamics becomes dominant for a strong laser pulse.

Figure~\ref{fig:figure7} shows the results of FHG spectra [panels (a)--(c)] and
sixth-order HG spectra [panels (d)--(f)]
for weak and strong laser pulses.
In panels (b), (c), (e), and (f), orange (blue) lines show the contributions from interband (intraband) dynamics,
and green lines show the full responses of interband and intraband dynamics as a function of the laser frequency $\Omega$.
One sees many peaks in the FHG and sixth-order HG spectra of panels (b), (c), (e), and (f).
Among them, the FHG peak at $\Omega=\mu/2$ [(I) in panel (b)] and the sixth-order HG one at $\Omega=\mu/3$ [(I) in panel (e)] can be understood by the perturbation theory: the former and latter correspond, respectively, to four- and six-photon absorption processes, described in the cartoons (I) of Figs.~\ref{fig:figure7}(a) and (d).
In the strong pulse cases corresponding to panels (c) and (f),
other peaks become grown at $\Omega=2\mu/3$ for the FHG and at $\Omega=2\mu/5$ and $\mu/2$ for the sixth-order HG.
For instance, the additional peak of FHG at $\Omega=2\mu/3$ can be governed by the four-photon process with an interband three-photon absorption and an intraband one-photon dynamics, as shown in the process (II) of Fig.~\ref{fig:figure7}(a).
In fact, we can observe an enhancement of
intraband contribution at $\Omega=2\mu/3$ with laser intensity increasing, by comparing Fig.~\ref{fig:figure7}(b) and (c).
Hereafter, we refer to a photo-excited process with interband $m$-photon absorption and intraband $n$-photon dynamics as an ``$m+n$ process''.
For the sixth-order HG,
the peaks of $\Omega=2\mu/5$ and $\Omega=\mu/2$ correspond to the $5+1$ process [process (II) of Fig.~\ref{fig:figure7}(d)] and the $4+2$ process [process (III) of Fig.~\ref{fig:figure7}(d)], respectively.
We stress that
these additional peaks in FHG and sixth-order HG are stronger than the typical perturbative peaks when the laser intensity is large enough ($e\eac a/t_0=0.05$).

At the end of this subsection, we briefly compare our numerical method with the analytic perturbation theory.
As discussed in Ref.~\cite{Bhalla2023Quantum}, the perturbative calculations make it possible to classify several laser-driven optical processes in a clear, analytic manner.
On the other hand, as we explained in this subsection, our numerical analysis classifies different laser-driven processes from the computed spectral peaks of the density matrix and physical quantities, and hence, it is generally difficult to classify the optical processes in more detail than the perturbation theory, especially when both interband and intraband processes are intertwined. However, our numerical approach has some advantages: The classification based on perturbative analytical approaches is limited to lower-order perturbation terms, whereas our numerical analysis can capture higher-order nonperturbative contributions. Therefore, these two strategies are complementary to each other.
In the present study, we utilize the numerical method based on the GKSL equation to describe the overall characteristics of the spectrum, including higher-order harmonics.

%%%%%%%%%%%%%%%%%%%%%%%%%%%%%%%%%
%%%%%%%%%%%%%%%%%%%%%%%%%%%%%%%%%
%%%%%%%%%%%%%%%%%%%%%%%%%%%%%%%%%
%%%%%%%%%%%%%%%%%%%%%%%%%%%%%%%%%
\subsection{Laser intensity dependence}\label{subsec:intensity}

\begin{figure}[tb]
    \centering
    \includegraphics[width=\linewidth]{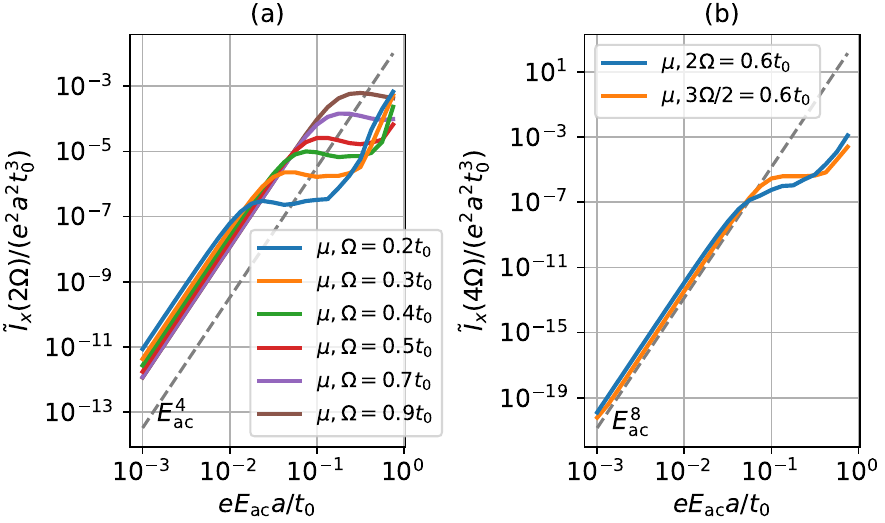}
    \caption{
        The $E_\ac$ dependence of (a) second-order harmonic generation (SHG) [$\tilde I_x(2\Omega)$] and (b) fourth-order harmonic generation (FHG) [$\tilde I_x(4\Omega)$] spectra in pulse-driven graphene.
        The gray dashed line represents the fitted line for a guide to eye: the line of panel (a) is $\propto E_\ac^4$ and that of (b) is $\propto E_\ac^8$.
        (a) SHG spectra at different values of chemical potentials $\mu/t_0=\{0.2,0.3,0.4,0.5,0.7,0.9\}$ under the condition of the laser frequency $\Omega=\mu$.
        (b) FHG at $\mu=\{2\Omega,3\Omega/2\}=0.6t_0$.
        The other parameters are set as follows: $\delta_\dc a=0.005$, $\gamma=0.1t_0$, and $\Delta=\epsilon=0$.
    }
    \label{fig:figure8}
\end{figure}

Next, we show a laser-intensity dependence of HHG spectra, focusing on SHG $\tilde I_x(2\Omega)$ and FHG $\tilde I_x(4\Omega)$, which appear only when DC current is applied.
Figure~\ref{fig:figure8}(a) shows the laser-intensity $E_\ac$ dependence of $\tilde I_x(2\Omega)$ for different chemical potential $\mu$ under the condition of $\Omega=\mu$,
in which the $2+0$ process is dominant as in Fig.~\ref{fig:figure2}(a).
The SHG spectra are proportional to $\eac^4$ for every $\mu$ for a weak enough laser pulse $e\eac a/t_0\lesssim10^{-2}$.
This feature is consistent with the second-order perturbation theory~\cite{Cheng2014DC, Takasan2021Currentinduced}.
However, when the laser pulse becomes strong,
the SHG intensity is no longer proportional to $\eac^4$, indicating the appearance of the nonperturbative region. Furthermore,
we can find that the laser intensity of the crossover from the perturbative to the nonperturbative regime is dependent on the chemical potential $\mu$ and laser frequency $\Omega=\mu$.
The laser intensity for the crossover is smaller with $\mu$ (and laser frequency $\Omega$) decreasing.
The laser-intensity dependence of the SHG spectra
is consistent with the result of
Figs.~\ref{fig:figure5}(a) and (b).

Figure~\ref{fig:figure8}(a) also indicates that the $\Omega$ and $\eac$ dependences of ${\tilde I}_x(2\Omega)$ are complicated and correlated in the nonperturbative regime of
$e\eac a/t_0 \gtrsim 10^{-1}$. For instance, in Fig.~\ref{fig:figure5}(b) and (d), we have observed the monotonically increasing $\Omega$ dependence for an intense laser $e\eac a/t_0 = 10^{-1}$, but Fig.~\ref{fig:figure8}(a) tells us that the $\Omega$ dependence further changes if we consider a more intense laser such as $e\eac a/t_0 = 10^{0}$.

The FHG spectra exhibit a similar feature as shown in Fig.~\ref{fig:figure8}(b).
Panel (b) shows the laser-intensity $E_\ac$ dependence of $\tilde I_x(4\Omega)$ for the condition of (I) $\Omega=\mu/2$ and (II) $\Omega=2\mu/3$, which corresponds to the processes (I) and (II) of Fig.~\ref{fig:figure7}(a), respectively.
In the weak laser pulse regime $e\eac a/t_0\lesssim10^{-2}$,
the FHG intensities are proportional to $\eac^8$,
which is consistent with the fourth-order perturbation viewpoint.
We also observe that the perturbative--nonperturbative crossover for condition (II) takes place at a higher laser intensity than for condition (I).
This feature explains the peak heights at $\Omega=\mu/2$ and $\Omega=2\mu/3$ in Figs.~\ref{fig:figure7}(b) and~\ref{fig:figure7}(c).

%%%%%%%%%%%%%%%%%%%%%%%%%%%%%%
%%%%%%%%%%%%%%%%%%%%%%%%%%%%%%
%%%%%%%%%%%%%%%%%%%%%%%%%%%%%%
%%%%%%%%%%%%%%%%%%%%%%%%%%%%%%
\subsection{Effect of a finite mass gap}\label{subsec:bandgap}

\begin{figure}[tb]
    \centering
    \includegraphics[width=0.9\linewidth]{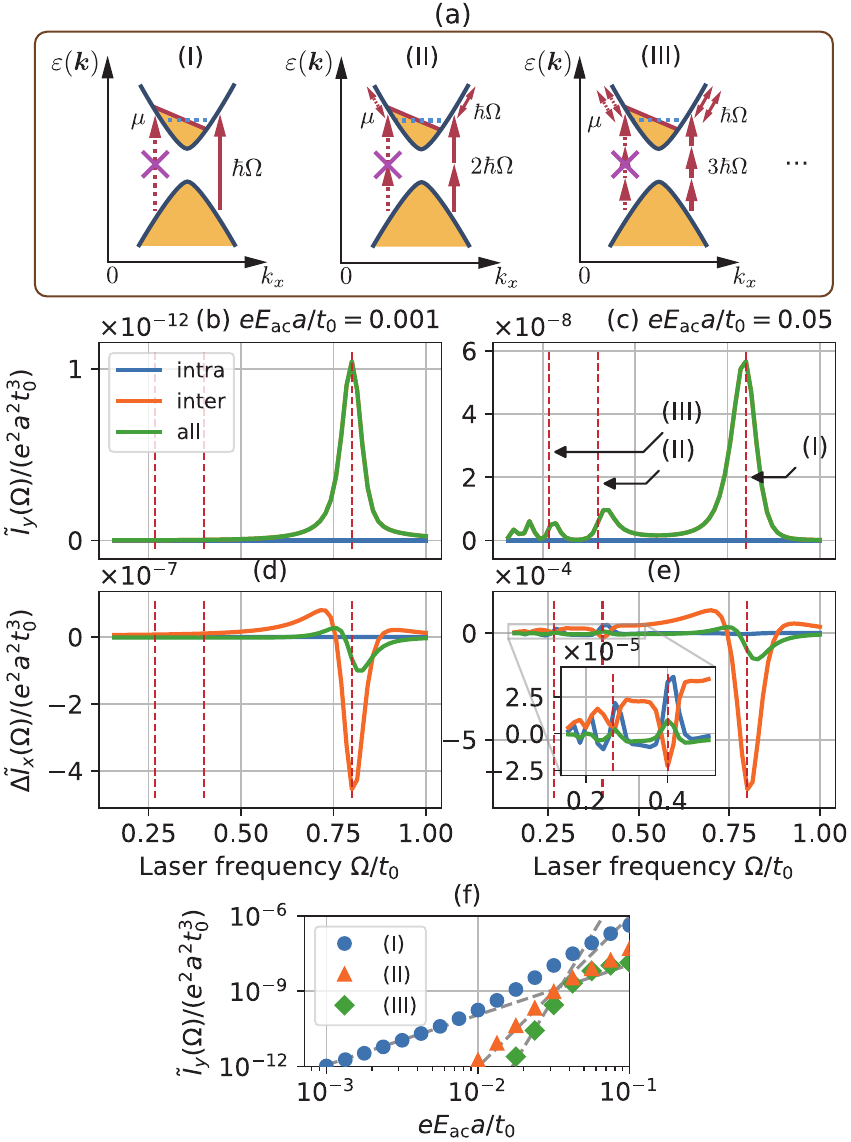}
    \caption{
        Several properties of linearly-polarized light driven $\tilde I_y(\Omega)$ in the gapped graphene model with a finite $\Delta$ under the application of a DC current.
        Directions of both the DC current and the electric field of the laser are parallel to the $x$ axis.
        (a) Schematic images of one photon absorption process (I) and multi-photon processes (II) and (III) with both interband and intraband dynamics in $\tilde I_{x,y}(\Omega)$. %in the gapped graphene model with a finite $\Delta$ under the application of a DC current.
        In the panels, we have revived the symbol $\hbar$.
            [(b),(c)] The $\Omega$ dependence of $\tilde I_y(\Omega)$ spectra in gapped graphene model at chemical potential $\mu/t_0=0.4$ under the irradiation of (b) a weak laser pulse with $e\eac a/t_0=0.001$ and (c) a strong one with $e\eac a/t_0=0.05$.
            [(d),(e)] The difference $\Delta\tilde I_x(\Omega)$ between $\tilde I_x(\Omega)$ under the condition of a finite DC current ($\delta_\dc a=0.005$) and that under zero DC current ($\delta_\dc a=0$).
        Panel (d) is the result of a weak laser pulse, while panel (e) is a strong pulse.
        (f) The $E_\ac$ dependence of the peak intensities of $\tilde I_y(\Omega)$ in the gapped graphene model.
        Blue (I), orange (II), and green (III) points, respectively, correspond to the peaks at positions (I), (II), and (III) of panel (c).
        Fitting dashed lines are a guide for the eye: lines for the conditions (I), (II), and (III), respectively, are proportional to $E_\ac^2$, $E_\ac^6$, and $E_\ac^{10}$.
        Due to the limitations of the accuracy of the numerical computation, intensities below $10^{-12}$ are not displayed.
        Other parameters are $\gamma=0.1t_0$, $\Delta=0.1t_0$, and $\epsilon=0$.
    }
    \label{fig:figure9}
\end{figure}

%\newpage
We have mainly focused on the graphene model with zero staggered potential $\Delta=0$ so far.
This subsection is devoted to a discussion on the effects of a staggered potential $\Delta$ under the application of a DC current.
Note that a finite $\Delta$ induces a mass gap at Dirac points $\bm K$ and $\bm K'$, and we focus on the situation with a Fermi surface, i.e., the chemical potential is significantly larger than the mass gap.
In this subsection, we fix $(\mu,\Delta)=(0.4t_0,0.1t_0)$ and then mainly see the $\Omega$ dependence of the HHG.
As we discussed in Sec.~\ref{subsec:DC-strength},
when the electric field of linear polarized light is along the $x$ axis,
the presence or absence of $\Delta$
affects the laser-driven current along the $y$ direction, especially the even-order harmonics.
Applying a DC current further induces odd-order harmonics of the current along the $y$ axis.
We therefore investigate the behavior of the fundamental frequency response (i.e., the first-order harmonics) $\tilde I_y(\Omega)$ among the DC-current driven odd-order harmonics.

Figures~\ref{fig:figure9}(b) and~\ref{fig:figure9}(c), respectively, show the $\Omega$ dependence of $\tilde I_y(\Omega)$ induced by weak and strong laser pulses.
The orange (blue) line shows the contribution from the interband transition (intraband dynamics),
and the green line shows overall intensities.
In Fig.~\ref{fig:figure9}(b),
there is a sharp peak of $\tilde I_y(\Omega)$ at $\Omega=2\mu$ for a weak pulse.
This peak is explained by the perturbation argument, in which a one-photon absorption as in the process (I) of Fig.~\ref{fig:figure9}(a) is dominant.
For a strong laser pulse,
additional peaks emerge around $\Omega=\mu$, $\Omega=2\mu/3$, $\Omega=\mu/2$, etc., as shown in Fig.~\ref{fig:figure9}(c).
These peak structures can be understood by the nonlinear optical responses, such as two-photon absorption, three-photon absorption, and others, whose images are given in processes (II) and (III) of Fig.~\ref{fig:figure9}(a).
Figures~\ref{fig:figure9}(d) and~\ref{fig:figure9}(e) show that
these interband transition (photon absorption) processes for $\tilde I_y(\Omega)$ %involve 
accompany the intraband dynamics for $\tilde I_x(\Omega)$.
Two panels (d) and (e) represent the difference $\Delta \tilde I_x(\Omega)$ between $\tilde I_x(\Omega)$ under a finite DC current and that under zero DC current.
Since the first-order response $\tilde I_x(\Omega)$ of the $x$-direction current exists even in the case without DC current, we introduce $\Delta \tilde I_x(\Omega)$.
Figure~\ref{fig:figure9}(d) shows that for a weak laser pulse, a single peak of $\Delta \tilde I_x(\Omega)$ appears around $\Omega=2\mu$ like $\tilde I_y(\Omega)$ of
Figs.~\ref{fig:figure9}(b). On the other hand, Fig.~\ref{fig:figure9}(e) tells us that when the laser pulse becomes strong, multiple peaks of $\Delta \tilde I_x(\Omega)$ appear, corresponding to those of $\tilde I_y(\Omega)$ in panel (c) and the contributions of not only interband but also intraband dynamics increase in $\Delta \tilde I_x(\Omega)$.
These results indicate that in the strong-laser regime, $\tilde I_y(\Omega)$ exhibits a complex $\Omega$ dependence accompanied by the intraband dynamics of $\tilde I_x(\Omega)$.

Figure~\ref{fig:figure9}(f) represents the laser intensity dependence of $\tilde I_y(\Omega)$ at peak positions corresponding to conditions (I), (II),
and (III) of panels (a) and (c).
Considering the laser frequency values at these peak positions and the fact that the intraband contribution of $\tilde I_y(\Omega)$ vanishes in panels (b) and (c), we can predict that the peaks (I), (II) and, (III) are viewed, respectively, as linear, third-order, and fifth-order nonlinear optical responses.
Namely, their intensities $\tilde I_y(\Omega)$ are expected to be proportional to $\eac^2$, $\eac^6$, and $\eac^{10}$ (remember that $\tilde I_y(\Omega)\propto \abs{J_y(\Omega)}^2$).
Figure~\ref{fig:figure9}(f) indeed shows that the intensities of $\tilde I_y(\Omega)$ at conditions (I), (II), and (III) follows the expected power laws in the region of the weak laser pulse.
Therefore, our predictions about the perturbative picture of each peak (I), (II), and (III) are consistent with the numerical results for the weak-laser regime.
Moreover, Fig.~\ref{fig:figure9}(f) illustrates the crossover from the perturbative to the nonperturbative regimes, demonstrating that our numerical method can capture a broad range from weak to strong lasers.

%%%%%%%%%%%%%%%%%%%%%%%%%%%%%%%%%%%%
%%%%%%%%%%%%%%%%%%%%%%%%%%%%%%%%%%%%
%%%%%%%%%%%%%%%%%%%%%%%%%%%%%%%%%%%%
%%%%%%%%%%%%%%%%%%%%%%%%%%%%%%%%%%%%
\subsection{Extremely strong laser fields}\label{subsec:strong-field}

\begin{figure}[tb]
    \centering
    \includegraphics[width=\linewidth]{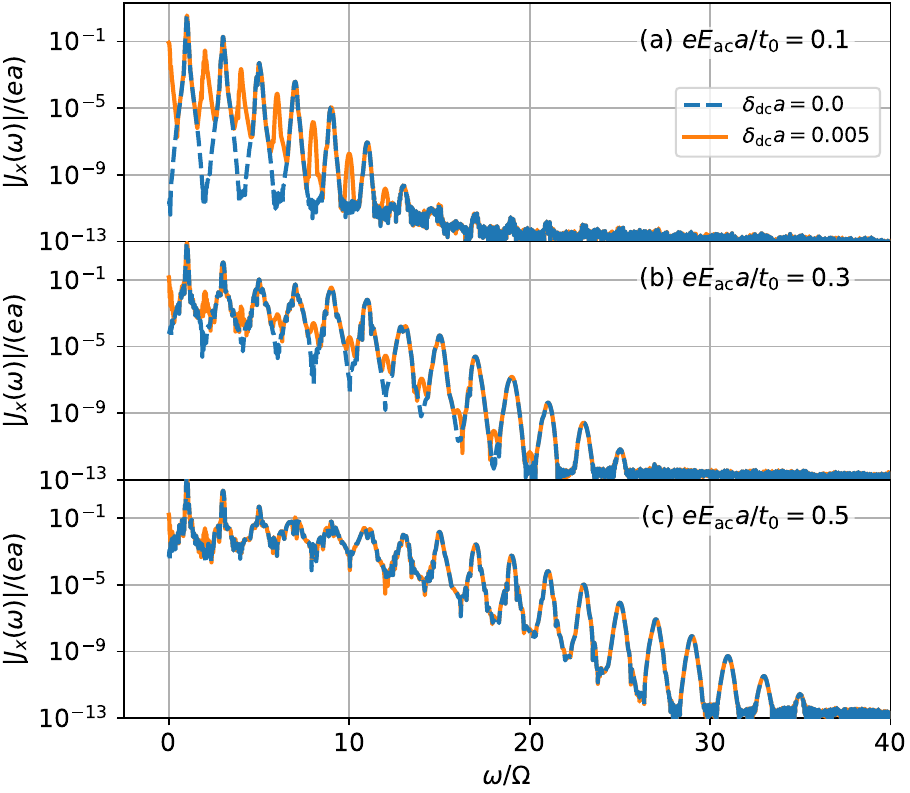}
    \caption{
        HHG spectra $|J_x(\omega)|$ in laser-pulse driven graphene with/without a DC electric field $\edc$ at the chemical potential $\mu=0.4t_0$ and the laser frequency $\Omega=\mu=0.4t_0$.
        Panels (a), (b), and (c) respectively correspond to $\abs{J_x(\omega)}$ under the irradiation of laser strength $e\eac a/t_0=0.1$, $e\eac a/t_0=0.3$, and $e\eac a/t_0=0.5$.
        Blue dotted and orange curves are, respectively, the results of $\ddc a=0$ (zero DC current) and $0.005$ (a finite DC current).
        The other parameters are chosen to $\gamma=0.1t_0$ and $\Delta=\epsilon=0$.
    }
    \label{fig:figure10}
\end{figure}

In this final subsection,
we focus on the case of extremely strong laser fields.
We show the HHG spectra $\abs{J_x(\omega)}$ at $\ddc a=0$ and $0.005$ under the irradiation of the strong laser pulses $e\eac a/t_0=(0.1,0.3,0.5)$ in Figs.~\ref{fig:figure10}(a)--\ref{fig:figure10}(c).
The effect of DC current clearly appears in Fig.~\ref{fig:figure10}(a),
which is shown as an activation of the even-order harmonics at $\edc\neq0$.
In the case of much stronger laser pulses of panels (b) and (c), we can find the plateau structure of $\abs{J_x(\omega)}$, and it does not depend well on the existence of DC current.
Here, ``plateau''~\cite{Ghimire2014Strongfield} means the frequency domain, in which the HHG intensity is roughly independent of the frequency $\omega$. In the present setup,
the plateau seems to continue up to 10th-order harmonics (i.e., $\omega< 10\Omega$) for both cases (b) and (c).

On the other hand, we also find that for strong laser cases of (b) and (c), the even-order harmonics driven by DC current are quite small compared to odd-order harmonics or almost invisible.
This is probably because the large line-width of each odd-order harmonics covers the peak of neighboring even-order harmonics.
Namely, this result indicates that we should apply a strong enough DC current to observe the DC-current-driven even-order harmonics in the case of the application of an extremely intense laser.
However, as we mentioned in Sec.~\ref{subsec:boltzmann}, we note that such a case with a large DC current is beyond the scope of the Boltzmann equation and the relaxation-time approximation.

%%%%%%%%%%%%%%%%%%%%%%%%%%%%%%%%
%%%%%%%%%%%%%%%%%%%%%%%%%%%%%%%%
%%%%%%%%%%%%%%%%%%%%%%%%%%%%%%%%
%%%%%%%%%%%%%%%%%%%%%%%%%%%%%%%%
\section{Conclusions}\label{sec:conclusions}

In the final section, we summarize and discuss the results of the present work.
This paper theoretically investigates HHG spectra in the graphene model subjected to a DC electric field.
Through the combination of the quantum master and Boltzmann equations,
we numerically compute the HHG spectra with high accuracy and reveal their dependence on laser frequency, laser intensity, and DC current strength
while accounting for the experimentally inevitable dissipation effects.
The DC current induces
an asymmetric shift of the Fermi surface, as shown in Fig.~\ref{fig:figure2}.
Our numerical method provides a generic way of computing the HHG spectra of DC-current-driven electron systems from weak laser (perturbative) to strong laser (nonperturbative) regimes.
The methodology is explained in Sec.~\ref{Sec:Model}.
Compared with the previous studies for HHG in current-driven systems, our method makes it possible to observe higher-order (more than third-order) harmonic generations and HHG spectra in the nonperturbative regime.

Section~\ref{sec:result} gives the numerical results of the present study.
Throughout this section, we mainly consider the setup in which DC current is applied along the $x$ axis, and the external laser is linearly polarized along the same $x$ direction.
In Sec.~\ref{subsec:DC-strength}, we first show the shape of HHG spectra in a wide frequency regime in Fig.~\ref{fig:figure3}.
The spectra, especially the presence or absence of $n$th-order harmonics,
drastically change by tuning DC current, the laser ellipticity $\epsilon$ of the laser, and staggered potential $\Delta$.
We find that this characteristic feature can be proved by the argument based on dynamical symmetry (see Table~\ref{tab:sym}).
These results clearly indicate that HHG spectra can be moderately controlled with external tuning parameters
such as DC current and the laser ellipticity.

In Sec.~\ref{subsec:frequency}, we discuss the laser-frequency dependence of HHG spectra.
The SHG peak displays a divergent behavior when the chemical potential and laser frequency approach zero, $\Omega=\mu\rightarrow0$, in the weak-laser (perturbative) regime.
However, in the strong-laser regime, the intensity tends to become stronger when $\Omega=\mu$ increases, as shown in Fig.~\ref{fig:figure5}.
This behavior in the nonperturbative regime was first observed through our numerical method.
We also argue that photo excitations at the wavevectors around the DC-current-driven shifted Fermi surface are essential when we consider DC-current-driven harmonics (see Fig.~\ref{fig:figure6}).

Furthermore, when observing the fourth- and sixth-order harmonics for strong laser pulses (see Fig.~\ref{fig:figure7}), we find that the spectra cannot be explained by taking only interband optical transition processes,
and the influence of intraband dynamics becomes more pronounced as the laser intensity increases.

In Sec.~\ref{subsec:bandgap}, we discuss the HHG spectra $\tilde I_y(\Omega)$ generated by the laser-driven current along the $y$ axis, which appears only in the case of a finite staggered potential $\Delta$.
We find that as the laser intensity increases, multiple peak structures appear like other HHG spectra for a strong laser.
In Sec.~\ref{subsec:strong-field}, we consider the case of extremely strong laser pulses.
We observe a plateau regime in the HHG spectra, and we discuss the possibility that the DC current effect in HHG spectra becomes invisible as the laser intensity is extremely strong.

The present and recent works~\cite{Ikeda2019Highharmonic, Sato2020Twophoton, Kanega2021Linear}
indicate that the theoretical analysis based on the quantum master (GKSL) equation offers a powerful tool to compute optical nonlinear responses of many-body systems in a broad parameter regime.

Finally, we again comment on the limitations of our present approach. As discussed in Sec.~\ref{subsec:boltzmann}, there are two important limitations.
Firstly, the chemical potential must be sufficiently far from the Dirac point energy, meaning a sufficiently large Fermi surface should exist. Secondly, our resultant HHG spectra are reliable only when they exhibit a linear behavior with respect to the DC electric field. These two conditions stem from the fact that the current-driven nonequilibrium steady state before the laser application is prepared by using the Boltzmann equation in our approach.
However, we note that many experiments of HHG in current-driven electron systems satisfy these conditions, and hence our approach is applicable to most such experiments.

%%%%%%%%%%%%%%%%%%%%%%%%%%%%%%%%%%%%%%%%%
%%%%%%%%%%%%%%%%%%%%%%%%%%%%%%%%%%%%%%%%%
%{\bf \textit{Acknowledgments}}\\
\begin{acknowledgments}
    M. K. was supported by JST, the establishment of university fellowships towards the creation of science technology innovation, Grant No. JPMJFS2107,
    and by JST SPRING, Grant No. JPMJSP2109.
    M. S. was supported by JSPS KAKENHI (Grants No. 17K05513, No. 20H01830, and No. 20H01849) and
    a Grant-in-Aid for Scientific Research on Innovative Areas
    ``Quantum Liquid Crystals'' (Grant No. JP19H05825) and ``Evolution of Chiral Materials Science using
    Helical Light Fields'' (Grants No. JP22H05131 and No.
    JP23H04576).
\end{acknowledgments}

%%%%%%%%%%%%%%%%%%%%%%%%%%%%%%%%%%%%%%%%%
%%%%%%%%%%%%%%%%%%%%%%%%%%%%%%%%%%%%%%%%%

\appendix

\section{Selection rules for HHG}\label{app:dynamical-symmetry}

Here, we derive the selection rules for HHG from dynamical symmetries.
These symmetries exactly hold in the case of a continuous-wave laser, i.e., $t_{\mathrm{FWHM}}\rightarrow \infty$,
whereas it is known that the selection rules derived from dynamical symmetries are often applicable even in laser-pulse cases of real experiments.
In fact, our numerical results for laser pulses are consistent with such selection rules (see Fig.~\ref{fig:figure3} and Table~\ref{tab:sym}).
The selection rules below are all realized only in the case of the absence of the DC current, i.e., $\ddc = 0$.

\subsection{LPL and \texorpdfstring{$x$}{x}-direction inversion}\label{app:selection-ruleA}

First, we consider the irradiation of linearly polarized light (LPL) whose electric field is along the $x$ direction.
Since graphene has the symmetry of mirror operation for the $x$ direction ($x\to -x$),
the Hamiltonian $\hat H(t)$ satisfies the following dynamical symmetry:
\begin{equation}
    \hat U_{\sigma_{yz}}\hat{H}(t+T/2)\hat U_{\sigma_{yz}}^\dag = \hat{H}(t),\label{eq:inversion-symmetry}
\end{equation}
where $\hat U_{\sigma_{yz}}$ is the unitary operator of mirror operation for the $x$ direction.
Similarly, the current operator $\hat{\bm J}(t)$ satisfies
\begin{align}
     & \hat U_{\sigma_{yz}}\hat{\bm J}(t+T/2)\hat U_{\sigma_{yz}}^\dag
    =\mqty(
    -\hat J_x(t)                                                       \\
    \hat J_y(t))
    .
\end{align}
As we will show soon later, these conditions are equivalent to
\begin{align}
     & J_x(2n\Omega)=0,     \label{eq:selection-ruleA1} \\
     & J_y((2n+1)\Omega)=0,
    \label{eq:selection-ruleA2}
\end{align}
where $n$ is an arbitrary integer.
That is, in graphene illuminated by the LPL,
the $x$-direction mirror symmetry results in the prohibition of even-order (odd-order) harmonics of $J_x$ ($J_y$). Below, we prove Eqs.~\eqref{eq:selection-ruleA1} and~\eqref{eq:selection-ruleA2}.

Through the Fourier transformation, the Hamiltonian and the current operator are represented as
\begin{align}
     & \hat H(t) = \sum_\bk \hat H_\bk(t),           \\
     & \hat{\bm J}(t) = \sum_\bk \hat{\bm J}_\bk(t).
\end{align}
These Fourier components with each wavevector $\bk$ satisfy
\begin{align}
     & \hat U_{\sigma_{yz}}\hat H_\bk(t+T/2)\hat U_{\sigma_{yz}}^\dag=\hat H_{\bk'}(t),        \\
     & \hat U_{\sigma_{yz}}\hat J_{x,\bk}(t+T/2)\hat U_{\sigma_{yz}}^\dag=-\hat J_{x,\bk'}(t), \\
     & \hat U_{\sigma_{yz}}\hat J_{y,\bk}(t+T/2)\hat U_{\sigma_{yz}}^\dag=\hat J_{y,\bk'}(t),
\end{align}
where $\bk'=(-k_x,k_y)$.
Next, we consider the density matrix.
Expressing the quantum master equation (Eq.\eqref{eq:quantum-master-eqre}) symbolically using the Liouvillian superoperator $\mathcal{L}_\bk(t)$, we have
\begin{equation}
    \dv{t}\hat{\rho}_\bk(t)=\mathcal{L}_\bk(t)\hat{\rho}_\bk(t).
    %   \label{eq:quantum-master-eqre}
\end{equation}
Since the jump operators in our setup satisfy
\begin{equation}
    \hat U_{\sigma_{yz}}\hat L_\bk \hat U_{\sigma_{yz}}^\dag=\hat L_{\bk'},
    %   \label{eq:lindbladSym}
\end{equation}
we obtain
\begin{align}
    \dv{t}\hat{\rho}_{\bk'}(t) & =\mathcal{L}_{\bk'}(t)\hat{\rho}_{\bk'}(t)\notag                                             \\
    %&
                               & =\hat U_{\sigma_{yz}}\mathcal{L}_{\bk}(t+T/2) \hat U_{\sigma_{yz}}^\dag\hat{\rho}_{\bk'}(t).
\end{align}
Therefore, we arrive at
\begin{equation}
    \dv{t}\hat U_{\sigma_{yz}}^\dag\hat{\rho}_{\bk'}(t)\hat U_{\sigma_{yz}}=\mathcal{L}_\bk(t+T/2) \hat U_{\sigma_{yz}}^\dag\hat{\rho}_{\bk'}(t)\hat U_{\sigma_{yz}}.
    \label{eq:dens-DS}
\end{equation}
By comparing Eq.~(\ref{eq:dens-DS}) with the original quantum master equation,
we find the equality,
\begin{equation}
    \hat U_{\sigma_{yz}}\hat{\rho}_{\bk}(t+T/2)\hat U_{\sigma_{yz}}^\dag=\hat{\rho}_{\bk'}(t).
    \label{eq:dens-unitaryre}
\end{equation}
This relation may be referred to as dynamical symmetry for the density matrix.
From this dynamical symmetry, the Fourier component of the $x$-direction current satisfies the following relation:
\begin{align}
    J_{x,\bk'}(t) & = \braket <\hat J_{x,\bk'}(t)>_t \notag                       \\
                  & =\Tr\ab[\hat{\rho}_{\bk'}(t)\hat J_{x,\bk'}(t)]  \notag       \\
                  & =-\Tr\ab[\hat{\rho}_{\bk}(t+T/2)\hat J_{x,\bk}(t+T/2)] \notag \\
                  & =-J_{x,\bk}(t+T/2).
\end{align}
Through a similar process for the $y$ component of current, we obtain
\begin{equation}
    J_{y,\bk'}(t) = J_{y,\bk}(t+T/2).
\end{equation}
Here, from the pair of $J_{\alpha,\bk}(t)$ and $J_{\alpha,\bk'}(t)$, we define
\begin{align}
     & \tilde J_{x,\bk}(t) \coloneqq J_{x,\bk}(t) + J_{x,\bk'}(t), \\
     & \tilde J_{y,\bk}(t) \coloneqq J_{y,\bk}(t) + J_{y,\bk'}(t).
\end{align}
We then find that they satisfy
\begin{align}
     & \tilde J_{x,\bk}(t) = -\tilde J_{x,\bk}(t+T/2), \\
     & \tilde J_{y,\bk}(t) = \tilde J_{y,\bk}(t+T/2).
\end{align}
These results directly lead to the selection rules for the HHG spectrum as follows.
The $x$ component of the current,
$\tilde J_{x,\bk}(n\Omega)$, is transformed as
\begin{align}
    \tilde J_{x,\bk}(n\Omega) & =\int_0^T\frac{dt}{T}\tilde J_{x,\bk}(t)e^{in\Omega t}\notag                                  \\
                              & =\int_{-T/2}^{T/2}\frac{dt}{T}\tilde J_{x,\bk}(t+T/2)e^{in\Omega(t+T/2)}\notag                \\
                              & =-e^{in\Omega\frac{T}{2}}\int_{-T/2}^{T/2}\frac{dt}{T}\tilde J_{x,\bk}(t)e^{in\Omega t}\notag \\
                              & =-e^{in\pi}\tilde J_{x,\bk}(n\Omega).
    %   \label{eq:polOdd}
\end{align}
Therefore, we have
\begin{equation}
    \tilde J_{x,\bk}(2m\Omega)=0\quad(m\in\mathbb Z).
\end{equation}
Similarly, for $\tilde J_y(n\Omega)$, we obtain
\begin{equation}
    \tilde J_{y,\bk}(n\Omega) = e^{in\pi}\tilde J_{y,\bk}(n\Omega)
\end{equation}
and
\begin{equation}
    \tilde J_{y,\bk}((2m+1)\Omega)=0\quad(m\in\mathbb Z).
\end{equation}
Taking the summation $\sum_{k_x>0}\tilde J_{\alpha,\bk}(n\Omega)$ in the positive-$k_x$ region of the Brillouin zone, we finally arrive at Eqs.~\eqref{eq:selection-ruleA1} and~\eqref{eq:selection-ruleA2}.

%%%%%%%%%%%%%%%%%%%%%%%%%%%%%%%%%%
%%%%%%%%%%%%%%%%%%%%%%%%%%%%%%%%%%
%%%%%%%%%%%%%%%%%%%%%%%%%%%%%%%%%%
\subsection{CPL and \texorpdfstring{$n$}{n}-fold rotation in the \texorpdfstring{$x-y$}{xy} plane}\label{app:selection-ruleB}

Next, we consider the case of the application of circularly polarized light (CPL) to a two-dimensional (2D) electron system on the $x$-$y$ plane.
Here, the electric field of CPL is in the same $x$-$y$ plane.
We assume that a CPL-driven 2D system in the $x$-$y$ plane satisfies the following symmetry relations
\begin{align}
     & \hat H_\bk(t) = \hat U_{n}\hat H_{\mathcal R_n\bk}(t+T/n)\hat U_{n}^\dag,                       \\
     & \hat{\bm J}_\bk(t) = \hat U_{n}\mathcal R_n\hat{\bm J}_{\mathcal R_n\bk}(t+T/n)\hat U_{n}^\dag,
\end{align}
where $\hat U_n$ is the unitary operator of $2\pi/n$-degree rotation in the $x$-$y$ plane and $\mathcal R_n$ is the $2\times2$ matrix for an in-plane rotation by $2\pi/n$, which acts on vector quantities.
In graphene models, as mentioned in Table~\ref{tab:sym} of the main text,
the above relations hold for $n=2$ and $n=3$.
We note that when 2D electron systems are irradiated by the CPL, its electric field rotates by $2\pi/n$ during the time interval $T/n$.

%%%%%%%%%%%%%%%%%%
For the above discrete-rotation symmetric system, we then consider the time evolution of the density matrix.
Similarly to the Appendix~\ref{app:selection-ruleA}, the quantum master equation (Eq.~\eqref{eq:quantum-master-eqre}) is expressed as
\begin{equation}
    \dv{t}\hat{\rho}_\bk(t)=\mathcal{L}_\bk(t)\hat{\rho}_\bk(t).
    %   \label{eq:quantum-master-eqre}
\end{equation}
Assuming the jump operators satisfy
\begin{equation}
    \hat L_\bk = \hat U_{n}\hat L_{\mathcal R_n\bk}\hat U_{n}^\dag,
    %   \label{eq:lindbladSym}
\end{equation}
we obtain
\begin{align}
    \dv{t}\hat{\rho}_\bk(t) & =\mathcal{L}_{\bk}(t)\hat{\rho}_\bk(t)\notag                                      \\
    %&
                            & =\hat U_{n}\mathcal{L}_{\mathcal R_n\bk}(t+T/n) \hat U_{n}^\dag\hat{\rho}_\bk(t).
\end{align}
Therefore, we find
\begin{equation}
    \dv{t}\hat U_{n}^\dag\hat{\rho}_\bk(t)\hat U_{n}=\mathcal{L}_{\mathcal R_n\bk}(t+T/n) \hat U_{n}^\dag\hat{\rho}_\bk(t)\hat U_{n}.
    %   \label{eq:dens-unitaryre}
\end{equation}
Comparing this result with the original quantum master equation,
we obtain
\begin{equation}
    \hat{\rho}_\bk(t) = \hat U_{n}\hat{\rho}_{\mathcal R_n\bk}(t+T/n)\hat U_{n}^\dag.
\end{equation}
Through this dynamical symmetry of the density matrix, the $\bk$ component of the current is computed as follows:
\begin{align}
    \bm J_\bk(t) = \braket <\hat{\bm J}_\bk(t)>_t & =\Tr\ab[\hat{\rho}_\bk(t)\hat{\bm J}_\bk(t)]  \notag                                                \\
                                                  & =\mathcal R_n\Tr\ab[\hat{\rho}_{\mathcal R_n\bk}(t+T/n)\hat{\bm J}_{\mathcal R_n\bk}(t+T/n)] \notag \\
                                                  & =\mathcal R_n\bm J_{\mathcal R_n\bk}(t+T/n).
\end{align}
Considering this relation about the rotating operations, we introduce a new quantity
\begin{align}
     & \tilde{\bm J}_\bk(t) \coloneqq \sum_{\ell=0}^{n-1}\bm J_{\mathcal (R_n)^\ell\bk}(t).
\end{align}
Then, we find that it satisfies
\begin{align}
    \tilde{\bm J}_\bk(t) & = \mathcal R_n\tilde{\bm J}_\bk(t+T/n) = (\mathcal R_n)^2\tilde{\bm J}_\bk(t+2T/n)\notag \\
                         & = \cdots = (\mathcal R_n)^{n-1}\tilde{\bm J}_\bk(t+(n-1)T/n).
    \label{eq:current_rot}
\end{align}
Equation~(\ref{eq:current_rot}) enables us to derive the selection rules for the HHG spectrum.
From the equation, the Fourier component of the current in time direction, $\bm J(m\Omega)$, is given by
\begin{align}
    \tilde{\bm J}_\bk(m\Omega) & =\mathcal R_ne^{i2m\pi/n}\tilde{\bm J}_\bk(m\Omega) = (\mathcal R_n)^2e^{i4m\pi/n}\tilde{\bm J}_\bk(m\Omega)\notag \\
                               & = \cdots = (\mathcal R_n)^{n-1}e^{i2(n-1)m\pi/n}\tilde{\bm J}_\bk(m\Omega).
\end{align}
If we consider the case of $m=n\ell\;\;(\ell\in\mathbb Z)$, we obtain
\begin{align}
     & \tilde{\bm J}_\bk(n\ell\Omega)\notag
    \\
     & = \frac{1}{n}\ab(1 + \mathcal R_n + (\mathcal R_n)^2 + \cdots + (\mathcal R_n)^{n-1})\tilde{\bm J}_\bk(n\ell\Omega).
\end{align}
Thus we have $\tilde{\bm J}_\bk(n\ell\Omega)=\bm 0$.
Taking the proper summation of $\tilde{\bm J}_\bk(n\ell\Omega)$ over the Brillouin zone,
we finally obtain
\begin{align}
    \bm J(n\ell\Omega)=\bm 0.
\end{align}
This means that the $n\ell$th-order harmonics with $n=2$ and $3$ disappear in graphene models under the irradiation of circularly polarized light.

%%%%%%%%%%%%%%%%%%%%%%%%%%%%%%%%%%
%%%%%%%%%%%%%%%%%%%%%%%%%%%%%%%%%%
%%%%%%%%%%%%%%%%%%%%%%%%%%%%%%%%%%
\subsection{Breakdown of dynamical symmetry by applying DC current}

The dynamical symmetry of the density matrix is broken by applying a DC current, and it usually accompanies the appearance of the harmonics forbidden by the dynamical symmetry.
In the case of Appendix~\ref{app:selection-ruleA}, the breakdown of the dynamical symmetry means the following inequality:
\begin{equation}
    \hat{\rho}_\bk(t) \neq \hat U_{\sigma_{yz}}\hat{\rho}_{\bk'}(t+T/2)\hat U_{\sigma_{yz}}^\dag.
    \label{eq:dens-unitaryre2}
\end{equation}
In this section, we discuss why this inequality is generally realized in the case of applying a DC current.

When we have a finite DC current,
the initial state before the application of a laser is given by
$\hat{\rho}_{\bk} = \dyad{g_\bkd}{g_\bkd}$. This density matrix of the NESS clearly breaks the mirror symmetry of $\hat U_{\sigma_{yz}}$, as shown in Fig.~\ref{fig:figure2}(a).
This could be a simple answer to why the inequality of Eq.~(\ref{eq:dens-unitaryre2}) holds.

We also try to construct a more serious argument for Eq.~(\ref{eq:dens-unitaryre2}).
We first assume that the density matrix at the initial time $t=t_0$ is given by $\hat{\rho}_{\bk} = \dyad{g_\bkd}{g_\bkd}$ ($\dyad{g_\bk}{g_\bk}$) for the case of a finite (zero) DC current.
Then, a continuous-wave laser ($t_{\mathrm{FWHM}}\to\infty$) is assumed to be adiabatically introduced. Under this setup, let us first consider the case without DC current.
As discussed in Appendix~\ref{app:selection-ruleA}, when $\ddc=0$,
\begin{equation}
    \mathcal{L}_{\bk}(t) = \hat U_{\sigma_{yz}}\mathcal{L}_{\bk'}(t+T/2) \hat U_{\sigma_{yz}}^\dag,
\end{equation}
holds.
Therefore, we can expect that the time-evolution superoperator $\mathcal{V}_\bk(t,t_0)=\mathcal{T}\exp(\int_{t_0}^{t}\mathcal{L}_{\bk}(s)\dd s)$ satisfies
\begin{equation}
    \mathcal{V}_{\bk}(t,t_0) = \hat U_{\sigma_{yz}}\mathcal{V}_{\bk'}(t+T/2,t_0) \hat U_{\sigma_{yz}}^\dag,
\end{equation}
if $t$ is sufficiently far from $t_0$. Here, $\mathcal{T}$ represents the time-ordered product.
This relation directly leads to Eq.~\eqref{eq:dens-unitaryre} as follows:
\begin{align}
     & \hat U_{\sigma_{yz}}\hat{\rho}_{\bk'}(t+T/2)\hat U_{\sigma_{yz}}^\dag \notag                                                                     \\
     & = \hat U_{\sigma_{yz}}\mathcal{V}_{\bk'}(t+T/2,t_0)\hat U_{\sigma_{yz}}^\dag\hat U_{\sigma_{yz}}\hat{\rho}_{\bk'}\hat U_{\sigma_{yz}}^\dag\notag \\
     & = \mathcal{V}_{\bk}(t,t_0)\hat{\rho}_{\bk} \notag
    \\
     & = \hat{\rho}_{\bk}(t).
\end{align}
Here, $\hat{\rho}_{\bk}\coloneqq\hat{\rho}_{\bk}(t_0)=\dyad{g_\bk}{g_\bk}$,
and we have used $\hat U_{\sigma_{yz}}\hat{\rho}_{\bk'}\hat U_{\sigma_{yz}}^\dag = \hat U_{\sigma_{yz}}\dyad{g_{\bk'}}{g_{\bk'}}\hat U_{\sigma_{yz}}^\dag = \dyad{g_\bk}{g_\bk} = \hat{\rho}_{\bk}$.

Next, we consider the case of $\ddc\neq0$, in which the density matrix at $t=t_0$ is given by $\hat{\rho}_{\bk} = \dyad{g_\bkd}{g_\bkd}$.
As we mentioned, this density matrix follows
\begin{align}
    \hat U_{\sigma_{yz}}\hat{\rho}_{\bk'}\hat U_{\sigma_{yz}}^\dag & = \hat U_{\sigma_{yz}}\dyad{g_{\bk'+\bm\delta_{\mathrm{dc}}}}{g_{\bk'+\bm\delta_{\mathrm{dc}}}}\hat U_{\sigma_{yz}}^\dag \notag \\
                                                                   & = \dyad{g_{\bk+\bm\delta_{\mathrm{dc}}'}}{g_{\bk+\bm\delta_{\mathrm{dc}}'}}\notag                                               \\
                                                                   & \neq \hat{\rho}_{\bk},
    \label{eq:inequal_density}
\end{align}
where we have introduced the new symbol $\bm\delta_{\mathrm{dc}}'=(-\ddc^x,\ddc^y)$.
In addition, for $\ddc\neq0$, we have
\begin{equation}
    \hat U_{\sigma_{yz}}\hat L_\bk \hat U_{\sigma_{yz}}^\dag\neq \hat L_{\bk'}.
    %   \label{eq:lindbladSym}
\end{equation}
This leads to
\begin{equation}
    \mathcal{V}_{\bk}(t,t_0) \neq \hat U_{\sigma_{yz}}\mathcal{V}_{\bk'}(t+T/2,t_0) \hat U_{\sigma_{yz}}^\dag.
    \label{eq:inequal_timeevolv}
\end{equation}
From these two inequalities of Eqs.~(\ref{eq:inequal_density}) and (\ref{eq:inequal_timeevolv}), we
can say that Eq.~\eqref{eq:dens-unitaryre2} generally holds when we apply a DC current.
In discrete-rotation symmetric systems in Appendix~\ref{app:selection-ruleB}, we can also make an argument in a similar way that the dynamical symmetry is generally broken by a DC current.

%%%%%%%%%%%%%%%%%%%%%%%%%%%%%%%%%%
%%%%%%%%%%%%%%%%%%%%%%%%%%%%%%%%%%
%%%%%%%%%%%%%%%%%%%%%%%%%%%%%%%%%%
\section{Relationship between the master equation and the Bloch equation}\label{app:relation}

This section briefly shows the relationship between the quantum master equation and the optical Bloch equation~\cite{Haug2008Quantum}.

We start from the master equation we have adopted:
\begin{align}
    \dv{\hat{\rho}_\bk(t)}{t}= & -i\comm{\hat{H}_\bk(t)}{\hat{\rho}_\bk(t)}\notag
    \\
                               & +\gamma\ab(\hat{L}_\bk\hat{\rho}_\bk(t)\hat{L}_\bk^\dag-\frac{1}{2}\acomm{\hat{L}_\bk^\dag\hat{L}_\bk}{\hat{\rho}_\bk(t)}).
    \label{eq:quantum-master-eqre2}
\end{align}
This is the equation for a $\bk$-diagonal
two-band system, including a simple relaxation process by the jump operator $\hat L_\bk=\dyad{g_\bk}{e_\bk}$.
In two-level systems,
any Hermitian operator $\hat A$ can be expressed using identity operator
$\hat I$ and Pauli operators
$\hat{\bm\sigma}=(\hat{\sigma}^x,\hat{\sigma}^y,\hat{\sigma}^z)$ as
\begin{equation}
    \hat A = A_0\hat I + \bm A\cdot\hat{\bm\sigma}
    \label{eq:2level}
\end{equation}
where $\bm A = (A_x, A_y, A_z)$ is a three-dimensional vector and $A_{0,x,y,z}\in\mathbb{R}$.
Hence, the product of two Hermitian operators is given by
\begin{equation}
    \hat A\hat B = (A_0B_0 + \bm A\cdot\bm B)\hat I + (A_0\bm B + B_0\bm A + i\bm A\times\bm B)\cdot\hat{\bm\sigma}.
    \label{eq:product}
\end{equation}
Since the Hamiltonian $\hat H_\bk(t)$ and the density matrix $\hat\rho_\bk(t)$ are both Hermitian,
they may be expressed as $\hat H_\bk(t) = \bm h_\bk(t)\cdot\hat{\bm\sigma}$ and $\hat\rho_\bk(t) = \tau^0_\bk(t)\hat I + \bm\tau_\bk(t)\cdot\hat{\bm\sigma}$.
Here, we have introduced real vector quantities
$\bm h_\bk(t)$ and $\bm\tau_\bk(t)$ and a real scalar quantity $\tau^0_\bk(t)$.
The jump operator is not generally a Hermitian operator,
but in the present model,
it can also be written by using Pauli
matrices as follows:
\begin{equation}
    \hat L_\bk = \bm L_\bk\cdot\hat{\bm\sigma},
    \label{eq:jump-operator}
\end{equation}
with the complex coefficients $\bm L_\bk = (L^x_\bk, L^y_\bk, L^z_\bk)$.
Since the jump operator we used is $\hat L_\bk = \dyad{g_\bk}{e_\bk} = (\hat\sigma^x -i\hat\sigma^y)/2$, the coefficient is given by $\bm L_\bk = (1/2,-i/2,0)$.

Substituting the $2\times 2$ forms of the Hamiltonian, the density matrix and the jump operator
into Eq.~\eqref{eq:quantum-master-eqre2} and then using Eq.~(\ref{eq:product}), we obtain
\begin{widetext}
    \begin{align}
        \dv{\tau^0_\bk}{t}\hat I
        + \ab(\dv{\bm\tau_\bk}{t}
        - 2\bm h_\bk\times\bm\tau_\bk
        - 2\gamma\ab[\mathrm{Re}\ab[(\bm L_\bk\cdot\bm\tau_\bk)\bm L^*_\bk]
            - \abs{\bm L_\bk}^2\bm\tau_\bk
            + i\tau^0_\bk\bm L_\bk\times\bm L^*_\bk]
        )\cdot\hat{\bm\sigma}
        = 0.
    \end{align}
\end{widetext}
This can be viewed as coupled differential equations for $\tau^0_\bk$ and $\bm\tau_\bk$.
Focusing on the coefficient of $\hat I$, we have $\dv{\tau^0_\bk}{t}=0$, and its solution is given by $\tau^0_\bk(t)=1$ because of the normalization of the density matrix.
The remaining vector $\bm\tau_\bk(t)$ satisfies
\begin{align}
    \dv{\bm\tau_\bk}{t} = &
    2\bm h_\bk\times\bm\tau_\bk \notag                                                        \\
    +
                          & 2\gamma\ab[\mathrm{Re}\ab[(\bm L_\bk\cdot\bm\tau_\bk)\bm L^*_\bk]
        - \abs{\bm L_\bk}^2\bm\tau_\bk
        + i\bm L_\bk\times\bm L^*_\bk].
\end{align}
Substituting $\bm L_\bk = (1/2,-i/2,0)$,
we arrive at
\begin{align}
    \dv{t}\bm\tau_\bk(t) =
    2\bm h_\bk(t)\times\bm\tau_\bk(t)
    - \mqty(\frac{\gamma}{2}\tau_{x,\bk}(t) \\ \frac{\gamma}{2}\tau_{y,\bk}(t) \\ \gamma(\tau_{z,\bk}(t)-\expval{\tau_{z,\bk}})),
\end{align}
where $\expval{\tau_{z,\bk}}=-1$.
This is nothing but a Bloch equation whose longitudinal and transverse relaxation times, $T_1$ and $T_2$, are given by
\begin{align}
    T_1 = \frac{1}{\gamma},
    \quad T_2 = \frac{2}{\gamma}.
\end{align}
Namely, our simple setup of the jump operator includes both longitudinal and transverse relaxation processes, satisfying the detailed balance condition at $T=0$.

In the present study, we have adopted a simple but realistic relaxation (jump) operator, while in general, the GKSL equation can describe wider sorts of dissipation processes rather than the Bloch equation~\cite{Breuer2007Theory,Haug2008Quantum,Tanaka2024Theory}.
Namely, in the mathematical sense, the Bloch equation is included within the framework of the GKSL equation.

We note that theoretical studies beyond the above relaxation-time approximation have also progressed~\cite{Passos2018Nonlinear, Michishita2021Effects, Terada2024Limitations}.
For instance, the quantum kinetic equation approach based on the Green's functions~\cite{Haug2008Quantum,Stefanucci2013Nonequilibrium} can describe the effects of various scattering processes, in principle, from the microscopic viewpoint.
However, it is generally difficult to simultaneously treat light-matter couplings, interactions, and impurity scatterings within the analytic quantum kinetic approach.
In fact, most of the previous quantum kinetic approaches for HHG take these effects into account in a perturbative manner, where only lower-order harmonics can be computed~\cite{Bhalla2020Resonant,Bhalla2022Resonant}.
On the other hand, our numerical approach using the GKSL equation with a phenomenological relaxation time can analyze weak-to-strong light-induced nonequilibrium phenomena, especially the intense-laser-driven nonperturbative effects.
As discussed in Sec.~\ref{subsec:frequency}, the analytic, perturbative approach and our numerical one are complementary with each other.


\begin{thebibliography}{94}%
    \makeatletter
    \providecommand \@ifxundefined [1]{%
        \@ifx{#1\undefined}
    }%
    \providecommand \@ifnum [1]{%
        \ifnum #1\expandafter \@firstoftwo
        \else \expandafter \@secondoftwo
        \fi
    }%
    \providecommand \@ifx [1]{%
        \ifx #1\expandafter \@firstoftwo
        \else \expandafter \@secondoftwo
        \fi
    }%
    \providecommand \natexlab [1]{#1}%
    \providecommand \enquote  [1]{``#1''}%
    \providecommand \bibnamefont  [1]{#1}%
    \providecommand \bibfnamefont [1]{#1}%
    \providecommand \citenamefont [1]{#1}%
    \providecommand \href@noop [0]{\@secondoftwo}%
    \providecommand \href [0]{\begingroup \@sanitize@url \@href}%
    \providecommand \@href[1]{\@@startlink{#1}\@@href}%
    \providecommand \@@href[1]{\endgroup#1\@@endlink}%
    \providecommand \@sanitize@url [0]{\catcode `\\12\catcode `\$12\catcode `\&12\catcode `\#12\catcode `\^12\catcode `\_12\catcode `\%12\relax}%
    \providecommand \@@startlink[1]{}%
    \providecommand \@@endlink[0]{}%
    \providecommand \url  [0]{\begingroup\@sanitize@url \@url }%
    \providecommand \@url [1]{\endgroup\@href {#1}{\urlprefix }}%
    \providecommand \urlprefix  [0]{URL }%
    \providecommand \Eprint [0]{\href }%
    \providecommand \doibase [0]{https://doi.org/}%
    \providecommand \selectlanguage [0]{\@gobble}%
    \providecommand \bibinfo  [0]{\@secondoftwo}%
    \providecommand \bibfield  [0]{\@secondoftwo}%
    \providecommand \translation [1]{[#1]}%
    \providecommand \BibitemOpen [0]{}%
    \providecommand \bibitemStop [0]{}%
    \providecommand \bibitemNoStop [0]{.\EOS\space}%
    \providecommand \EOS [0]{\spacefactor3000\relax}%
    \providecommand \BibitemShut  [1]{\csname bibitem#1\endcsname}%
    \let\auto@bib@innerbib\@empty
    %</preamble>
    \bibitem [{\citenamefont {Ghimire}\ and\ \citenamefont {Reis}(2019)}]{Ghimire2019Highharmonic}%
    \BibitemOpen
    \bibfield  {author} {\bibinfo {author} {\bibfnamefont {S.}~\bibnamefont {Ghimire}}\ and\ \bibinfo {author} {\bibfnamefont {D.~A.}\ \bibnamefont {Reis}},\ }\bibfield  {title} {\bibinfo {title} {High-harmonic generation from solids},\ }\href {https://doi.org/10.1038/s41567-018-0315-5} {\bibfield  {journal} {\bibinfo  {journal} {Nature Phys}\ }\textbf {\bibinfo {volume} {15}},\ \bibinfo {pages} {10} (\bibinfo {year} {2019})}\BibitemShut {NoStop}%
    \bibitem [{\citenamefont {Yue}\ and\ \citenamefont {Gaarde}(2022)}]{Yue2022Introduction}%
    \BibitemOpen
    \bibfield  {author} {\bibinfo {author} {\bibfnamefont {L.}~\bibnamefont {Yue}}\ and\ \bibinfo {author} {\bibfnamefont {M.~B.}\ \bibnamefont {Gaarde}},\ }\bibfield  {title} {\bibinfo {title} {Introduction to theory of high-harmonic generation in solids: Tutorial},\ }\href {https://doi.org/10.1364/JOSAB.448602} {\bibfield  {journal} {\bibinfo  {journal} {J. Opt. Soc. Am. B, JOSAB}\ }\textbf {\bibinfo {volume} {39}},\ \bibinfo {pages} {535} (\bibinfo {year} {2022})}\BibitemShut {NoStop}%
    \bibitem [{\citenamefont {Li}\ \emph {et~al.}(2023)\citenamefont {Li}, \citenamefont {Lan}, \citenamefont {Zhu},\ and\ \citenamefont {Lu}}]{Li2023High}%
    \BibitemOpen
    \bibfield  {author} {\bibinfo {author} {\bibfnamefont {L.}~\bibnamefont {Li}}, \bibinfo {author} {\bibfnamefont {P.}~\bibnamefont {Lan}}, \bibinfo {author} {\bibfnamefont {X.}~\bibnamefont {Zhu}},\ and\ \bibinfo {author} {\bibfnamefont {P.}~\bibnamefont {Lu}},\ }\bibfield  {title} {\bibinfo {title} {High harmonic generation in solids: Particle and wave perspectives},\ }\href {https://doi.org/10.1088/1361-6633/acf144} {\bibfield  {journal} {\bibinfo  {journal} {Rep. Prog. Phys.}\ }\textbf {\bibinfo {volume} {86}},\ \bibinfo {pages} {116401} (\bibinfo {year} {2023})}\BibitemShut {NoStop}%
    \bibitem [{\citenamefont {Bhattacharya}\ \emph {et~al.}(2023)\citenamefont {Bhattacharya}, \citenamefont {Lamprou}, \citenamefont {Maxwell}, \citenamefont {Ord{\'o}{\~n}ez}, \citenamefont {Pisanty}, \citenamefont {{Rivera-Dean}}, \citenamefont {Stammer}, \citenamefont {Ciappina}, \citenamefont {Lewenstein},\ and\ \citenamefont {Tzallas}}]{Bhattacharya2023Strong}%
    \BibitemOpen
    \bibfield  {author} {\bibinfo {author} {\bibfnamefont {U.}~\bibnamefont {Bhattacharya}}, \bibinfo {author} {\bibfnamefont {T.}~\bibnamefont {Lamprou}}, \bibinfo {author} {\bibfnamefont {A.~S.}\ \bibnamefont {Maxwell}}, \bibinfo {author} {\bibfnamefont {A.}~\bibnamefont {Ord{\'o}{\~n}ez}}, \bibinfo {author} {\bibfnamefont {E.}~\bibnamefont {Pisanty}}, \bibinfo {author} {\bibfnamefont {J.}~\bibnamefont {{Rivera-Dean}}}, \bibinfo {author} {\bibfnamefont {P.}~\bibnamefont {Stammer}}, \bibinfo {author} {\bibfnamefont {M.~F.}\ \bibnamefont {Ciappina}}, \bibinfo {author} {\bibfnamefont {M.}~\bibnamefont {Lewenstein}},\ and\ \bibinfo {author} {\bibfnamefont {P.}~\bibnamefont {Tzallas}},\ }\bibfield  {title} {\bibinfo {title} {Strong--laser--field physics, non--classical light states and quantum information science},\ }\href {https://doi.org/10.1088/1361-6633/acea31} {\bibfield  {journal} {\bibinfo  {journal} {Rep. Prog. Phys.}\ }\textbf {\bibinfo {volume} {86}},\ \bibinfo {pages} {094401} (\bibinfo {year} {2023})}\BibitemShut {NoStop}%
    \bibitem [{\citenamefont {Hirori}\ \emph {et~al.}(2024)\citenamefont {Hirori}, \citenamefont {Sato},\ and\ \citenamefont {Kanemitsu}}]{Hirori2024HighOrder}%
    \BibitemOpen
    \bibfield  {author} {\bibinfo {author} {\bibfnamefont {H.}~\bibnamefont {Hirori}}, \bibinfo {author} {\bibfnamefont {S.~A.}\ \bibnamefont {Sato}},\ and\ \bibinfo {author} {\bibfnamefont {Y.}~\bibnamefont {Kanemitsu}},\ }\bibfield  {title} {\bibinfo {title} {High-{{Order Harmonic Generation}} in {{Solids}}: {{The Role}} of {{Intraband Transitions}} in {{Extreme Nonlinear Optics}}},\ }\href {https://doi.org/10.1021/acs.jpclett.3c03415} {\bibfield  {journal} {\bibinfo  {journal} {J. Phys. Chem. Lett.}\ }\textbf {\bibinfo {volume} {15}},\ \bibinfo {pages} {2184} (\bibinfo {year} {2024})}\BibitemShut {NoStop}%
    \bibitem [{\citenamefont {Young}\ and\ \citenamefont {Rappe}(2012)}]{Young2012First}%
    \BibitemOpen
    \bibfield  {author} {\bibinfo {author} {\bibfnamefont {S.~M.}\ \bibnamefont {Young}}\ and\ \bibinfo {author} {\bibfnamefont {A.~M.}\ \bibnamefont {Rappe}},\ }\bibfield  {title} {\bibinfo {title} {First {{Principles Calculation}} of the {{Shift Current Photovoltaic Effect}} in {{Ferroelectrics}}},\ }\href {https://doi.org/10.1103/PhysRevLett.109.116601} {\bibfield  {journal} {\bibinfo  {journal} {Phys. Rev. Lett.}\ }\textbf {\bibinfo {volume} {109}},\ \bibinfo {pages} {116601} (\bibinfo {year} {2012})}\BibitemShut {NoStop}%
    \bibitem [{\citenamefont {Tan}\ \emph {et~al.}(2016)\citenamefont {Tan}, \citenamefont {Zheng}, \citenamefont {Young}, \citenamefont {Wang}, \citenamefont {Liu},\ and\ \citenamefont {Rappe}}]{Tan2016Shift}%
    \BibitemOpen
    \bibfield  {author} {\bibinfo {author} {\bibfnamefont {L.~Z.}\ \bibnamefont {Tan}}, \bibinfo {author} {\bibfnamefont {F.}~\bibnamefont {Zheng}}, \bibinfo {author} {\bibfnamefont {S.~M.}\ \bibnamefont {Young}}, \bibinfo {author} {\bibfnamefont {F.}~\bibnamefont {Wang}}, \bibinfo {author} {\bibfnamefont {S.}~\bibnamefont {Liu}},\ and\ \bibinfo {author} {\bibfnamefont {A.~M.}\ \bibnamefont {Rappe}},\ }\bibfield  {title} {\bibinfo {title} {Shift current bulk photovoltaic effect in polar materials---hybrid and oxide perovskites and beyond},\ }\href {https://doi.org/10.1038/npjcompumats.2016.26} {\bibfield  {journal} {\bibinfo  {journal} {npj Comput Mater}\ }\textbf {\bibinfo {volume} {2}},\ \bibinfo {pages} {1} (\bibinfo {year} {2016})}\BibitemShut {NoStop}%
    \bibitem [{\citenamefont {Morimoto}\ and\ \citenamefont {Nagaosa}(2016)}]{Morimoto2016Topological}%
    \BibitemOpen
    \bibfield  {author} {\bibinfo {author} {\bibfnamefont {T.}~\bibnamefont {Morimoto}}\ and\ \bibinfo {author} {\bibfnamefont {N.}~\bibnamefont {Nagaosa}},\ }\bibfield  {title} {\bibinfo {title} {Topological nature of nonlinear optical effects in solids},\ }\href {https://doi.org/10.1126/sciadv.1501524} {\bibfield  {journal} {\bibinfo  {journal} {Sci. Adv.}\ }\textbf {\bibinfo {volume} {2}},\ \bibinfo {pages} {e1501524} (\bibinfo {year} {2016})}\BibitemShut {NoStop}%
    \bibitem [{\citenamefont {Tokura}\ and\ \citenamefont {Nagaosa}(2018)}]{Tokura2018Nonreciprocal}%
    \BibitemOpen
    \bibfield  {author} {\bibinfo {author} {\bibfnamefont {Y.}~\bibnamefont {Tokura}}\ and\ \bibinfo {author} {\bibfnamefont {N.}~\bibnamefont {Nagaosa}},\ }\bibfield  {title} {\bibinfo {title} {Nonreciprocal responses from non-centrosymmetric quantum materials},\ }\href {https://doi.org/10.1038/s41467-018-05759-4} {\bibfield  {journal} {\bibinfo  {journal} {Nat Commun}\ }\textbf {\bibinfo {volume} {9}},\ \bibinfo {pages} {3740} (\bibinfo {year} {2018})}\BibitemShut {NoStop}%
    \bibitem [{\citenamefont {Ishizuka}\ and\ \citenamefont {Sato}(2019)}]{Ishizuka2019Rectification}%
    \BibitemOpen
    \bibfield  {author} {\bibinfo {author} {\bibfnamefont {H.}~\bibnamefont {Ishizuka}}\ and\ \bibinfo {author} {\bibfnamefont {M.}~\bibnamefont {Sato}},\ }\bibfield  {title} {\bibinfo {title} {Rectification of {{Spin Current}} in {{Inversion-Asymmetric Magnets}} with {{Linearly Polarized Electromagnetic Waves}}},\ }\href {https://doi.org/10.1103/PhysRevLett.122.197702} {\bibfield  {journal} {\bibinfo  {journal} {Phys. Rev. Lett.}\ }\textbf {\bibinfo {volume} {122}},\ \bibinfo {pages} {197702} (\bibinfo {year} {2019})}\BibitemShut {NoStop}%
    \bibitem [{\citenamefont {Sturman}\ and\ \citenamefont {Fridkin}(1992)}]{Sturman1992Photovoltaic}%
    \BibitemOpen
    \bibfield  {author} {\bibinfo {author} {\bibfnamefont {B.~I.}\ \bibnamefont {Sturman}}\ and\ \bibinfo {author} {\bibfnamefont {V.~M.}\ \bibnamefont {Fridkin}},\ }\href {https://doi.org/10.1201/9780203743416} {\emph {\bibinfo {title} {The {{Photovoltaic}} and {{Photorefractive Effects}} in {{Noncentrosymmetric Materials}}}}},\ \bibinfo {edition} {1st}\ ed.\ (\bibinfo  {publisher} {Routledge},\ \bibinfo {address} {London},\ \bibinfo {year} {1992})\BibitemShut {NoStop}%
    \bibitem [{\citenamefont {Watanabe}\ and\ \citenamefont {Yanase}(2021)}]{Watanabe2021Chiral}%
    \BibitemOpen
    \bibfield  {author} {\bibinfo {author} {\bibfnamefont {H.}~\bibnamefont {Watanabe}}\ and\ \bibinfo {author} {\bibfnamefont {Y.}~\bibnamefont {Yanase}},\ }\bibfield  {title} {\bibinfo {title} {Chiral {{Photocurrent}} in {{Parity-Violating Magnet}} and {{Enhanced Response}} in {{Topological Antiferromagnet}}},\ }\href {https://doi.org/10.1103/PhysRevX.11.011001} {\bibfield  {journal} {\bibinfo  {journal} {Phys. Rev. X}\ }\textbf {\bibinfo {volume} {11}},\ \bibinfo {pages} {011001} (\bibinfo {year} {2021})}\BibitemShut {NoStop}%
    \bibitem [{\citenamefont {Ishizuka}\ and\ \citenamefont {Sato}(2024)}]{Ishizuka2024Peltier}%
    \BibitemOpen
    \bibfield  {author} {\bibinfo {author} {\bibfnamefont {H.}~\bibnamefont {Ishizuka}}\ and\ \bibinfo {author} {\bibfnamefont {M.}~\bibnamefont {Sato}},\ }\bibfield  {title} {\bibinfo {title} {Peltier effect of phonons driven by electromagnetic waves},\ }\href {https://doi.org/10.1103/PhysRevB.110.L020303} {\bibfield  {journal} {\bibinfo  {journal} {Phys. Rev. B}\ }\textbf {\bibinfo {volume} {110}},\ \bibinfo {pages} {L020303} (\bibinfo {year} {2024})}\BibitemShut {NoStop}%
    \bibitem [{\citenamefont {Eckardt}\ and\ \citenamefont {Anisimovas}(2015)}]{Eckardt2015Highfrequency}%
    \BibitemOpen
    \bibfield  {author} {\bibinfo {author} {\bibfnamefont {A.}~\bibnamefont {Eckardt}}\ and\ \bibinfo {author} {\bibfnamefont {E.}~\bibnamefont {Anisimovas}},\ }\bibfield  {title} {\bibinfo {title} {High-frequency approximation for periodically driven quantum systems from a {{Floquet-space}} perspective},\ }\href {https://doi.org/10.1088/1367-2630/17/9/093039} {\bibfield  {journal} {\bibinfo  {journal} {New J. Phys.}\ }\textbf {\bibinfo {volume} {17}},\ \bibinfo {pages} {093039} (\bibinfo {year} {2015})}\BibitemShut {NoStop}%
    \bibitem [{\citenamefont {Mikami}\ \emph {et~al.}(2016)\citenamefont {Mikami}, \citenamefont {Kitamura}, \citenamefont {Yasuda}, \citenamefont {Tsuji}, \citenamefont {Oka},\ and\ \citenamefont {Aoki}}]{Mikami2016BrillouinWigner}%
    \BibitemOpen
    \bibfield  {author} {\bibinfo {author} {\bibfnamefont {T.}~\bibnamefont {Mikami}}, \bibinfo {author} {\bibfnamefont {S.}~\bibnamefont {Kitamura}}, \bibinfo {author} {\bibfnamefont {K.}~\bibnamefont {Yasuda}}, \bibinfo {author} {\bibfnamefont {N.}~\bibnamefont {Tsuji}}, \bibinfo {author} {\bibfnamefont {T.}~\bibnamefont {Oka}},\ and\ \bibinfo {author} {\bibfnamefont {H.}~\bibnamefont {Aoki}},\ }\bibfield  {title} {\bibinfo {title} {Brillouin-{{Wigner}} theory for high-frequency expansion in periodically driven systems: {{Application}} to {{Floquet}} topological insulators},\ }\href {https://doi.org/10.1103/PhysRevB.93.144307} {\bibfield  {journal} {\bibinfo  {journal} {Phys. Rev. B}\ }\textbf {\bibinfo {volume} {93}},\ \bibinfo {pages} {144307} (\bibinfo {year} {2016})}\BibitemShut {NoStop}%
    \bibitem [{\citenamefont {Mori}\ \emph {et~al.}(2016)\citenamefont {Mori}, \citenamefont {Kuwahara},\ and\ \citenamefont {Saito}}]{Mori2016Rigorous}%
    \BibitemOpen
    \bibfield  {author} {\bibinfo {author} {\bibfnamefont {T.}~\bibnamefont {Mori}}, \bibinfo {author} {\bibfnamefont {T.}~\bibnamefont {Kuwahara}},\ and\ \bibinfo {author} {\bibfnamefont {K.}~\bibnamefont {Saito}},\ }\bibfield  {title} {\bibinfo {title} {Rigorous {{Bound}} on {{Energy Absorption}} and {{Generic Relaxation}} in {{Periodically Driven Quantum Systems}}},\ }\href {https://doi.org/10.1103/PhysRevLett.116.120401} {\bibfield  {journal} {\bibinfo  {journal} {Phys. Rev. Lett.}\ }\textbf {\bibinfo {volume} {116}},\ \bibinfo {pages} {120401} (\bibinfo {year} {2016})}\BibitemShut {NoStop}%
    \bibitem [{\citenamefont {Kuwahara}\ \emph {et~al.}(2016)\citenamefont {Kuwahara}, \citenamefont {Mori},\ and\ \citenamefont {Saito}}]{Kuwahara2016Floquet}%
    \BibitemOpen
    \bibfield  {author} {\bibinfo {author} {\bibfnamefont {T.}~\bibnamefont {Kuwahara}}, \bibinfo {author} {\bibfnamefont {T.}~\bibnamefont {Mori}},\ and\ \bibinfo {author} {\bibfnamefont {K.}~\bibnamefont {Saito}},\ }\bibfield  {title} {\bibinfo {title} {Floquet--{{Magnus}} theory and generic transient dynamics in periodically driven many-body quantum systems},\ }\href {https://doi.org/10.1016/j.aop.2016.01.012} {\bibfield  {journal} {\bibinfo  {journal} {Ann. Phys.}\ }\textbf {\bibinfo {volume} {367}},\ \bibinfo {pages} {96} (\bibinfo {year} {2016})}\BibitemShut {NoStop}%
    \bibitem [{\citenamefont {Eckardt}(2017)}]{Eckardt2017Colloquium}%
    \BibitemOpen
    \bibfield  {author} {\bibinfo {author} {\bibfnamefont {A.}~\bibnamefont {Eckardt}},\ }\bibfield  {title} {\bibinfo {title} {Colloquium: {{Atomic}} quantum gases in periodically driven optical lattices},\ }\href {https://doi.org/10.1103/RevModPhys.89.011004} {\bibfield  {journal} {\bibinfo  {journal} {Rev. Mod. Phys.}\ }\textbf {\bibinfo {volume} {89}},\ \bibinfo {pages} {011004} (\bibinfo {year} {2017})}\BibitemShut {NoStop}%
    \bibitem [{\citenamefont {Oka}\ and\ \citenamefont {Kitamura}(2019)}]{Oka2019Floquet}%
    \BibitemOpen
    \bibfield  {author} {\bibinfo {author} {\bibfnamefont {T.}~\bibnamefont {Oka}}\ and\ \bibinfo {author} {\bibfnamefont {S.}~\bibnamefont {Kitamura}},\ }\bibfield  {title} {\bibinfo {title} {Floquet {{Engineering}} of {{Quantum Materials}}},\ }\href {https://doi.org/10.1146/annurev-conmatphys-031218-013423} {\bibfield  {journal} {\bibinfo  {journal} {Annu. Rev. Condens. Matter Phys.}\ }\textbf {\bibinfo {volume} {10}},\ \bibinfo {pages} {387} (\bibinfo {year} {2019})}\BibitemShut {NoStop}%
    \bibitem [{\citenamefont {Sato}(2021)}]{Sato2021Floquet}%
    \BibitemOpen
    \bibfield  {author} {\bibinfo {author} {\bibfnamefont {M.}~\bibnamefont {Sato}},\ }\bibfield  {title} {\bibinfo {title} {Floquet {{Theory}} and {{Ultrafast Control}} of {{Magnetism}}},\ }in\ \href {https://doi.org/10.1007/978-3-030-62844-4_11} {\emph {\bibinfo {booktitle} {Chirality, {{Magnetism}} and {{Magnetoelectricity}}}}},\ Vol.\ \bibinfo {volume} {138},\ \bibinfo {editor} {edited by\ \bibinfo {editor} {\bibfnamefont {E.}~\bibnamefont {Kamenetskii}}}\ (\bibinfo  {publisher} {Springer International},\ \bibinfo {address} {Cham},\ \bibinfo {year} {2021})\ pp.\ \bibinfo {pages} {265--286}\BibitemShut {NoStop}%
    \bibitem [{\citenamefont {McPherson}\ \emph {et~al.}(1987)\citenamefont {McPherson}, \citenamefont {Gibson}, \citenamefont {Jara}, \citenamefont {Johann}, \citenamefont {Luk}, \citenamefont {McIntyre}, \citenamefont {Boyer},\ and\ \citenamefont {Rhodes}}]{McPherson1987Studies}%
    \BibitemOpen
    \bibfield  {author} {\bibinfo {author} {\bibfnamefont {A.}~\bibnamefont {McPherson}}, \bibinfo {author} {\bibfnamefont {G.}~\bibnamefont {Gibson}}, \bibinfo {author} {\bibfnamefont {H.}~\bibnamefont {Jara}}, \bibinfo {author} {\bibfnamefont {U.}~\bibnamefont {Johann}}, \bibinfo {author} {\bibfnamefont {T.~S.}\ \bibnamefont {Luk}}, \bibinfo {author} {\bibfnamefont {I.~A.}\ \bibnamefont {McIntyre}}, \bibinfo {author} {\bibfnamefont {K.}~\bibnamefont {Boyer}},\ and\ \bibinfo {author} {\bibfnamefont {C.~K.}\ \bibnamefont {Rhodes}},\ }\bibfield  {title} {\bibinfo {title} {Studies of multiphoton production of vacuum-ultraviolet radiation in the rare gases},\ }\href {https://doi.org/10.1364/JOSAB.4.000595} {\bibfield  {journal} {\bibinfo  {journal} {J. Opt. Soc. Am. B, JOSAB}\ }\textbf {\bibinfo {volume} {4}},\ \bibinfo {pages} {595} (\bibinfo {year} {1987})}\BibitemShut {NoStop}%
    \bibitem [{\citenamefont {Ferray}\ \emph {et~al.}(1988)\citenamefont {Ferray}, \citenamefont {L'Huillier}, \citenamefont {Li}, \citenamefont {Lompre}, \citenamefont {Mainfray},\ and\ \citenamefont {Manus}}]{Ferray1988Multipleharmonic}%
    \BibitemOpen
    \bibfield  {author} {\bibinfo {author} {\bibfnamefont {M.}~\bibnamefont {Ferray}}, \bibinfo {author} {\bibfnamefont {A.}~\bibnamefont {L'Huillier}}, \bibinfo {author} {\bibfnamefont {X.~F.}\ \bibnamefont {Li}}, \bibinfo {author} {\bibfnamefont {L.~A.}\ \bibnamefont {Lompre}}, \bibinfo {author} {\bibfnamefont {G.}~\bibnamefont {Mainfray}},\ and\ \bibinfo {author} {\bibfnamefont {C.}~\bibnamefont {Manus}},\ }\bibfield  {title} {\bibinfo {title} {Multiple-harmonic conversion of 1064 nm radiation in rare gases},\ }\href {https://doi.org/10.1088/0953-4075/21/3/001} {\bibfield  {journal} {\bibinfo  {journal} {J. Phys. B}\ }\textbf {\bibinfo {volume} {21}},\ \bibinfo {pages} {L31} (\bibinfo {year} {1988})}\BibitemShut {NoStop}%
    \bibitem [{\citenamefont {Krause}\ \emph {et~al.}(1992)\citenamefont {Krause}, \citenamefont {Schafer},\ and\ \citenamefont {Kulander}}]{Krause1992Highorder}%
    \BibitemOpen
    \bibfield  {author} {\bibinfo {author} {\bibfnamefont {J.~L.}\ \bibnamefont {Krause}}, \bibinfo {author} {\bibfnamefont {K.~J.}\ \bibnamefont {Schafer}},\ and\ \bibinfo {author} {\bibfnamefont {K.~C.}\ \bibnamefont {Kulander}},\ }\bibfield  {title} {\bibinfo {title} {High-order harmonic generation from atoms and ions in the high intensity regime},\ }\href {https://doi.org/10.1103/PhysRevLett.68.3535} {\bibfield  {journal} {\bibinfo  {journal} {Phys. Rev. Lett.}\ }\textbf {\bibinfo {volume} {68}},\ \bibinfo {pages} {3535} (\bibinfo {year} {1992})}\BibitemShut {NoStop}%
    \bibitem [{\citenamefont {Corkum}(1993)}]{Corkum1993Plasma}%
    \BibitemOpen
    \bibfield  {author} {\bibinfo {author} {\bibfnamefont {P.~B.}\ \bibnamefont {Corkum}},\ }\bibfield  {title} {\bibinfo {title} {Plasma perspective on strong field multiphoton ionization},\ }\href {https://doi.org/10.1103/PhysRevLett.71.1994} {\bibfield  {journal} {\bibinfo  {journal} {Phys. Rev. Lett.}\ }\textbf {\bibinfo {volume} {71}},\ \bibinfo {pages} {1994} (\bibinfo {year} {1993})}\BibitemShut {NoStop}%
    \bibitem [{\citenamefont {Schafer}\ \emph {et~al.}(1993)\citenamefont {Schafer}, \citenamefont {Yang}, \citenamefont {DiMauro},\ and\ \citenamefont {Kulander}}]{Schafer1993threshold}%
    \BibitemOpen
    \bibfield  {author} {\bibinfo {author} {\bibfnamefont {K.~J.}\ \bibnamefont {Schafer}}, \bibinfo {author} {\bibfnamefont {B.}~\bibnamefont {Yang}}, \bibinfo {author} {\bibfnamefont {L.~F.}\ \bibnamefont {DiMauro}},\ and\ \bibinfo {author} {\bibfnamefont {K.~C.}\ \bibnamefont {Kulander}},\ }\bibfield  {title} {\bibinfo {title} {Above threshold ionization beyond the high harmonic cutoff},\ }\href {https://doi.org/10.1103/PhysRevLett.70.1599} {\bibfield  {journal} {\bibinfo  {journal} {Phys. Rev. Lett.}\ }\textbf {\bibinfo {volume} {70}},\ \bibinfo {pages} {1599} (\bibinfo {year} {1993})}\BibitemShut {NoStop}%
    \bibitem [{\citenamefont {Macklin}\ \emph {et~al.}(1993)\citenamefont {Macklin}, \citenamefont {Kmetec},\ and\ \citenamefont {Gordon}}]{Macklin1993Highorder}%
    \BibitemOpen
    \bibfield  {author} {\bibinfo {author} {\bibfnamefont {J.~J.}\ \bibnamefont {Macklin}}, \bibinfo {author} {\bibfnamefont {J.~D.}\ \bibnamefont {Kmetec}},\ and\ \bibinfo {author} {\bibfnamefont {C.~L.}\ \bibnamefont {Gordon}},\ }\bibfield  {title} {\bibinfo {title} {High-order harmonic generation using intense femtosecond pulses},\ }\href {https://doi.org/10.1103/PhysRevLett.70.766} {\bibfield  {journal} {\bibinfo  {journal} {Phys. Rev. Lett.}\ }\textbf {\bibinfo {volume} {70}},\ \bibinfo {pages} {766} (\bibinfo {year} {1993})}\BibitemShut {NoStop}%
    \bibitem [{\citenamefont {Ghimire}\ \emph {et~al.}(2011)\citenamefont {Ghimire}, \citenamefont {DiChiara}, \citenamefont {Sistrunk}, \citenamefont {Agostini}, \citenamefont {DiMauro},\ and\ \citenamefont {Reis}}]{Ghimire2011Observation}%
    \BibitemOpen
    \bibfield  {author} {\bibinfo {author} {\bibfnamefont {S.}~\bibnamefont {Ghimire}}, \bibinfo {author} {\bibfnamefont {A.~D.}\ \bibnamefont {DiChiara}}, \bibinfo {author} {\bibfnamefont {E.}~\bibnamefont {Sistrunk}}, \bibinfo {author} {\bibfnamefont {P.}~\bibnamefont {Agostini}}, \bibinfo {author} {\bibfnamefont {L.~F.}\ \bibnamefont {DiMauro}},\ and\ \bibinfo {author} {\bibfnamefont {D.~A.}\ \bibnamefont {Reis}},\ }\bibfield  {title} {\bibinfo {title} {Observation of high-order harmonic generation in a bulk crystal},\ }\href {https://doi.org/10.1038/nphys1847} {\bibfield  {journal} {\bibinfo  {journal} {Nature Phys}\ }\textbf {\bibinfo {volume} {7}},\ \bibinfo {pages} {138} (\bibinfo {year} {2011})}\BibitemShut {NoStop}%
    \bibitem [{\citenamefont {Schubert}\ \emph {et~al.}(2014)\citenamefont {Schubert}, \citenamefont {Hohenleutner}, \citenamefont {Langer}, \citenamefont {Urbanek}, \citenamefont {Lange}, \citenamefont {Huttner}, \citenamefont {Golde}, \citenamefont {Meier}, \citenamefont {Kira}, \citenamefont {Koch},\ and\ \citenamefont {Huber}}]{Schubert2014Subcycle}%
    \BibitemOpen
    \bibfield  {author} {\bibinfo {author} {\bibfnamefont {O.}~\bibnamefont {Schubert}}, \bibinfo {author} {\bibfnamefont {M.}~\bibnamefont {Hohenleutner}}, \bibinfo {author} {\bibfnamefont {F.}~\bibnamefont {Langer}}, \bibinfo {author} {\bibfnamefont {B.}~\bibnamefont {Urbanek}}, \bibinfo {author} {\bibfnamefont {C.}~\bibnamefont {Lange}}, \bibinfo {author} {\bibfnamefont {U.}~\bibnamefont {Huttner}}, \bibinfo {author} {\bibfnamefont {D.}~\bibnamefont {Golde}}, \bibinfo {author} {\bibfnamefont {T.}~\bibnamefont {Meier}}, \bibinfo {author} {\bibfnamefont {M.}~\bibnamefont {Kira}}, \bibinfo {author} {\bibfnamefont {S.~W.}\ \bibnamefont {Koch}},\ and\ \bibinfo {author} {\bibfnamefont {R.}~\bibnamefont {Huber}},\ }\bibfield  {title} {\bibinfo {title} {Sub-cycle control of terahertz high-harmonic generation by dynamical {{Bloch}} oscillations},\ }\href {https://doi.org/10.1038/nphoton.2013.349} {\bibfield  {journal} {\bibinfo  {journal} {Nat. Photonics}\ }\textbf {\bibinfo {volume} {8}},\ \bibinfo {pages} {119} (\bibinfo {year} {2014})}\BibitemShut {NoStop}%
    \bibitem [{\citenamefont {Vampa}\ \emph {et~al.}(2015)\citenamefont {Vampa}, \citenamefont {McDonald}, \citenamefont {Orlando}, \citenamefont {Corkum},\ and\ \citenamefont {Brabec}}]{Vampa2015Semiclassical}%
    \BibitemOpen
    \bibfield  {author} {\bibinfo {author} {\bibfnamefont {G.}~\bibnamefont {Vampa}}, \bibinfo {author} {\bibfnamefont {C.~R.}\ \bibnamefont {McDonald}}, \bibinfo {author} {\bibfnamefont {G.}~\bibnamefont {Orlando}}, \bibinfo {author} {\bibfnamefont {P.~B.}\ \bibnamefont {Corkum}},\ and\ \bibinfo {author} {\bibfnamefont {T.}~\bibnamefont {Brabec}},\ }\bibfield  {title} {\bibinfo {title} {Semiclassical analysis of high harmonic generation in bulk crystals},\ }\href {https://doi.org/10.1103/PhysRevB.91.064302} {\bibfield  {journal} {\bibinfo  {journal} {Phys. Rev. B}\ }\textbf {\bibinfo {volume} {91}},\ \bibinfo {pages} {064302} (\bibinfo {year} {2015})}\BibitemShut {NoStop}%
    \bibitem [{\citenamefont {Luu}\ \emph {et~al.}(2015)\citenamefont {Luu}, \citenamefont {Garg}, \citenamefont {Kruchinin}, \citenamefont {Moulet}, \citenamefont {Hassan},\ and\ \citenamefont {Goulielmakis}}]{Luu2015Extreme}%
    \BibitemOpen
    \bibfield  {author} {\bibinfo {author} {\bibfnamefont {T.~T.}\ \bibnamefont {Luu}}, \bibinfo {author} {\bibfnamefont {M.}~\bibnamefont {Garg}}, \bibinfo {author} {\bibfnamefont {S.~Y.}\ \bibnamefont {Kruchinin}}, \bibinfo {author} {\bibfnamefont {A.}~\bibnamefont {Moulet}}, \bibinfo {author} {\bibfnamefont {M.~T.}\ \bibnamefont {Hassan}},\ and\ \bibinfo {author} {\bibfnamefont {E.}~\bibnamefont {Goulielmakis}},\ }\bibfield  {title} {\bibinfo {title} {Extreme ultraviolet high-harmonic spectroscopy of solids},\ }\href {https://doi.org/10.1038/nature14456} {\bibfield  {journal} {\bibinfo  {journal} {Nature}\ }\textbf {\bibinfo {volume} {521}},\ \bibinfo {pages} {498} (\bibinfo {year} {2015})}\BibitemShut {NoStop}%
    \bibitem [{\citenamefont {Liu}\ \emph {et~al.}(2017)\citenamefont {Liu}, \citenamefont {Bromberger}, \citenamefont {Cartella}, \citenamefont {Gebert}, \citenamefont {F{\"o}rst},\ and\ \citenamefont {Cavalleri}}]{Liu2017Generation}%
    \BibitemOpen
    \bibfield  {author} {\bibinfo {author} {\bibfnamefont {B.}~\bibnamefont {Liu}}, \bibinfo {author} {\bibfnamefont {H.}~\bibnamefont {Bromberger}}, \bibinfo {author} {\bibfnamefont {A.}~\bibnamefont {Cartella}}, \bibinfo {author} {\bibfnamefont {T.}~\bibnamefont {Gebert}}, \bibinfo {author} {\bibfnamefont {M.}~\bibnamefont {F{\"o}rst}},\ and\ \bibinfo {author} {\bibfnamefont {A.}~\bibnamefont {Cavalleri}},\ }\bibfield  {title} {\bibinfo {title} {Generation of narrowband, high-intensity, carrier-envelope phase-stable pulses tunable between 4 and 18 {{THz}}},\ }\href {https://doi.org/10.1364/OL.42.000129} {\bibfield  {journal} {\bibinfo  {journal} {Opt. Lett.}\ }\textbf {\bibinfo {volume} {42}},\ \bibinfo {pages} {129} (\bibinfo {year} {2017})}\BibitemShut {NoStop}%
    \bibitem [{\citenamefont {You}\ \emph {et~al.}(2017)\citenamefont {You}, \citenamefont {Reis},\ and\ \citenamefont {Ghimire}}]{You2017Anisotropic}%
    \BibitemOpen
    \bibfield  {author} {\bibinfo {author} {\bibfnamefont {Y.~S.}\ \bibnamefont {You}}, \bibinfo {author} {\bibfnamefont {D.~A.}\ \bibnamefont {Reis}},\ and\ \bibinfo {author} {\bibfnamefont {S.}~\bibnamefont {Ghimire}},\ }\bibfield  {title} {\bibinfo {title} {Anisotropic high-harmonic generation in bulk crystals},\ }\href {https://doi.org/10.1038/nphys3955} {\bibfield  {journal} {\bibinfo  {journal} {Nature Phys}\ }\textbf {\bibinfo {volume} {13}},\ \bibinfo {pages} {345} (\bibinfo {year} {2017})}\BibitemShut {NoStop}%
    \bibitem [{\citenamefont {Vampa}\ \emph {et~al.}(2018)\citenamefont {Vampa}, \citenamefont {Hammond}, \citenamefont {Taucer}, \citenamefont {Ding}, \citenamefont {Ropagnol}, \citenamefont {Ozaki}, \citenamefont {Delprat}, \citenamefont {Chaker}, \citenamefont {Thir{\'e}}, \citenamefont {Schmidt}, \citenamefont {L{\'e}gar{\'e}}, \citenamefont {Klug}, \citenamefont {Naumov}, \citenamefont {Villeneuve}, \citenamefont {Staudte},\ and\ \citenamefont {Corkum}}]{Vampa2018Strongfield}%
    \BibitemOpen
    \bibfield  {author} {\bibinfo {author} {\bibfnamefont {G.}~\bibnamefont {Vampa}}, \bibinfo {author} {\bibfnamefont {T.~J.}\ \bibnamefont {Hammond}}, \bibinfo {author} {\bibfnamefont {M.}~\bibnamefont {Taucer}}, \bibinfo {author} {\bibfnamefont {X.}~\bibnamefont {Ding}}, \bibinfo {author} {\bibfnamefont {X.}~\bibnamefont {Ropagnol}}, \bibinfo {author} {\bibfnamefont {T.}~\bibnamefont {Ozaki}}, \bibinfo {author} {\bibfnamefont {S.}~\bibnamefont {Delprat}}, \bibinfo {author} {\bibfnamefont {M.}~\bibnamefont {Chaker}}, \bibinfo {author} {\bibfnamefont {N.}~\bibnamefont {Thir{\'e}}}, \bibinfo {author} {\bibfnamefont {B.~E.}\ \bibnamefont {Schmidt}}, \bibinfo {author} {\bibfnamefont {F.}~\bibnamefont {L{\'e}gar{\'e}}}, \bibinfo {author} {\bibfnamefont {D.~D.}\ \bibnamefont {Klug}}, \bibinfo {author} {\bibfnamefont {A.~Y.}\ \bibnamefont {Naumov}}, \bibinfo {author} {\bibfnamefont {D.~M.}\ \bibnamefont {Villeneuve}}, \bibinfo {author} {\bibfnamefont {A.}~\bibnamefont {Staudte}},\ and\ \bibinfo {author} {\bibfnamefont {P.~B.}\ \bibnamefont {Corkum}},\ }\bibfield  {title} {\bibinfo {title} {Strong-field optoelectronics in solids},\ }\href {https://doi.org/10.1038/s41566-018-0193-5} {\bibfield  {journal} {\bibinfo  {journal} {Nature Photon}\ }\textbf {\bibinfo {volume} {12}},\ \bibinfo {pages} {465} (\bibinfo {year} {2018})}\BibitemShut {NoStop}%
    \bibitem [{\citenamefont {Xia}\ \emph {et~al.}(2018)\citenamefont {Xia}, \citenamefont {Kim}, \citenamefont {Lu}, \citenamefont {Kanai}, \citenamefont {Akiyama}, \citenamefont {Itatani},\ and\ \citenamefont {Ishii}}]{Xia2018Nonlinear}%
    \BibitemOpen
    \bibfield  {author} {\bibinfo {author} {\bibfnamefont {P.}~\bibnamefont {Xia}}, \bibinfo {author} {\bibfnamefont {C.}~\bibnamefont {Kim}}, \bibinfo {author} {\bibfnamefont {F.}~\bibnamefont {Lu}}, \bibinfo {author} {\bibfnamefont {T.}~\bibnamefont {Kanai}}, \bibinfo {author} {\bibfnamefont {H.}~\bibnamefont {Akiyama}}, \bibinfo {author} {\bibfnamefont {J.}~\bibnamefont {Itatani}},\ and\ \bibinfo {author} {\bibfnamefont {N.}~\bibnamefont {Ishii}},\ }\bibfield  {title} {\bibinfo {title} {Nonlinear propagation effects in high harmonic generation in reflection and transmission from gallium arsenide},\ }\href {https://doi.org/10.1364/OE.26.029393} {\bibfield  {journal} {\bibinfo  {journal} {Opt. Express}\ }\textbf {\bibinfo {volume} {26}},\ \bibinfo {pages} {29393} (\bibinfo {year} {2018})}\BibitemShut {NoStop}%
    \bibitem [{\citenamefont {Yoshikawa}\ \emph {et~al.}(2019)\citenamefont {Yoshikawa}, \citenamefont {Nagai}, \citenamefont {Uchida}, \citenamefont {Takaguchi}, \citenamefont {Sasaki}, \citenamefont {Miyata},\ and\ \citenamefont {Tanaka}}]{Yoshikawa2019Interband}%
    \BibitemOpen
    \bibfield  {author} {\bibinfo {author} {\bibfnamefont {N.}~\bibnamefont {Yoshikawa}}, \bibinfo {author} {\bibfnamefont {K.}~\bibnamefont {Nagai}}, \bibinfo {author} {\bibfnamefont {K.}~\bibnamefont {Uchida}}, \bibinfo {author} {\bibfnamefont {Y.}~\bibnamefont {Takaguchi}}, \bibinfo {author} {\bibfnamefont {S.}~\bibnamefont {Sasaki}}, \bibinfo {author} {\bibfnamefont {Y.}~\bibnamefont {Miyata}},\ and\ \bibinfo {author} {\bibfnamefont {K.}~\bibnamefont {Tanaka}},\ }\bibfield  {title} {\bibinfo {title} {Interband resonant high-harmonic generation by valley polarized electron--hole pairs},\ }\href {https://doi.org/10.1038/s41467-019-11697-6} {\bibfield  {journal} {\bibinfo  {journal} {Nat Commun}\ }\textbf {\bibinfo {volume} {10}},\ \bibinfo {pages} {3709} (\bibinfo {year} {2019})}\BibitemShut {NoStop}%
    \bibitem [{\citenamefont {Lakhotia}\ \emph {et~al.}(2020)\citenamefont {Lakhotia}, \citenamefont {Kim}, \citenamefont {Zhan}, \citenamefont {Hu}, \citenamefont {Meng},\ and\ \citenamefont {Goulielmakis}}]{Lakhotia2020Laser}%
    \BibitemOpen
    \bibfield  {author} {\bibinfo {author} {\bibfnamefont {H.}~\bibnamefont {Lakhotia}}, \bibinfo {author} {\bibfnamefont {H.~Y.}\ \bibnamefont {Kim}}, \bibinfo {author} {\bibfnamefont {M.}~\bibnamefont {Zhan}}, \bibinfo {author} {\bibfnamefont {S.}~\bibnamefont {Hu}}, \bibinfo {author} {\bibfnamefont {S.}~\bibnamefont {Meng}},\ and\ \bibinfo {author} {\bibfnamefont {E.}~\bibnamefont {Goulielmakis}},\ }\bibfield  {title} {\bibinfo {title} {Laser picoscopy of valence electrons in solids},\ }\href {https://doi.org/10.1038/s41586-020-2429-z} {\bibfield  {journal} {\bibinfo  {journal} {Nature}\ }\textbf {\bibinfo {volume} {583}},\ \bibinfo {pages} {55} (\bibinfo {year} {2020})}\BibitemShut {NoStop}%
    \bibitem [{\citenamefont {Matsunaga}\ \emph {et~al.}(2014)\citenamefont {Matsunaga}, \citenamefont {Tsuji}, \citenamefont {Fujita}, \citenamefont {Sugioka}, \citenamefont {Makise}, \citenamefont {Uzawa}, \citenamefont {Terai}, \citenamefont {Wang}, \citenamefont {Aoki},\ and\ \citenamefont {Shimano}}]{Matsunaga2014Lightinduced}%
    \BibitemOpen
    \bibfield  {author} {\bibinfo {author} {\bibfnamefont {R.}~\bibnamefont {Matsunaga}}, \bibinfo {author} {\bibfnamefont {N.}~\bibnamefont {Tsuji}}, \bibinfo {author} {\bibfnamefont {H.}~\bibnamefont {Fujita}}, \bibinfo {author} {\bibfnamefont {A.}~\bibnamefont {Sugioka}}, \bibinfo {author} {\bibfnamefont {K.}~\bibnamefont {Makise}}, \bibinfo {author} {\bibfnamefont {Y.}~\bibnamefont {Uzawa}}, \bibinfo {author} {\bibfnamefont {H.}~\bibnamefont {Terai}}, \bibinfo {author} {\bibfnamefont {Z.}~\bibnamefont {Wang}}, \bibinfo {author} {\bibfnamefont {H.}~\bibnamefont {Aoki}},\ and\ \bibinfo {author} {\bibfnamefont {R.}~\bibnamefont {Shimano}},\ }\bibfield  {title} {\bibinfo {title} {Light-induced collective pseudospin precession resonating with {{Higgs}} mode in a superconductor},\ }\href {https://doi.org/10.1126/science.1254697} {\bibfield  {journal} {\bibinfo  {journal} {Science}\ }\textbf {\bibinfo {volume} {345}},\ \bibinfo {pages} {1145} (\bibinfo {year} {2014})}\BibitemShut {NoStop}%
    \bibitem [{\citenamefont {Yoshikawa}\ \emph {et~al.}(2017)\citenamefont {Yoshikawa}, \citenamefont {Tamaya},\ and\ \citenamefont {Tanaka}}]{Yoshikawa2017Highharmonic}%
    \BibitemOpen
    \bibfield  {author} {\bibinfo {author} {\bibfnamefont {N.}~\bibnamefont {Yoshikawa}}, \bibinfo {author} {\bibfnamefont {T.}~\bibnamefont {Tamaya}},\ and\ \bibinfo {author} {\bibfnamefont {K.}~\bibnamefont {Tanaka}},\ }\bibfield  {title} {\bibinfo {title} {High-harmonic generation in graphene enhanced by elliptically polarized light excitation},\ }\href {https://doi.org/10.1126/science.aam8861} {\bibfield  {journal} {\bibinfo  {journal} {Science}\ }\textbf {\bibinfo {volume} {356}},\ \bibinfo {pages} {736} (\bibinfo {year} {2017})}\BibitemShut {NoStop}%
    \bibitem [{\citenamefont {Hafez}\ \emph {et~al.}(2018)\citenamefont {Hafez}, \citenamefont {Kovalev}, \citenamefont {Deinert}, \citenamefont {Mics}, \citenamefont {Green}, \citenamefont {Awari}, \citenamefont {Chen}, \citenamefont {Germanskiy}, \citenamefont {Lehnert}, \citenamefont {Teichert}, \citenamefont {Wang}, \citenamefont {Tielrooij}, \citenamefont {Liu}, \citenamefont {Chen}, \citenamefont {Narita}, \citenamefont {M{\"u}llen}, \citenamefont {Bonn}, \citenamefont {Gensch},\ and\ \citenamefont {Turchinovich}}]{Hafez2018Extremely}%
    \BibitemOpen
    \bibfield  {author} {\bibinfo {author} {\bibfnamefont {H.~A.}\ \bibnamefont {Hafez}}, \bibinfo {author} {\bibfnamefont {S.}~\bibnamefont {Kovalev}}, \bibinfo {author} {\bibfnamefont {J.-C.}\ \bibnamefont {Deinert}}, \bibinfo {author} {\bibfnamefont {Z.}~\bibnamefont {Mics}}, \bibinfo {author} {\bibfnamefont {B.}~\bibnamefont {Green}}, \bibinfo {author} {\bibfnamefont {N.}~\bibnamefont {Awari}}, \bibinfo {author} {\bibfnamefont {M.}~\bibnamefont {Chen}}, \bibinfo {author} {\bibfnamefont {S.}~\bibnamefont {Germanskiy}}, \bibinfo {author} {\bibfnamefont {U.}~\bibnamefont {Lehnert}}, \bibinfo {author} {\bibfnamefont {J.}~\bibnamefont {Teichert}}, \bibinfo {author} {\bibfnamefont {Z.}~\bibnamefont {Wang}}, \bibinfo {author} {\bibfnamefont {K.-J.}\ \bibnamefont {Tielrooij}}, \bibinfo {author} {\bibfnamefont {Z.}~\bibnamefont {Liu}}, \bibinfo {author} {\bibfnamefont {Z.}~\bibnamefont {Chen}}, \bibinfo {author} {\bibfnamefont {A.}~\bibnamefont {Narita}}, \bibinfo {author} {\bibfnamefont {K.}~\bibnamefont {M{\"u}llen}}, \bibinfo {author} {\bibfnamefont {M.}~\bibnamefont {Bonn}}, \bibinfo {author} {\bibfnamefont {M.}~\bibnamefont {Gensch}},\ and\ \bibinfo {author} {\bibfnamefont {D.}~\bibnamefont {Turchinovich}},\ }\bibfield  {title} {\bibinfo {title} {Extremely efficient terahertz high-harmonic generation in graphene by hot {{Dirac}} fermions},\ }\href {https://doi.org/10.1038/s41586-018-0508-1} {\bibfield  {journal} {\bibinfo  {journal} {Nature}\ }\textbf {\bibinfo {volume} {561}},\ \bibinfo {pages} {507} (\bibinfo {year} {2018})}\BibitemShut {NoStop}%
    \bibitem [{\citenamefont {Kovalev}\ \emph {et~al.}(2020)\citenamefont {Kovalev}, \citenamefont {Dantas}, \citenamefont {Germanskiy}, \citenamefont {Deinert}, \citenamefont {Green}, \citenamefont {Ilyakov}, \citenamefont {Awari}, \citenamefont {Chen}, \citenamefont {Bawatna}, \citenamefont {Ling}, \citenamefont {Xiu}, \citenamefont {Van~Loosdrecht}, \citenamefont {Sur{\'o}wka}, \citenamefont {Oka},\ and\ \citenamefont {Wang}}]{Kovalev2020Nonperturbative}%
    \BibitemOpen
    \bibfield  {author} {\bibinfo {author} {\bibfnamefont {S.}~\bibnamefont {Kovalev}}, \bibinfo {author} {\bibfnamefont {R.~M.~A.}\ \bibnamefont {Dantas}}, \bibinfo {author} {\bibfnamefont {S.}~\bibnamefont {Germanskiy}}, \bibinfo {author} {\bibfnamefont {J.-C.}\ \bibnamefont {Deinert}}, \bibinfo {author} {\bibfnamefont {B.}~\bibnamefont {Green}}, \bibinfo {author} {\bibfnamefont {I.}~\bibnamefont {Ilyakov}}, \bibinfo {author} {\bibfnamefont {N.}~\bibnamefont {Awari}}, \bibinfo {author} {\bibfnamefont {M.}~\bibnamefont {Chen}}, \bibinfo {author} {\bibfnamefont {M.}~\bibnamefont {Bawatna}}, \bibinfo {author} {\bibfnamefont {J.}~\bibnamefont {Ling}}, \bibinfo {author} {\bibfnamefont {F.}~\bibnamefont {Xiu}}, \bibinfo {author} {\bibfnamefont {P.~H.~M.}\ \bibnamefont {Van~Loosdrecht}}, \bibinfo {author} {\bibfnamefont {P.}~\bibnamefont {Sur{\'o}wka}}, \bibinfo {author} {\bibfnamefont {T.}~\bibnamefont {Oka}},\ and\ \bibinfo {author} {\bibfnamefont {Z.}~\bibnamefont {Wang}},\ }\bibfield  {title} {\bibinfo {title} {Non-perturbative terahertz high-harmonic generation in the three-dimensional {{Dirac}} semimetal {{Cd3As2}}},\ }\href {https://doi.org/10.1038/s41467-020-16133-8} {\bibfield  {journal} {\bibinfo  {journal} {Nat Commun}\ }\textbf {\bibinfo {volume} {11}},\ \bibinfo {pages} {2451} (\bibinfo {year} {2020})}\BibitemShut {NoStop}%
    \bibitem [{\citenamefont {Cheng}\ \emph {et~al.}(2020)\citenamefont {Cheng}, \citenamefont {Kanda}, \citenamefont {Ikeda}, \citenamefont {Matsuda}, \citenamefont {Xia}, \citenamefont {Schumann}, \citenamefont {Stemmer}, \citenamefont {Itatani}, \citenamefont {Armitage},\ and\ \citenamefont {Matsunaga}}]{Cheng2020Efficient}%
    \BibitemOpen
    \bibfield  {author} {\bibinfo {author} {\bibfnamefont {B.}~\bibnamefont {Cheng}}, \bibinfo {author} {\bibfnamefont {N.}~\bibnamefont {Kanda}}, \bibinfo {author} {\bibfnamefont {T.~N.}\ \bibnamefont {Ikeda}}, \bibinfo {author} {\bibfnamefont {T.}~\bibnamefont {Matsuda}}, \bibinfo {author} {\bibfnamefont {P.}~\bibnamefont {Xia}}, \bibinfo {author} {\bibfnamefont {T.}~\bibnamefont {Schumann}}, \bibinfo {author} {\bibfnamefont {S.}~\bibnamefont {Stemmer}}, \bibinfo {author} {\bibfnamefont {J.}~\bibnamefont {Itatani}}, \bibinfo {author} {\bibfnamefont {N.~P.}\ \bibnamefont {Armitage}},\ and\ \bibinfo {author} {\bibfnamefont {R.}~\bibnamefont {Matsunaga}},\ }\bibfield  {title} {\bibinfo {title} {Efficient {{Terahertz Harmonic Generation}} with {{Coherent Acceleration}} of {{Electrons}} in the {{Dirac Semimetal Cd3As2}}},\ }\href {https://doi.org/10.1103/PhysRevLett.124.117402} {\bibfield  {journal} {\bibinfo  {journal} {Phys. Rev. Lett.}\ }\textbf {\bibinfo {volume} {124}},\ \bibinfo {pages} {117402} (\bibinfo {year} {2020})}\BibitemShut {NoStop}%
    \bibitem [{\citenamefont {Murakami}\ \emph {et~al.}(2018)\citenamefont {Murakami}, \citenamefont {Eckstein},\ and\ \citenamefont {Werner}}]{Murakami2018HighHarmonic}%
    \BibitemOpen
    \bibfield  {author} {\bibinfo {author} {\bibfnamefont {Y.}~\bibnamefont {Murakami}}, \bibinfo {author} {\bibfnamefont {M.}~\bibnamefont {Eckstein}},\ and\ \bibinfo {author} {\bibfnamefont {P.}~\bibnamefont {Werner}},\ }\bibfield  {title} {\bibinfo {title} {High-{{Harmonic Generation}} in {{Mott Insulators}}},\ }\href {https://doi.org/10.1103/PhysRevLett.121.057405} {\bibfield  {journal} {\bibinfo  {journal} {Phys. Rev. Lett.}\ }\textbf {\bibinfo {volume} {121}},\ \bibinfo {pages} {057405} (\bibinfo {year} {2018})}\BibitemShut {NoStop}%
    \bibitem [{\citenamefont {Bionta}\ \emph {et~al.}(2021)\citenamefont {Bionta}, \citenamefont {Haddad}, \citenamefont {Leblanc}, \citenamefont {Gruson}, \citenamefont {Lassonde}, \citenamefont {Ibrahim}, \citenamefont {Chaillou}, \citenamefont {{\'E}mond}, \citenamefont {Otto}, \citenamefont {{Jim{\'e}nez-Gal{\'a}n}}, \citenamefont {Silva}, \citenamefont {Ivanov}, \citenamefont {Siwick}, \citenamefont {Chaker},\ and\ \citenamefont {L{\'e}gar{\'e}}}]{Bionta2021Tracking}%
    \BibitemOpen
    \bibfield  {author} {\bibinfo {author} {\bibfnamefont {M.~R.}\ \bibnamefont {Bionta}}, \bibinfo {author} {\bibfnamefont {E.}~\bibnamefont {Haddad}}, \bibinfo {author} {\bibfnamefont {A.}~\bibnamefont {Leblanc}}, \bibinfo {author} {\bibfnamefont {V.}~\bibnamefont {Gruson}}, \bibinfo {author} {\bibfnamefont {P.}~\bibnamefont {Lassonde}}, \bibinfo {author} {\bibfnamefont {H.}~\bibnamefont {Ibrahim}}, \bibinfo {author} {\bibfnamefont {J.}~\bibnamefont {Chaillou}}, \bibinfo {author} {\bibfnamefont {N.}~\bibnamefont {{\'E}mond}}, \bibinfo {author} {\bibfnamefont {M.~R.}\ \bibnamefont {Otto}}, \bibinfo {author} {\bibfnamefont {{\'A}.}~\bibnamefont {{Jim{\'e}nez-Gal{\'a}n}}}, \bibinfo {author} {\bibfnamefont {R.~E.~F.}\ \bibnamefont {Silva}}, \bibinfo {author} {\bibfnamefont {M.}~\bibnamefont {Ivanov}}, \bibinfo {author} {\bibfnamefont {B.~J.}\ \bibnamefont {Siwick}}, \bibinfo {author} {\bibfnamefont {M.}~\bibnamefont {Chaker}},\ and\ \bibinfo {author} {\bibfnamefont {F.}~\bibnamefont {L{\'e}gar{\'e}}},\ }\bibfield  {title} {\bibinfo {title} {Tracking ultrafast solid-state dynamics using high harmonic spectroscopy},\ }\href {https://doi.org/10.1103/PhysRevResearch.3.023250} {\bibfield  {journal} {\bibinfo  {journal} {Phys. Rev. Res.}\ }\textbf {\bibinfo {volume} {3}},\ \bibinfo {pages} {023250} (\bibinfo {year} {2021})}\BibitemShut {NoStop}%
    \bibitem [{\citenamefont {Gr{\aa}n{\"a}s}\ \emph {et~al.}(2022)\citenamefont {Gr{\aa}n{\"a}s}, \citenamefont {Vaskivskyi}, \citenamefont {Wang}, \citenamefont {Thunstr{\"o}m}, \citenamefont {Ghimire}, \citenamefont {Knut}, \citenamefont {S{\"o}derstr{\"o}m}, \citenamefont {Kjellsson}, \citenamefont {Turenne}, \citenamefont {Engel}, \citenamefont {Beye}, \citenamefont {Lu}, \citenamefont {Higley}, \citenamefont {Reid}, \citenamefont {Schlotter}, \citenamefont {Coslovich}, \citenamefont {Hoffmann}, \citenamefont {Kolesov}, \citenamefont {{Sch{\"u}{\ss}ler-Langeheine}}, \citenamefont {Styervoyedov}, \citenamefont {{Tancogne-Dejean}}, \citenamefont {Sentef}, \citenamefont {Reis}, \citenamefont {Rubio}, \citenamefont {Parkin}, \citenamefont {Karis}, \citenamefont {Rubensson}, \citenamefont {Eriksson},\ and\ \citenamefont {D{\"u}rr}}]{Granas2022Ultrafast}%
    \BibitemOpen
    \bibfield  {author} {\bibinfo {author} {\bibfnamefont {O.}~\bibnamefont {Gr{\aa}n{\"a}s}}, \bibinfo {author} {\bibfnamefont {I.}~\bibnamefont {Vaskivskyi}}, \bibinfo {author} {\bibfnamefont {X.}~\bibnamefont {Wang}}, \bibinfo {author} {\bibfnamefont {P.}~\bibnamefont {Thunstr{\"o}m}}, \bibinfo {author} {\bibfnamefont {S.}~\bibnamefont {Ghimire}}, \bibinfo {author} {\bibfnamefont {R.}~\bibnamefont {Knut}}, \bibinfo {author} {\bibfnamefont {J.}~\bibnamefont {S{\"o}derstr{\"o}m}}, \bibinfo {author} {\bibfnamefont {L.}~\bibnamefont {Kjellsson}}, \bibinfo {author} {\bibfnamefont {D.}~\bibnamefont {Turenne}}, \bibinfo {author} {\bibfnamefont {R.~Y.}\ \bibnamefont {Engel}}, \bibinfo {author} {\bibfnamefont {M.}~\bibnamefont {Beye}}, \bibinfo {author} {\bibfnamefont {J.}~\bibnamefont {Lu}}, \bibinfo {author} {\bibfnamefont {D.~J.}\ \bibnamefont {Higley}}, \bibinfo {author} {\bibfnamefont {A.~H.}\ \bibnamefont {Reid}}, \bibinfo {author} {\bibfnamefont {W.}~\bibnamefont {Schlotter}}, \bibinfo {author} {\bibfnamefont {G.}~\bibnamefont {Coslovich}}, \bibinfo {author} {\bibfnamefont {M.}~\bibnamefont {Hoffmann}}, \bibinfo {author} {\bibfnamefont {G.}~\bibnamefont {Kolesov}}, \bibinfo {author} {\bibfnamefont {C.}~\bibnamefont {{Sch{\"u}{\ss}ler-Langeheine}}}, \bibinfo {author} {\bibfnamefont {A.}~\bibnamefont {Styervoyedov}}, \bibinfo {author} {\bibfnamefont {N.}~\bibnamefont {{Tancogne-Dejean}}}, \bibinfo {author} {\bibfnamefont {M.~A.}\ \bibnamefont {Sentef}}, \bibinfo {author} {\bibfnamefont {D.~A.}\ \bibnamefont {Reis}}, \bibinfo {author} {\bibfnamefont {A.}~\bibnamefont {Rubio}}, \bibinfo {author} {\bibfnamefont {S.~S.~P.}\ \bibnamefont {Parkin}}, \bibinfo {author} {\bibfnamefont {O.}~\bibnamefont {Karis}}, \bibinfo {author} {\bibfnamefont {J.-E.}\ \bibnamefont {Rubensson}}, \bibinfo {author} {\bibfnamefont {O.}~\bibnamefont {Eriksson}},\ and\ \bibinfo {author} {\bibfnamefont {H.~A.}\ \bibnamefont {D{\"u}rr}},\ }\bibfield  {title} {\bibinfo {title} {Ultrafast modification of the electronic structure of a correlated insulator},\ }\href {https://doi.org/10.1103/PhysRevResearch.4.L032030} {\bibfield  {journal} {\bibinfo  {journal} {Phys. Rev. Res.}\ }\textbf {\bibinfo {volume} {4}},\ \bibinfo {pages} {L032030} (\bibinfo {year} {2022})}\BibitemShut {NoStop}%
    \bibitem [{\citenamefont {Uchida}\ \emph {et~al.}(2022)\citenamefont {Uchida}, \citenamefont {Mattoni}, \citenamefont {Yonezawa}, \citenamefont {Nakamura}, \citenamefont {Maeno},\ and\ \citenamefont {Tanaka}}]{Uchida2022HighOrder}%
    \BibitemOpen
    \bibfield  {author} {\bibinfo {author} {\bibfnamefont {K.}~\bibnamefont {Uchida}}, \bibinfo {author} {\bibfnamefont {G.}~\bibnamefont {Mattoni}}, \bibinfo {author} {\bibfnamefont {S.}~\bibnamefont {Yonezawa}}, \bibinfo {author} {\bibfnamefont {F.}~\bibnamefont {Nakamura}}, \bibinfo {author} {\bibfnamefont {Y.}~\bibnamefont {Maeno}},\ and\ \bibinfo {author} {\bibfnamefont {K.}~\bibnamefont {Tanaka}},\ }\bibfield  {title} {\bibinfo {title} {High-{{Order Harmonic Generation}} and {{Its Unconventional Scaling Law}} in the {{Mott-Insulating}} \$\{{\textbackslash}mathrm\{\vphantom{\}\}}{{Ca}}\vphantom\{\}\vphantom\{\}\_\{2\}\{{\textbackslash}mathrm\{\vphantom{\}\}}{{RuO}}\vphantom\{\}\vphantom\{\}\_\{4\}\$},\ }\href {https://doi.org/10.1103/PhysRevLett.128.127401} {\bibfield  {journal} {\bibinfo  {journal} {Phys. Rev. Lett.}\ }\textbf {\bibinfo {volume} {128}},\ \bibinfo {pages} {127401} (\bibinfo {year} {2022})}\BibitemShut {NoStop}%
    \bibitem [{\citenamefont {Baierl}\ \emph {et~al.}(2016)\citenamefont {Baierl}, \citenamefont {Mentink}, \citenamefont {Hohenleutner}, \citenamefont {Braun}, \citenamefont {Do}, \citenamefont {Lange}, \citenamefont {Sell}, \citenamefont {Fiebig}, \citenamefont {Woltersdorf}, \citenamefont {Kampfrath},\ and\ \citenamefont {Huber}}]{Baierl2016TerahertzDriven}%
    \BibitemOpen
    \bibfield  {author} {\bibinfo {author} {\bibfnamefont {S.}~\bibnamefont {Baierl}}, \bibinfo {author} {\bibfnamefont {J.~H.}\ \bibnamefont {Mentink}}, \bibinfo {author} {\bibfnamefont {M.}~\bibnamefont {Hohenleutner}}, \bibinfo {author} {\bibfnamefont {L.}~\bibnamefont {Braun}}, \bibinfo {author} {\bibfnamefont {T.-M.}\ \bibnamefont {Do}}, \bibinfo {author} {\bibfnamefont {C.}~\bibnamefont {Lange}}, \bibinfo {author} {\bibfnamefont {A.}~\bibnamefont {Sell}}, \bibinfo {author} {\bibfnamefont {M.}~\bibnamefont {Fiebig}}, \bibinfo {author} {\bibfnamefont {G.}~\bibnamefont {Woltersdorf}}, \bibinfo {author} {\bibfnamefont {T.}~\bibnamefont {Kampfrath}},\ and\ \bibinfo {author} {\bibfnamefont {R.}~\bibnamefont {Huber}},\ }\bibfield  {title} {\bibinfo {title} {Terahertz-{{Driven Nonlinear Spin Response}} of {{Antiferromagnetic Nickel Oxide}}},\ }\href {https://doi.org/10.1103/PhysRevLett.117.197201} {\bibfield  {journal} {\bibinfo  {journal} {Phys. Rev. Lett.}\ }\textbf {\bibinfo {volume} {117}},\ \bibinfo {pages} {197201} (\bibinfo {year} {2016})}\BibitemShut {NoStop}%
    \bibitem [{\citenamefont {Lu}\ \emph {et~al.}(2017)\citenamefont {Lu}, \citenamefont {Li}, \citenamefont {Hwang}, \citenamefont {{Ofori-Okai}}, \citenamefont {Kurihara}, \citenamefont {Suemoto},\ and\ \citenamefont {Nelson}}]{Lu2017Coherent}%
    \BibitemOpen
    \bibfield  {author} {\bibinfo {author} {\bibfnamefont {J.}~\bibnamefont {Lu}}, \bibinfo {author} {\bibfnamefont {X.}~\bibnamefont {Li}}, \bibinfo {author} {\bibfnamefont {H.~Y.}\ \bibnamefont {Hwang}}, \bibinfo {author} {\bibfnamefont {B.~K.}\ \bibnamefont {{Ofori-Okai}}}, \bibinfo {author} {\bibfnamefont {T.}~\bibnamefont {Kurihara}}, \bibinfo {author} {\bibfnamefont {T.}~\bibnamefont {Suemoto}},\ and\ \bibinfo {author} {\bibfnamefont {K.~A.}\ \bibnamefont {Nelson}},\ }\bibfield  {title} {\bibinfo {title} {Coherent {{Two-Dimensional Terahertz Magnetic Resonance Spectroscopy}} of {{Collective Spin Waves}}},\ }\href {https://doi.org/10.1103/PhysRevLett.118.207204} {\bibfield  {journal} {\bibinfo  {journal} {Phys. Rev. Lett.}\ }\textbf {\bibinfo {volume} {118}},\ \bibinfo {pages} {207204} (\bibinfo {year} {2017})}\BibitemShut {NoStop}%
    \bibitem [{\citenamefont {Ikeda}\ and\ \citenamefont {Sato}(2019)}]{Ikeda2019Highharmonic}%
    \BibitemOpen
    \bibfield  {author} {\bibinfo {author} {\bibfnamefont {T.~N.}\ \bibnamefont {Ikeda}}\ and\ \bibinfo {author} {\bibfnamefont {M.}~\bibnamefont {Sato}},\ }\bibfield  {title} {\bibinfo {title} {High-harmonic generation by electric polarization, spin current, and magnetization},\ }\href {https://doi.org/10.1103/PhysRevB.100.214424} {\bibfield  {journal} {\bibinfo  {journal} {Phys. Rev. B}\ }\textbf {\bibinfo {volume} {100}},\ \bibinfo {pages} {214424} (\bibinfo {year} {2019})}\BibitemShut {NoStop}%
    \bibitem [{\citenamefont {Kanega}\ \emph {et~al.}(2021)\citenamefont {Kanega}, \citenamefont {Ikeda},\ and\ \citenamefont {Sato}}]{Kanega2021Linear}%
    \BibitemOpen
    \bibfield  {author} {\bibinfo {author} {\bibfnamefont {M.}~\bibnamefont {Kanega}}, \bibinfo {author} {\bibfnamefont {T.~N.}\ \bibnamefont {Ikeda}},\ and\ \bibinfo {author} {\bibfnamefont {M.}~\bibnamefont {Sato}},\ }\bibfield  {title} {\bibinfo {title} {Linear and nonlinear optical responses in {{Kitaev}} spin liquids},\ }\href {https://doi.org/10.1103/PhysRevResearch.3.L032024} {\bibfield  {journal} {\bibinfo  {journal} {Phys. Rev. Research}\ }\textbf {\bibinfo {volume} {3}},\ \bibinfo {pages} {L032024} (\bibinfo {year} {2021})}\BibitemShut {NoStop}%
    \bibitem [{\citenamefont {Zhang}\ \emph {et~al.}(2023)\citenamefont {Zhang}, \citenamefont {Sekiguchi}, \citenamefont {Moriyama}, \citenamefont {Furuya}, \citenamefont {Sato}, \citenamefont {Satoh}, \citenamefont {Mukai}, \citenamefont {Tanaka}, \citenamefont {Yamamoto}, \citenamefont {Kageyama}, \citenamefont {Kanemitsu},\ and\ \citenamefont {Hirori}}]{Zhang2023Generation}%
    \BibitemOpen
    \bibfield  {author} {\bibinfo {author} {\bibfnamefont {Z.}~\bibnamefont {Zhang}}, \bibinfo {author} {\bibfnamefont {F.}~\bibnamefont {Sekiguchi}}, \bibinfo {author} {\bibfnamefont {T.}~\bibnamefont {Moriyama}}, \bibinfo {author} {\bibfnamefont {S.~C.}\ \bibnamefont {Furuya}}, \bibinfo {author} {\bibfnamefont {M.}~\bibnamefont {Sato}}, \bibinfo {author} {\bibfnamefont {T.}~\bibnamefont {Satoh}}, \bibinfo {author} {\bibfnamefont {Y.}~\bibnamefont {Mukai}}, \bibinfo {author} {\bibfnamefont {K.}~\bibnamefont {Tanaka}}, \bibinfo {author} {\bibfnamefont {T.}~\bibnamefont {Yamamoto}}, \bibinfo {author} {\bibfnamefont {H.}~\bibnamefont {Kageyama}}, \bibinfo {author} {\bibfnamefont {Y.}~\bibnamefont {Kanemitsu}},\ and\ \bibinfo {author} {\bibfnamefont {H.}~\bibnamefont {Hirori}},\ }\bibfield  {title} {\bibinfo {title} {Generation of third-harmonic spin oscillation from strong spin precession induced by terahertz magnetic near fields},\ }\href {https://doi.org/10.1038/s41467-023-37473-1} {\bibfield  {journal} {\bibinfo  {journal} {Nat Commun}\ }\textbf {\bibinfo {volume} {14}},\ \bibinfo {pages} {1795} (\bibinfo {year} {2023})}\BibitemShut {NoStop}%
    \bibitem [{\citenamefont {Takayoshi}\ \emph {et~al.}(2019)\citenamefont {Takayoshi}, \citenamefont {Murakami},\ and\ \citenamefont {Werner}}]{Takayoshi2019Highharmonic}%
    \BibitemOpen
    \bibfield  {author} {\bibinfo {author} {\bibfnamefont {S.}~\bibnamefont {Takayoshi}}, \bibinfo {author} {\bibfnamefont {Y.}~\bibnamefont {Murakami}},\ and\ \bibinfo {author} {\bibfnamefont {P.}~\bibnamefont {Werner}},\ }\bibfield  {title} {\bibinfo {title} {High-harmonic generation in quantum spin systems},\ }\href {https://doi.org/10.1103/PhysRevB.99.184303} {\bibfield  {journal} {\bibinfo  {journal} {Phys. Rev. B}\ }\textbf {\bibinfo {volume} {99}},\ \bibinfo {pages} {184303} (\bibinfo {year} {2019})}\BibitemShut {NoStop}%
    \bibitem [{\citenamefont {Neufeld}\ \emph {et~al.}(2019)\citenamefont {Neufeld}, \citenamefont {Podolsky},\ and\ \citenamefont {Cohen}}]{Neufeld2019Floquet}%
    \BibitemOpen
    \bibfield  {author} {\bibinfo {author} {\bibfnamefont {O.}~\bibnamefont {Neufeld}}, \bibinfo {author} {\bibfnamefont {D.}~\bibnamefont {Podolsky}},\ and\ \bibinfo {author} {\bibfnamefont {O.}~\bibnamefont {Cohen}},\ }\bibfield  {title} {\bibinfo {title} {Floquet group theory and its application to selection rules in harmonic generation},\ }\href {https://doi.org/10.1038/s41467-018-07935-y} {\bibfield  {journal} {\bibinfo  {journal} {Nat. Commun.}\ }\textbf {\bibinfo {volume} {10}},\ \bibinfo {pages} {1} (\bibinfo {year} {2019})}\BibitemShut {NoStop}%
    \bibitem [{\citenamefont {Sze}\ and\ \citenamefont {Ng}(2006)}]{Sze2006Physics}%
    \BibitemOpen
    \bibfield  {author} {\bibinfo {author} {\bibfnamefont {S.}~\bibnamefont {Sze}}\ and\ \bibinfo {author} {\bibfnamefont {K.~K.}\ \bibnamefont {Ng}},\ }\href {https://doi.org/10.1002/0470068329} {\emph {\bibinfo {title} {Physics of {{Semiconductor Devices}}}}},\ \bibinfo {edition} {3rd}\ ed.\ (\bibinfo  {publisher} {Wiley},\ \bibinfo {address} {New York},\ \bibinfo {year} {2006})\BibitemShut {NoStop}%
    \bibitem [{\citenamefont {Li}\ \emph {et~al.}(2021)\citenamefont {Li}, \citenamefont {Li}, \citenamefont {Shen}, \citenamefont {Han}, \citenamefont {Chen}, \citenamefont {Dong}, \citenamefont {Chen}, \citenamefont {Zhou},\ and\ \citenamefont {Wang}}]{Li2021Photoferroelectric}%
    \BibitemOpen
    \bibfield  {author} {\bibinfo {author} {\bibfnamefont {H.}~\bibnamefont {Li}}, \bibinfo {author} {\bibfnamefont {F.}~\bibnamefont {Li}}, \bibinfo {author} {\bibfnamefont {Z.}~\bibnamefont {Shen}}, \bibinfo {author} {\bibfnamefont {S.-T.}\ \bibnamefont {Han}}, \bibinfo {author} {\bibfnamefont {J.}~\bibnamefont {Chen}}, \bibinfo {author} {\bibfnamefont {C.}~\bibnamefont {Dong}}, \bibinfo {author} {\bibfnamefont {C.}~\bibnamefont {Chen}}, \bibinfo {author} {\bibfnamefont {Y.}~\bibnamefont {Zhou}},\ and\ \bibinfo {author} {\bibfnamefont {M.}~\bibnamefont {Wang}},\ }\bibfield  {title} {\bibinfo {title} {Photoferroelectric perovskite solar cells: {{Principles}}, advances and insights},\ }\href {https://doi.org/10.1016/j.nantod.2020.101062} {\bibfield  {journal} {\bibinfo  {journal} {Nano Today}\ }\textbf {\bibinfo {volume} {37}},\ \bibinfo {pages} {101062} (\bibinfo {year} {2021})}\BibitemShut {NoStop}%
    \bibitem [{\citenamefont {Wu}\ \emph {et~al.}(2017)\citenamefont {Wu}, \citenamefont {Patankar}, \citenamefont {Morimoto}, \citenamefont {Nair}, \citenamefont {Thewalt}, \citenamefont {Little}, \citenamefont {Analytis}, \citenamefont {Moore},\ and\ \citenamefont {Orenstein}}]{Wu2017Giant}%
    \BibitemOpen
    \bibfield  {author} {\bibinfo {author} {\bibfnamefont {L.}~\bibnamefont {Wu}}, \bibinfo {author} {\bibfnamefont {S.}~\bibnamefont {Patankar}}, \bibinfo {author} {\bibfnamefont {T.}~\bibnamefont {Morimoto}}, \bibinfo {author} {\bibfnamefont {N.~L.}\ \bibnamefont {Nair}}, \bibinfo {author} {\bibfnamefont {E.}~\bibnamefont {Thewalt}}, \bibinfo {author} {\bibfnamefont {A.}~\bibnamefont {Little}}, \bibinfo {author} {\bibfnamefont {J.~G.}\ \bibnamefont {Analytis}}, \bibinfo {author} {\bibfnamefont {J.~E.}\ \bibnamefont {Moore}},\ and\ \bibinfo {author} {\bibfnamefont {J.}~\bibnamefont {Orenstein}},\ }\bibfield  {title} {\bibinfo {title} {Giant anisotropic nonlinear optical response in transition metal monopnictide {{Weyl}} semimetals},\ }\href {https://doi.org/10.1038/nphys3969} {\bibfield  {journal} {\bibinfo  {journal} {Nature Phys}\ }\textbf {\bibinfo {volume} {13}},\ \bibinfo {pages} {350} (\bibinfo {year} {2017})}\BibitemShut {NoStop}%
    \bibitem [{\citenamefont {Patankar}\ \emph {et~al.}(2018)\citenamefont {Patankar}, \citenamefont {Wu}, \citenamefont {Lu}, \citenamefont {Rai}, \citenamefont {Tran}, \citenamefont {Morimoto}, \citenamefont {Parker}, \citenamefont {Grushin}, \citenamefont {Nair}, \citenamefont {Analytis}, \citenamefont {Moore}, \citenamefont {Orenstein},\ and\ \citenamefont {Torchinsky}}]{Patankar2018Resonanceenhanced}%
    \BibitemOpen
    \bibfield  {author} {\bibinfo {author} {\bibfnamefont {S.}~\bibnamefont {Patankar}}, \bibinfo {author} {\bibfnamefont {L.}~\bibnamefont {Wu}}, \bibinfo {author} {\bibfnamefont {B.}~\bibnamefont {Lu}}, \bibinfo {author} {\bibfnamefont {M.}~\bibnamefont {Rai}}, \bibinfo {author} {\bibfnamefont {J.~D.}\ \bibnamefont {Tran}}, \bibinfo {author} {\bibfnamefont {T.}~\bibnamefont {Morimoto}}, \bibinfo {author} {\bibfnamefont {D.~E.}\ \bibnamefont {Parker}}, \bibinfo {author} {\bibfnamefont {A.~G.}\ \bibnamefont {Grushin}}, \bibinfo {author} {\bibfnamefont {N.~L.}\ \bibnamefont {Nair}}, \bibinfo {author} {\bibfnamefont {J.~G.}\ \bibnamefont {Analytis}}, \bibinfo {author} {\bibfnamefont {J.~E.}\ \bibnamefont {Moore}}, \bibinfo {author} {\bibfnamefont {J.}~\bibnamefont {Orenstein}},\ and\ \bibinfo {author} {\bibfnamefont {D.~H.}\ \bibnamefont {Torchinsky}},\ }\bibfield  {title} {\bibinfo {title} {Resonance-enhanced optical nonlinearity in the {{Weyl}} semimetal {{TaAs}}},\ }\href {https://doi.org/10.1103/PhysRevB.98.165113} {\bibfield  {journal} {\bibinfo  {journal} {Phys. Rev. B}\ }\textbf {\bibinfo {volume} {98}},\ \bibinfo {pages} {165113} (\bibinfo {year} {2018})}\BibitemShut {NoStop}%
    \bibitem [{\citenamefont {Osterhoudt}\ \emph {et~al.}(2019)\citenamefont {Osterhoudt}, \citenamefont {Diebel}, \citenamefont {Gray}, \citenamefont {Yang}, \citenamefont {Stanco}, \citenamefont {Huang}, \citenamefont {Shen}, \citenamefont {Ni}, \citenamefont {Moll}, \citenamefont {Ran},\ and\ \citenamefont {Burch}}]{Osterhoudt2019Colossal}%
    \BibitemOpen
    \bibfield  {author} {\bibinfo {author} {\bibfnamefont {G.~B.}\ \bibnamefont {Osterhoudt}}, \bibinfo {author} {\bibfnamefont {L.~K.}\ \bibnamefont {Diebel}}, \bibinfo {author} {\bibfnamefont {M.~J.}\ \bibnamefont {Gray}}, \bibinfo {author} {\bibfnamefont {X.}~\bibnamefont {Yang}}, \bibinfo {author} {\bibfnamefont {J.}~\bibnamefont {Stanco}}, \bibinfo {author} {\bibfnamefont {X.}~\bibnamefont {Huang}}, \bibinfo {author} {\bibfnamefont {B.}~\bibnamefont {Shen}}, \bibinfo {author} {\bibfnamefont {N.}~\bibnamefont {Ni}}, \bibinfo {author} {\bibfnamefont {P.~J.~W.}\ \bibnamefont {Moll}}, \bibinfo {author} {\bibfnamefont {Y.}~\bibnamefont {Ran}},\ and\ \bibinfo {author} {\bibfnamefont {K.~S.}\ \bibnamefont {Burch}},\ }\bibfield  {title} {\bibinfo {title} {Colossal mid-infrared bulk photovoltaic effect in a type-{{I Weyl}} semimetal},\ }\href {https://doi.org/10.1038/s41563-019-0297-4} {\bibfield  {journal} {\bibinfo  {journal} {Nat. Mater.}\ }\textbf {\bibinfo {volume} {18}},\ \bibinfo {pages} {471} (\bibinfo {year} {2019})}\BibitemShut {NoStop}%
    \bibitem [{\citenamefont {Sirica}\ \emph {et~al.}(2019)\citenamefont {Sirica}, \citenamefont {Tobey}, \citenamefont {Zhao}, \citenamefont {Chen}, \citenamefont {Xu}, \citenamefont {Yang}, \citenamefont {Shen}, \citenamefont {Yarotski}, \citenamefont {Bowlan}, \citenamefont {Trugman}, \citenamefont {Zhu}, \citenamefont {Dai}, \citenamefont {Azad}, \citenamefont {Ni}, \citenamefont {Qiu}, \citenamefont {Taylor},\ and\ \citenamefont {Prasankumar}}]{Sirica2019Tracking}%
    \BibitemOpen
    \bibfield  {author} {\bibinfo {author} {\bibfnamefont {N.}~\bibnamefont {Sirica}}, \bibinfo {author} {\bibfnamefont {R.~I.}\ \bibnamefont {Tobey}}, \bibinfo {author} {\bibfnamefont {L.~X.}\ \bibnamefont {Zhao}}, \bibinfo {author} {\bibfnamefont {G.~F.}\ \bibnamefont {Chen}}, \bibinfo {author} {\bibfnamefont {B.}~\bibnamefont {Xu}}, \bibinfo {author} {\bibfnamefont {R.}~\bibnamefont {Yang}}, \bibinfo {author} {\bibfnamefont {B.}~\bibnamefont {Shen}}, \bibinfo {author} {\bibfnamefont {D.~A.}\ \bibnamefont {Yarotski}}, \bibinfo {author} {\bibfnamefont {P.}~\bibnamefont {Bowlan}}, \bibinfo {author} {\bibfnamefont {S.~A.}\ \bibnamefont {Trugman}}, \bibinfo {author} {\bibfnamefont {J.-X.}\ \bibnamefont {Zhu}}, \bibinfo {author} {\bibfnamefont {Y.~M.}\ \bibnamefont {Dai}}, \bibinfo {author} {\bibfnamefont {A.~K.}\ \bibnamefont {Azad}}, \bibinfo {author} {\bibfnamefont {N.}~\bibnamefont {Ni}}, \bibinfo {author} {\bibfnamefont {X.~G.}\ \bibnamefont {Qiu}}, \bibinfo {author} {\bibfnamefont {A.~J.}\ \bibnamefont {Taylor}},\ and\ \bibinfo {author} {\bibfnamefont {R.~P.}\ \bibnamefont {Prasankumar}},\ }\bibfield  {title} {\bibinfo {title} {Tracking {{Ultrafast Photocurrents}} in the {{Weyl Semimetal TaAs Using THz Emission Spectroscopy}}},\ }\href {https://doi.org/10.1103/PhysRevLett.122.197401} {\bibfield  {journal} {\bibinfo  {journal} {Phys. Rev. Lett.}\ }\textbf {\bibinfo {volume} {122}},\ \bibinfo {pages} {197401} (\bibinfo {year} {2019})}\BibitemShut {NoStop}%
    \bibitem [{\citenamefont {Khurgin}(1995)}]{Khurgin1995Current}%
    \BibitemOpen
    \bibfield  {author} {\bibinfo {author} {\bibfnamefont {J.~B.}\ \bibnamefont {Khurgin}},\ }\bibfield  {title} {\bibinfo {title} {Current induced second harmonic generation in semiconductors},\ }\href {https://doi.org/10.1063/1.114978} {\bibfield  {journal} {\bibinfo  {journal} {Appl. Phys. Lett.}\ }\textbf {\bibinfo {volume} {67}},\ \bibinfo {pages} {1113} (\bibinfo {year} {1995})}\BibitemShut {NoStop}%
    \bibitem [{\citenamefont {Wu}\ \emph {et~al.}(2012)\citenamefont {Wu}, \citenamefont {Mao}, \citenamefont {Jones}, \citenamefont {Yao}, \citenamefont {Zhang},\ and\ \citenamefont {Xu}}]{Wu2012QuantumEnhanced}%
    \BibitemOpen
    \bibfield  {author} {\bibinfo {author} {\bibfnamefont {S.}~\bibnamefont {Wu}}, \bibinfo {author} {\bibfnamefont {L.}~\bibnamefont {Mao}}, \bibinfo {author} {\bibfnamefont {A.~M.}\ \bibnamefont {Jones}}, \bibinfo {author} {\bibfnamefont {W.}~\bibnamefont {Yao}}, \bibinfo {author} {\bibfnamefont {C.}~\bibnamefont {Zhang}},\ and\ \bibinfo {author} {\bibfnamefont {X.}~\bibnamefont {Xu}},\ }\bibfield  {title} {\bibinfo {title} {Quantum-{{Enhanced Tunable Second-Order Optical Nonlinearity}} in {{Bilayer Graphene}}},\ }\href {https://doi.org/10.1021/nl300084j} {\bibfield  {journal} {\bibinfo  {journal} {Nano Lett.}\ }\textbf {\bibinfo {volume} {12}},\ \bibinfo {pages} {2032} (\bibinfo {year} {2012})}\BibitemShut {NoStop}%
    \bibitem [{\citenamefont {Cheng}\ \emph {et~al.}(2014)\citenamefont {Cheng}, \citenamefont {Vermeulen},\ and\ \citenamefont {Sipe}}]{Cheng2014DC}%
    \BibitemOpen
    \bibfield  {author} {\bibinfo {author} {\bibfnamefont {J.~L.}\ \bibnamefont {Cheng}}, \bibinfo {author} {\bibfnamefont {N.}~\bibnamefont {Vermeulen}},\ and\ \bibinfo {author} {\bibfnamefont {J.~E.}\ \bibnamefont {Sipe}},\ }\bibfield  {title} {\bibinfo {title} {{{DC}} current induced second order optical nonlinearity in graphene},\ }\href {https://doi.org/10.1364/OE.22.015868} {\bibfield  {journal} {\bibinfo  {journal} {Opt. Express}\ }\textbf {\bibinfo {volume} {22}},\ \bibinfo {pages} {15868} (\bibinfo {year} {2014})}\BibitemShut {NoStop}%
    \bibitem [{\citenamefont {Takasan}\ \emph {et~al.}(2021)\citenamefont {Takasan}, \citenamefont {Morimoto}, \citenamefont {Orenstein},\ and\ \citenamefont {Moore}}]{Takasan2021Currentinduced}%
    \BibitemOpen
    \bibfield  {author} {\bibinfo {author} {\bibfnamefont {K.}~\bibnamefont {Takasan}}, \bibinfo {author} {\bibfnamefont {T.}~\bibnamefont {Morimoto}}, \bibinfo {author} {\bibfnamefont {J.}~\bibnamefont {Orenstein}},\ and\ \bibinfo {author} {\bibfnamefont {J.~E.}\ \bibnamefont {Moore}},\ }\bibfield  {title} {\bibinfo {title} {Current-induced second harmonic generation in inversion-symmetric {{Dirac}} and {{Weyl}} semimetals},\ }\href {https://doi.org/10.1103/PhysRevB.104.L161202} {\bibfield  {journal} {\bibinfo  {journal} {Phys. Rev. B}\ }\textbf {\bibinfo {volume} {104}},\ \bibinfo {pages} {L161202} (\bibinfo {year} {2021})}\BibitemShut {NoStop}%
    \bibitem [{\citenamefont {Gao}\ and\ \citenamefont {Zhang}(2021)}]{Gao2021Currentinduced}%
    \BibitemOpen
    \bibfield  {author} {\bibinfo {author} {\bibfnamefont {Y.}~\bibnamefont {Gao}}\ and\ \bibinfo {author} {\bibfnamefont {F.}~\bibnamefont {Zhang}},\ }\bibfield  {title} {\bibinfo {title} {Current-induced second harmonic generation of {{Dirac}} or {{Weyl}} semimetals in a strong magnetic field},\ }\href {https://doi.org/10.1103/PhysRevB.103.L041301} {\bibfield  {journal} {\bibinfo  {journal} {Phys. Rev. B}\ }\textbf {\bibinfo {volume} {103}},\ \bibinfo {pages} {L041301} (\bibinfo {year} {2021})}\BibitemShut {NoStop}%
    \bibitem [{\citenamefont {Aktsipetrov}\ \emph {et~al.}(2009)\citenamefont {Aktsipetrov}, \citenamefont {Bessonov}, \citenamefont {Fedyanin},\ and\ \citenamefont {Val'dner}}]{Aktsipetrov2009DCinduced}%
    \BibitemOpen
    \bibfield  {author} {\bibinfo {author} {\bibfnamefont {O.~A.}\ \bibnamefont {Aktsipetrov}}, \bibinfo {author} {\bibfnamefont {V.~O.}\ \bibnamefont {Bessonov}}, \bibinfo {author} {\bibfnamefont {A.~A.}\ \bibnamefont {Fedyanin}},\ and\ \bibinfo {author} {\bibfnamefont {V.~O.}\ \bibnamefont {Val'dner}},\ }\bibfield  {title} {\bibinfo {title} {{{DC-induced}} generation of the reflected second harmonic in silicon},\ }\href {https://doi.org/10.1134/S0021364009020027} {\bibfield  {journal} {\bibinfo  {journal} {JETP Lett.}\ }\textbf {\bibinfo {volume} {89}},\ \bibinfo {pages} {58} (\bibinfo {year} {2009})}\BibitemShut {NoStop}%
    \bibitem [{\citenamefont {Ruzicka}\ \emph {et~al.}(2012)\citenamefont {Ruzicka}, \citenamefont {Werake}, \citenamefont {Xu}, \citenamefont {Khurgin}, \citenamefont {Sherman}, \citenamefont {Wu},\ and\ \citenamefont {Zhao}}]{Ruzicka2012SecondHarmonic}%
    \BibitemOpen
    \bibfield  {author} {\bibinfo {author} {\bibfnamefont {B.~A.}\ \bibnamefont {Ruzicka}}, \bibinfo {author} {\bibfnamefont {L.~K.}\ \bibnamefont {Werake}}, \bibinfo {author} {\bibfnamefont {G.}~\bibnamefont {Xu}}, \bibinfo {author} {\bibfnamefont {J.~B.}\ \bibnamefont {Khurgin}}, \bibinfo {author} {\bibfnamefont {E.~{\relax Ya}.}\ \bibnamefont {Sherman}}, \bibinfo {author} {\bibfnamefont {J.~Z.}\ \bibnamefont {Wu}},\ and\ \bibinfo {author} {\bibfnamefont {H.}~\bibnamefont {Zhao}},\ }\bibfield  {title} {\bibinfo {title} {Second-{{Harmonic Generation Induced}} by {{Electric Currents}} in {{GaAs}}},\ }\href {https://doi.org/10.1103/PhysRevLett.108.077403} {\bibfield  {journal} {\bibinfo  {journal} {Phys. Rev. Lett.}\ }\textbf {\bibinfo {volume} {108}},\ \bibinfo {pages} {077403} (\bibinfo {year} {2012})}\BibitemShut {NoStop}%
    \bibitem [{\citenamefont {Bykov}\ \emph {et~al.}(2012)\citenamefont {Bykov}, \citenamefont {Murzina}, \citenamefont {Rybin},\ and\ \citenamefont {Obraztsova}}]{Bykov2012Second}%
    \BibitemOpen
    \bibfield  {author} {\bibinfo {author} {\bibfnamefont {A.~Y.}\ \bibnamefont {Bykov}}, \bibinfo {author} {\bibfnamefont {T.~V.}\ \bibnamefont {Murzina}}, \bibinfo {author} {\bibfnamefont {M.~G.}\ \bibnamefont {Rybin}},\ and\ \bibinfo {author} {\bibfnamefont {E.~D.}\ \bibnamefont {Obraztsova}},\ }\bibfield  {title} {\bibinfo {title} {Second harmonic generation in multilayer graphene induced by direct electric current},\ }\href {https://doi.org/10.1103/PhysRevB.85.121413} {\bibfield  {journal} {\bibinfo  {journal} {Phys. Rev. B}\ }\textbf {\bibinfo {volume} {85}},\ \bibinfo {pages} {121413} (\bibinfo {year} {2012})}\BibitemShut {NoStop}%
    \bibitem [{\citenamefont {An}\ \emph {et~al.}(2013)\citenamefont {An}, \citenamefont {Nelson}, \citenamefont {Lee},\ and\ \citenamefont {Diebold}}]{An2013Enhanced}%
    \BibitemOpen
    \bibfield  {author} {\bibinfo {author} {\bibfnamefont {Y.~Q.}\ \bibnamefont {An}}, \bibinfo {author} {\bibfnamefont {F.}~\bibnamefont {Nelson}}, \bibinfo {author} {\bibfnamefont {J.~U.}\ \bibnamefont {Lee}},\ and\ \bibinfo {author} {\bibfnamefont {A.~C.}\ \bibnamefont {Diebold}},\ }\bibfield  {title} {\bibinfo {title} {Enhanced {{Optical Second-Harmonic Generation}} from the {{Current-Biased Graphene}}/{{SiO2}}/{{Si}}(001) {{Structure}}},\ }\href {https://doi.org/10.1021/NL4004514} {\bibfield  {journal} {\bibinfo  {journal} {Nano Lett.}\ }\textbf {\bibinfo {volume} {13}},\ \bibinfo {pages} {2104} (\bibinfo {year} {2013})}\BibitemShut {NoStop}%
    \bibitem [{\citenamefont {Nakamura}\ \emph {et~al.}(2020)\citenamefont {Nakamura}, \citenamefont {Katsumi}, \citenamefont {Terai},\ and\ \citenamefont {Shimano}}]{Nakamura2020Nonreciprocal}%
    \BibitemOpen
    \bibfield  {author} {\bibinfo {author} {\bibfnamefont {S.}~\bibnamefont {Nakamura}}, \bibinfo {author} {\bibfnamefont {K.}~\bibnamefont {Katsumi}}, \bibinfo {author} {\bibfnamefont {H.}~\bibnamefont {Terai}},\ and\ \bibinfo {author} {\bibfnamefont {R.}~\bibnamefont {Shimano}},\ }\bibfield  {title} {\bibinfo {title} {Nonreciprocal {{Terahertz Second-Harmonic Generation}} in {{Superconducting NbN}} under {{Supercurrent Injection}}},\ }\href {https://doi.org/10.1103/PhysRevLett.125.097004} {\bibfield  {journal} {\bibinfo  {journal} {Phys. Rev. Lett.}\ }\textbf {\bibinfo {volume} {125}},\ \bibinfo {pages} {097004} (\bibinfo {year} {2020})}\BibitemShut {NoStop}%
    \bibitem [{\citenamefont {Gorini}\ \emph {et~al.}(1976)\citenamefont {Gorini}, \citenamefont {Kossakowski},\ and\ \citenamefont {Sudarshan}}]{Gorini1976Completely}%
    \BibitemOpen
    \bibfield  {author} {\bibinfo {author} {\bibfnamefont {V.}~\bibnamefont {Gorini}}, \bibinfo {author} {\bibfnamefont {A.}~\bibnamefont {Kossakowski}},\ and\ \bibinfo {author} {\bibfnamefont {E.~C.~G.}\ \bibnamefont {Sudarshan}},\ }\bibfield  {title} {\bibinfo {title} {Completely positive dynamical semigroups of {{{\emph{N}}}} -level systems},\ }\href {https://doi.org/10.1063/1.522979} {\bibfield  {journal} {\bibinfo  {journal} {J. Math. Phys.}\ }\textbf {\bibinfo {volume} {17}},\ \bibinfo {pages} {821} (\bibinfo {year} {1976})}\BibitemShut {NoStop}%
    \bibitem [{\citenamefont {Lindblad}(1976)}]{Lindblad1976generators}%
    \BibitemOpen
    \bibfield  {author} {\bibinfo {author} {\bibfnamefont {G.}~\bibnamefont {Lindblad}},\ }\bibfield  {title} {\bibinfo {title} {On the generators of quantum dynamical semigroups},\ }\href {https://doi.org/10.1007/BF01608499} {\bibfield  {journal} {\bibinfo  {journal} {Commun.Math. Phys.}\ }\textbf {\bibinfo {volume} {48}},\ \bibinfo {pages} {119} (\bibinfo {year} {1976})}\BibitemShut {NoStop}%
    \bibitem [{\citenamefont {Breuer}\ and\ \citenamefont {Petruccione}(2007)}]{Breuer2007Theory}%
    \BibitemOpen
    \bibfield  {author} {\bibinfo {author} {\bibfnamefont {H.-P.}\ \bibnamefont {Breuer}}\ and\ \bibinfo {author} {\bibfnamefont {F.}~\bibnamefont {Petruccione}},\ }\href {https://doi.org/10.1093/acprof:oso/9780199213900.001.0001} {\emph {\bibinfo {title} {The {{Theory}} of {{Open Quantum Systems}}}}},\ \bibinfo {edition} {1st}\ ed.\ (\bibinfo  {publisher} {Oxford University PressOxford},\ \bibinfo {year} {2007})\BibitemShut {NoStop}%
    \bibitem [{\citenamefont {Sato}\ and\ \citenamefont {Morisaku}(2020)}]{Sato2020Twophoton}%
    \BibitemOpen
    \bibfield  {author} {\bibinfo {author} {\bibfnamefont {M.}~\bibnamefont {Sato}}\ and\ \bibinfo {author} {\bibfnamefont {Y.}~\bibnamefont {Morisaku}},\ }\bibfield  {title} {\bibinfo {title} {Two-photon driven magnon-pair resonance as a signature of spin-nematic order},\ }\href {https://doi.org/10.1103/PhysRevB.102.060401} {\bibfield  {journal} {\bibinfo  {journal} {Phys. Rev. B}\ }\textbf {\bibinfo {volume} {102}},\ \bibinfo {pages} {060401} (\bibinfo {year} {2020})}\BibitemShut {NoStop}%
    \bibitem [{\citenamefont {Abrikosov}(2017)}]{Abrikosov2017Fundamentals}%
    \BibitemOpen
    \bibfield  {author} {\bibinfo {author} {\bibfnamefont {A.~A.}\ \bibnamefont {Abrikosov}},\ }\href@noop {} {\emph {\bibinfo {title} {Fundamentals of the Theory of Metals}}},\ \bibinfo {edition} {dover edition}\ ed.\ (\bibinfo  {publisher} {Dover Publications, Inc.},\ \bibinfo {address} {Mineola, New York},\ \bibinfo {year} {2017})\BibitemShut {NoStop}%
    \bibitem [{\citenamefont {Castro~Neto}\ \emph {et~al.}(2009)\citenamefont {Castro~Neto}, \citenamefont {Guinea}, \citenamefont {Peres}, \citenamefont {Novoselov},\ and\ \citenamefont {Geim}}]{CastroNeto2009electronic}%
    \BibitemOpen
    \bibfield  {author} {\bibinfo {author} {\bibfnamefont {A.~H.}\ \bibnamefont {Castro~Neto}}, \bibinfo {author} {\bibfnamefont {F.}~\bibnamefont {Guinea}}, \bibinfo {author} {\bibfnamefont {N.~M.~R.}\ \bibnamefont {Peres}}, \bibinfo {author} {\bibfnamefont {K.~S.}\ \bibnamefont {Novoselov}},\ and\ \bibinfo {author} {\bibfnamefont {A.~K.}\ \bibnamefont {Geim}},\ }\bibfield  {title} {\bibinfo {title} {The electronic properties of graphene},\ }\href {https://doi.org/10.1103/RevModPhys.81.109} {\bibfield  {journal} {\bibinfo  {journal} {Rev. Mod. Phys.}\ }\textbf {\bibinfo {volume} {81}},\ \bibinfo {pages} {109} (\bibinfo {year} {2009})}\BibitemShut {NoStop}%
    \bibitem [{\citenamefont {Aoki}\ and\ \citenamefont {S.~Dresselhaus}(2014)}]{Aoki2014Physics}%
    \BibitemOpen
    \bibinfo {editor} {\bibfnamefont {H.}~\bibnamefont {Aoki}}\ and\ \bibinfo {editor} {\bibfnamefont {M.}~\bibnamefont {S.~Dresselhaus}},\ eds.,\ \href {https://doi.org/10.1007/978-3-319-02633-6} {\emph {\bibinfo {title} {Physics of {{Graphene}}}}},\ {{NanoScience}} and {{Technology}}\ (\bibinfo  {publisher} {Springer International},\ \bibinfo {address} {Cham},\ \bibinfo {year} {2014})\BibitemShut {NoStop}%
    \bibitem [{\citenamefont {Reich}\ \emph {et~al.}(2002)\citenamefont {Reich}, \citenamefont {Maultzsch}, \citenamefont {Thomsen},\ and\ \citenamefont {Ordej{\'o}n}}]{Reich2002Tightbinding}%
    \BibitemOpen
    \bibfield  {author} {\bibinfo {author} {\bibfnamefont {S.}~\bibnamefont {Reich}}, \bibinfo {author} {\bibfnamefont {J.}~\bibnamefont {Maultzsch}}, \bibinfo {author} {\bibfnamefont {C.}~\bibnamefont {Thomsen}},\ and\ \bibinfo {author} {\bibfnamefont {P.}~\bibnamefont {Ordej{\'o}n}},\ }\bibfield  {title} {\bibinfo {title} {Tight-binding description of graphene},\ }\href {https://doi.org/10.1103/PhysRevB.66.035412} {\bibfield  {journal} {\bibinfo  {journal} {Phys. Rev. B}\ }\textbf {\bibinfo {volume} {66}},\ \bibinfo {pages} {035412} (\bibinfo {year} {2002})}\BibitemShut {NoStop}%
    \bibitem [{\citenamefont {Jackson}(1998)}]{Jackson1998}%
    \BibitemOpen
    \bibfield  {author} {\bibinfo {author} {\bibfnamefont {{\relax John}.~{\relax David}.}\ \bibnamefont {Jackson}},\ }\href@noop {} {\emph {\bibinfo {title} {Classical {{Electrodynamics}}}}}\ (\bibinfo  {publisher} {Wiley},\ \bibinfo {address} {Weinheim, Germany},\ \bibinfo {year} {1998})\BibitemShut {NoStop}%
    \bibitem [{\citenamefont {Hwang}\ and\ \citenamefont {Das~Sarma}(2008)}]{Hwang2008Singleparticle}%
    \BibitemOpen
    \bibfield  {author} {\bibinfo {author} {\bibfnamefont {E.~H.}\ \bibnamefont {Hwang}}\ and\ \bibinfo {author} {\bibfnamefont {S.}~\bibnamefont {Das~Sarma}},\ }\bibfield  {title} {\bibinfo {title} {Single-particle relaxation time versus transport scattering time in a two-dimensional graphene layer},\ }\href {https://doi.org/10.1103/PhysRevB.77.195412} {\bibfield  {journal} {\bibinfo  {journal} {Phys. Rev. B}\ }\textbf {\bibinfo {volume} {77}},\ \bibinfo {pages} {195412} (\bibinfo {year} {2008})}\BibitemShut {NoStop}%
    \bibitem [{\citenamefont {Nair}\ \emph {et~al.}(2008)\citenamefont {Nair}, \citenamefont {Blake}, \citenamefont {Grigorenko}, \citenamefont {Novoselov}, \citenamefont {Booth}, \citenamefont {Stauber}, \citenamefont {Peres},\ and\ \citenamefont {Geim}}]{Nair2008Fine}%
    \BibitemOpen
    \bibfield  {author} {\bibinfo {author} {\bibfnamefont {R.~R.}\ \bibnamefont {Nair}}, \bibinfo {author} {\bibfnamefont {P.}~\bibnamefont {Blake}}, \bibinfo {author} {\bibfnamefont {A.~N.}\ \bibnamefont {Grigorenko}}, \bibinfo {author} {\bibfnamefont {K.~S.}\ \bibnamefont {Novoselov}}, \bibinfo {author} {\bibfnamefont {T.~J.}\ \bibnamefont {Booth}}, \bibinfo {author} {\bibfnamefont {T.}~\bibnamefont {Stauber}}, \bibinfo {author} {\bibfnamefont {N.~M.~R.}\ \bibnamefont {Peres}},\ and\ \bibinfo {author} {\bibfnamefont {A.~K.}\ \bibnamefont {Geim}},\ }\bibfield  {title} {\bibinfo {title} {Fine {{Structure Constant Defines Visual Transparency}} of {{Graphene}}},\ }\href {https://doi.org/10.1126/science.1156965} {\bibfield  {journal} {\bibinfo  {journal} {Science}\ }\textbf {\bibinfo {volume} {320}},\ \bibinfo {pages} {1308} (\bibinfo {year} {2008})}\BibitemShut {NoStop}%
    \bibitem [{\citenamefont {Moser}\ \emph {et~al.}(2007)\citenamefont {Moser}, \citenamefont {Barreiro},\ and\ \citenamefont {Bachtold}}]{Moser2007Currentinduced}%
    \BibitemOpen
    \bibfield  {author} {\bibinfo {author} {\bibfnamefont {J.}~\bibnamefont {Moser}}, \bibinfo {author} {\bibfnamefont {A.}~\bibnamefont {Barreiro}},\ and\ \bibinfo {author} {\bibfnamefont {A.}~\bibnamefont {Bachtold}},\ }\bibfield  {title} {\bibinfo {title} {Current-induced cleaning of graphene},\ }\href {https://doi.org/10.1063/1.2789673} {\bibfield  {journal} {\bibinfo  {journal} {Appl. Phys. Lett.}\ }\textbf {\bibinfo {volume} {91}},\ \bibinfo {pages} {163513} (\bibinfo {year} {2007})}\BibitemShut {NoStop}%
    \bibitem [{\citenamefont {Alon}\ \emph {et~al.}(1998)\citenamefont {Alon}, \citenamefont {Averbukh},\ and\ \citenamefont {Moiseyev}}]{Alon1998Selection}%
    \BibitemOpen
    \bibfield  {author} {\bibinfo {author} {\bibfnamefont {O.~E.}\ \bibnamefont {Alon}}, \bibinfo {author} {\bibfnamefont {V.}~\bibnamefont {Averbukh}},\ and\ \bibinfo {author} {\bibfnamefont {N.}~\bibnamefont {Moiseyev}},\ }\bibfield  {title} {\bibinfo {title} {Selection {{Rules}} for the {{High Harmonic Generation Spectra}}},\ }\href {https://doi.org/10.1103/PhysRevLett.80.3743} {\bibfield  {journal} {\bibinfo  {journal} {Phys. Rev. Lett.}\ }\textbf {\bibinfo {volume} {80}},\ \bibinfo {pages} {3743} (\bibinfo {year} {1998})}\BibitemShut {NoStop}%
    \bibitem [{\citenamefont {Morimoto}\ \emph {et~al.}(2017)\citenamefont {Morimoto}, \citenamefont {Po},\ and\ \citenamefont {Vishwanath}}]{Morimoto2017Floquet}%
    \BibitemOpen
    \bibfield  {author} {\bibinfo {author} {\bibfnamefont {T.}~\bibnamefont {Morimoto}}, \bibinfo {author} {\bibfnamefont {H.~C.}\ \bibnamefont {Po}},\ and\ \bibinfo {author} {\bibfnamefont {A.}~\bibnamefont {Vishwanath}},\ }\bibfield  {title} {\bibinfo {title} {Floquet topological phases protected by time glide symmetry},\ }\href {https://doi.org/10.1103/PhysRevB.95.195155} {\bibfield  {journal} {\bibinfo  {journal} {Phys. Rev. B}\ }\textbf {\bibinfo {volume} {95}},\ \bibinfo {pages} {195155} (\bibinfo {year} {2017})}\BibitemShut {NoStop}%
    \bibitem [{\citenamefont {Chinzei}\ and\ \citenamefont {Ikeda}(2020)}]{Chinzei2020Time}%
    \BibitemOpen
    \bibfield  {author} {\bibinfo {author} {\bibfnamefont {K.}~\bibnamefont {Chinzei}}\ and\ \bibinfo {author} {\bibfnamefont {T.~N.}\ \bibnamefont {Ikeda}},\ }\bibfield  {title} {\bibinfo {title} {Time {{Crystals Protected}} by {{Floquet Dynamical Symmetry}} in {{Hubbard Models}}},\ }\href {https://doi.org/10.1103/PhysRevLett.125.060601} {\bibfield  {journal} {\bibinfo  {journal} {Phys. Rev. Lett.}\ }\textbf {\bibinfo {volume} {125}},\ \bibinfo {pages} {060601} (\bibinfo {year} {2020})}\BibitemShut {NoStop}%
    \bibitem [{\citenamefont {Ikeda}(2020)}]{Ikeda2020Highorder}%
    \BibitemOpen
    \bibfield  {author} {\bibinfo {author} {\bibfnamefont {T.~N.}\ \bibnamefont {Ikeda}},\ }\bibfield  {title} {\bibinfo {title} {High-order nonlinear optical response of a twisted bilayer graphene},\ }\href {https://doi.org/10.1103/PhysRevResearch.2.032015} {\bibfield  {journal} {\bibinfo  {journal} {Phys. Rev. Res.}\ }\textbf {\bibinfo {volume} {2}},\ \bibinfo {pages} {032015} (\bibinfo {year} {2020})}\BibitemShut {NoStop}%
    \bibitem [{\citenamefont {Bhalla}\ \emph {et~al.}(2023)\citenamefont {Bhalla}, \citenamefont {Das}, \citenamefont {Agarwal},\ and\ \citenamefont {Culcer}}]{Bhalla2023Quantum}%
    \BibitemOpen
    \bibfield  {author} {\bibinfo {author} {\bibfnamefont {P.}~\bibnamefont {Bhalla}}, \bibinfo {author} {\bibfnamefont {K.}~\bibnamefont {Das}}, \bibinfo {author} {\bibfnamefont {A.}~\bibnamefont {Agarwal}},\ and\ \bibinfo {author} {\bibfnamefont {D.}~\bibnamefont {Culcer}},\ }\bibfield  {title} {\bibinfo {title} {Quantum kinetic theory of nonlinear optical currents: {{Finite Fermi}} surface and {{Fermi}} sea contributions},\ }\href {https://doi.org/10.1103/PhysRevB.107.165131} {\bibfield  {journal} {\bibinfo  {journal} {Phys. Rev. B}\ }\textbf {\bibinfo {volume} {107}},\ \bibinfo {pages} {165131} (\bibinfo {year} {2023})}\BibitemShut {NoStop}%
    \bibitem [{\citenamefont {Ghimire}\ \emph {et~al.}(2014)\citenamefont {Ghimire}, \citenamefont {Ndabashimiye}, \citenamefont {DiChiara}, \citenamefont {Sistrunk}, \citenamefont {Stockman}, \citenamefont {Agostini}, \citenamefont {DiMauro},\ and\ \citenamefont {Reis}}]{Ghimire2014Strongfield}%
    \BibitemOpen
    \bibfield  {author} {\bibinfo {author} {\bibfnamefont {S.}~\bibnamefont {Ghimire}}, \bibinfo {author} {\bibfnamefont {G.}~\bibnamefont {Ndabashimiye}}, \bibinfo {author} {\bibfnamefont {A.~D.}\ \bibnamefont {DiChiara}}, \bibinfo {author} {\bibfnamefont {E.}~\bibnamefont {Sistrunk}}, \bibinfo {author} {\bibfnamefont {M.~I.}\ \bibnamefont {Stockman}}, \bibinfo {author} {\bibfnamefont {P.}~\bibnamefont {Agostini}}, \bibinfo {author} {\bibfnamefont {L.~F.}\ \bibnamefont {DiMauro}},\ and\ \bibinfo {author} {\bibfnamefont {D.~A.}\ \bibnamefont {Reis}},\ }\bibfield  {title} {\bibinfo {title} {Strong-field and attosecond physics in solids},\ }\href {https://doi.org/10.1088/0953-4075/47/20/204030} {\bibfield  {journal} {\bibinfo  {journal} {J. Phys. B: At. Mol. Opt. Phys.}\ }\textbf {\bibinfo {volume} {47}},\ \bibinfo {pages} {204030} (\bibinfo {year} {2014})}\BibitemShut {NoStop}%
    \bibitem [{\citenamefont {Haug}\ and\ \citenamefont {Jauho}(2008)}]{Haug2008Quantum}%
    \BibitemOpen
    \bibfield  {author} {\bibinfo {author} {\bibfnamefont {H.}~\bibnamefont {Haug}}\ and\ \bibinfo {author} {\bibfnamefont {A.-P.}\ \bibnamefont {Jauho}},\ }\href {https://doi.org/10.1007/978-3-540-73564-9} {\emph {\bibinfo {title} {Quantum {{Kinetics}} in {{Transport}} and {{Optics}} of {{Semiconductors}}}}},\ \bibinfo {series} {Solid-{{State Sciences}}}, Vol.\ \bibinfo {volume} {123}\ (\bibinfo  {publisher} {Springer Berlin Heidelberg},\ \bibinfo {address} {Berlin, Heidelberg},\ \bibinfo {year} {2008})\BibitemShut {NoStop}%
    \bibitem [{\citenamefont {Tanaka}\ and\ \citenamefont {Sato}(2024)}]{Tanaka2024Theory}%
    \BibitemOpen
    \bibfield  {author} {\bibinfo {author} {\bibfnamefont {M.}~\bibnamefont {Tanaka}}\ and\ \bibinfo {author} {\bibfnamefont {M.}~\bibnamefont {Sato}},\ }\bibfield  {title} {\bibinfo {title} {Theory of the inverse {{Faraday}} effect in dissipative {{Rashba}} electron systems: {{Floquet}} engineering perspective},\ }\href {https://doi.org/10.1103/PhysRevB.110.045204} {\bibfield  {journal} {\bibinfo  {journal} {Phys. Rev. B}\ }\textbf {\bibinfo {volume} {110}},\ \bibinfo {pages} {045204} (\bibinfo {year} {2024})}\BibitemShut {NoStop}%
    \bibitem [{\citenamefont {Passos}\ \emph {et~al.}(2018)\citenamefont {Passos}, \citenamefont {Ventura}, \citenamefont {Lopes}, \citenamefont {dos Santos},\ and\ \citenamefont {Peres}}]{Passos2018Nonlinear}%
    \BibitemOpen
    \bibfield  {author} {\bibinfo {author} {\bibfnamefont {D.~J.}\ \bibnamefont {Passos}}, \bibinfo {author} {\bibfnamefont {G.~B.}\ \bibnamefont {Ventura}}, \bibinfo {author} {\bibfnamefont {J.~M. V.~P.}\ \bibnamefont {Lopes}}, \bibinfo {author} {\bibfnamefont {J.~M. B.~L.}\ \bibnamefont {dos Santos}},\ and\ \bibinfo {author} {\bibfnamefont {N.~M.~R.}\ \bibnamefont {Peres}},\ }\bibfield  {title} {\bibinfo {title} {Nonlinear optical responses of crystalline systems: {{Results}} from a velocity gauge analysis},\ }\href {https://doi.org/10.1103/PhysRevB.97.235446} {\bibfield  {journal} {\bibinfo  {journal} {Phys. Rev. B}\ }\textbf {\bibinfo {volume} {97}},\ \bibinfo {pages} {235446} (\bibinfo {year} {2018})}\BibitemShut {NoStop}%
    \bibitem [{\citenamefont {Michishita}\ and\ \citenamefont {Peters}(2021)}]{Michishita2021Effects}%
    \BibitemOpen
    \bibfield  {author} {\bibinfo {author} {\bibfnamefont {Y.}~\bibnamefont {Michishita}}\ and\ \bibinfo {author} {\bibfnamefont {R.}~\bibnamefont {Peters}},\ }\bibfield  {title} {\bibinfo {title} {Effects of renormalization and non-{{Hermiticity}} on nonlinear responses in strongly correlated electron systems},\ }\href {https://doi.org/10.1103/PhysRevB.103.195133} {\bibfield  {journal} {\bibinfo  {journal} {Phys. Rev. B}\ }\textbf {\bibinfo {volume} {103}},\ \bibinfo {pages} {195133} (\bibinfo {year} {2021})}\BibitemShut {NoStop}%
    \bibitem [{\citenamefont {Terada}\ \emph {et~al.}(2024)\citenamefont {Terada}, \citenamefont {Kitamura}, \citenamefont {Watanabe},\ and\ \citenamefont {Ikeda}}]{Terada2024Limitations}%
    \BibitemOpen
    \bibfield  {author} {\bibinfo {author} {\bibfnamefont {I.}~\bibnamefont {Terada}}, \bibinfo {author} {\bibfnamefont {S.}~\bibnamefont {Kitamura}}, \bibinfo {author} {\bibfnamefont {H.}~\bibnamefont {Watanabe}},\ and\ \bibinfo {author} {\bibfnamefont {H.}~\bibnamefont {Ikeda}},\ }\bibfield  {title} {\bibinfo {title} {Limitations and improvements of the relaxation time approximation in the quantum master equation: {{Linear}} conductivity in insulating systems},\ }\href {https://doi.org/10.1103/PhysRevB.109.L180302} {\bibfield  {journal} {\bibinfo  {journal} {Phys. Rev. B}\ }\textbf {\bibinfo {volume} {109}},\ \bibinfo {pages} {L180302} (\bibinfo {year} {2024})}\BibitemShut {NoStop}%
    \bibitem [{\citenamefont {Stefanucci}\ and\ \citenamefont {Van~Leeuwen}(2013)}]{Stefanucci2013Nonequilibrium}%
    \BibitemOpen
    \bibfield  {author} {\bibinfo {author} {\bibfnamefont {G.}~\bibnamefont {Stefanucci}}\ and\ \bibinfo {author} {\bibfnamefont {R.}~\bibnamefont {Van~Leeuwen}},\ }\href {https://doi.org/10.1017/CBO9781139023979} {\emph {\bibinfo {title} {Nonequilibrium {{Many-Body Theory}} of {{Quantum Systems}}: {{A Modern Introduction}}}}},\ \bibinfo {edition} {1st}\ ed.\ (\bibinfo  {publisher} {Cambridge University Press},\ \bibinfo {address} {Cambridge},\ \bibinfo {year} {2013})\BibitemShut {NoStop}%
    \bibitem [{\citenamefont {Bhalla}\ \emph {et~al.}(2020)\citenamefont {Bhalla}, \citenamefont {MacDonald},\ and\ \citenamefont {Culcer}}]{Bhalla2020Resonant}%
    \BibitemOpen
    \bibfield  {author} {\bibinfo {author} {\bibfnamefont {P.}~\bibnamefont {Bhalla}}, \bibinfo {author} {\bibfnamefont {A.~H.}\ \bibnamefont {MacDonald}},\ and\ \bibinfo {author} {\bibfnamefont {D.}~\bibnamefont {Culcer}},\ }\bibfield  {title} {\bibinfo {title} {Resonant {{Photovoltaic Effect}} in {{Doped Magnetic Semiconductors}}},\ }\href {https://doi.org/10.1103/PhysRevLett.124.087402} {\bibfield  {journal} {\bibinfo  {journal} {Phys. Rev. Lett.}\ }\textbf {\bibinfo {volume} {124}},\ \bibinfo {pages} {087402} (\bibinfo {year} {2020})}\BibitemShut {NoStop}%
    \bibitem [{\citenamefont {Bhalla}\ \emph {et~al.}(2022)\citenamefont {Bhalla}, \citenamefont {Das}, \citenamefont {Culcer},\ and\ \citenamefont {Agarwal}}]{Bhalla2022Resonant}%
    \BibitemOpen
    \bibfield  {author} {\bibinfo {author} {\bibfnamefont {P.}~\bibnamefont {Bhalla}}, \bibinfo {author} {\bibfnamefont {K.}~\bibnamefont {Das}}, \bibinfo {author} {\bibfnamefont {D.}~\bibnamefont {Culcer}},\ and\ \bibinfo {author} {\bibfnamefont {A.}~\bibnamefont {Agarwal}},\ }\bibfield  {title} {\bibinfo {title} {Resonant {{Second-Harmonic Generation}} as a {{Probe}} of {{Quantum Geometry}}},\ }\href {https://doi.org/10.1103/PhysRevLett.129.227401} {\bibfield  {journal} {\bibinfo  {journal} {Phys. Rev. Lett.}\ }\textbf {\bibinfo {volume} {129}},\ \bibinfo {pages} {227401} (\bibinfo {year} {2022})}\BibitemShut {NoStop}%
\end{thebibliography}
\end{document}